\documentclass[preprint2, tighten]{aastex6}
\begin{document}

\title{A statistical analysis of the accuracy of the digitized magnitudes of photometric plates on the time scale of decades with an application to the century-long light curve of KIC 8462852}
\shorttitle{A statistical analysis of the stability of photometric plates}
\shortauthors{Hippke et al.}

\author{Michael Hippke}
\email{hippke@ifda.eu}
\affil{Luiter Stra{\ss}e 21b, 47506 Neukirchen-Vluyn, Germany}

\author{Daniel Angerhausen}
\affil{NASA Postdoctoral Program Fellow, NASA Goddard Space Flight Center, Exoplanets \& Stellar Astrophysics Laboratory, Code 667, Greenbelt, MD 20771}
\email{daniel.angerhausen@nasa.gov}

\author{Michael B. Lund}
\affil{Department of Physics and Astronomy, Vanderbilt University, Nashville, TN 37235, USA}
\email{michael.b.lund@vanderbilt.edu}

\author{Joshua Pepper}
\affil{Department of Physics, Lehigh University, 16 Memorial Drive East, Bethlehem, PA 18015, USA}
\email{joshua.pepper@lehigh.edu}

\author{Keivan G. Stassun}
\affil{Department of Physics and Astronomy, Vanderbilt University, 6301 Stevenson Center, Nashville, TN 37235, USA}
\email{keivan.stassun@vanderbilt.edu}

\begin{abstract}
We present a statistical analysis of the accuracy of the digitized magnitudes of photometric plates on the time scale of decades. In our examination of archival Johnson B photometry from the Harvard DASCH archive, we find a median RMS scatter of lightcurves of order 0.15mag over the range $B\sim9-17$ for all calibrations. Slight underlying systematics (trends or flux discontinuities) are on a level of $\lesssim0.2$mag per century (1889--1990) for the majority of constant stars. These historic data can be unambiguously used for processes that happen on scales of magnitudes, and need to be carefully examined in cases approaching the noise floor. The characterization of these limits in photometric stability may guide future studies in their use of plate archives. We explain these limitations for the example case of KIC8462852, which has been claimed to dim by $0.16$mag per century, and show that this trend cannot be considered as significant.
\end{abstract}

\section{Introduction}
From 1886 to 1992, the Harvard College Observatory observed large portions of the sky repeatedly, with the observations saved on over half a million photographic plates produced by different telescopes and observatories. The photographs constitute of glass plates with an emulsion most sensitive in the blue, which consitute the basis for the later Johnson B magnitude system \citep{1953ApJ...117..313J}. The plates were digitized with a high speed scanner \citep{2006SPIE.6312E..17S}. The initial photometry and astrometry are described in \citet{2010AJ....140.1062L} and \citet{2011ASPC..442..273S}, and the current progress is summarized in \citet{2013PASP..125..857T}.

The digitization of the Harvard Astronomical Plate Collection (DASCH, Digital Access to a Sky Century @ Harvard) \citep{2009ASPC..410..101G, 2010AJ....140.1062L,2012IAUS..285..243G,2012IAUS..285...29G}) is an extraordinarily important project. The long-term light curves are and will be invaluable for comparison with data from ongoing missions such as K2, and upcoming space-based photometry missions such as TESS and PLATO, as the long time baseline of DASCH reference gives leverage in time like no other data set. Great efforts have been made by the DASCH team to digitize and calibrate the glass plates.  The volume and the quality achieved is nothing short of impressive. As of 02 February 2016, 136,949 plates have been scanned; and 8,540,081,000 magnitudes have been measured. This work is unprecedented and of the greatest importance for the astrophysical community. Recent studies have made use of these data in the fields of GRBs \citep{2013AcPol..53c..27H}, RR Lyrae \citep{2015OEJV..173....1L}, LBVs \citep{2015A&A...581A..12B, 2014ATel.6295....1W}, AGN flares \citep{2015AAS...22532003G}, stellar activity cycles \citep{2014CoSka..43..411O}, novae \citep{2012ApJ...751...99T}, dust accretion events \citep{2011ApJ...738....7T}, long-period eclipsing binary systems \citep{2016arXiv160100135R}, or long-term amplitudes in K giants \citep{2010ApJ...710L..77T}. 

These studies focus on processes that happen on scales of magnitudes, and there is as of yet no dedicated study to explore the long-term accuracy of these data. In this article, we examine the stability and distribution of brightness measurements. As an immediate application, we compare our limits and uncertainties with a recent study on KIC8462852, which has been claimed to dim by $0.16$mag per century \citep{2016arXiv160103256S} (in the following: S16). This F3 main-sequence star KIC8462852 was observed by the NASA Kepler mission from April 2009 to May 2013, and an analysis by \citet{2016MNRAS.457.3988B} shows unusual series of brightness dips up to 20\%.  This behavior has been theorized to originate from a family of large comets \citep{2015arXiv151108821B}, or signs of a Dyson sphere \citep{2016ApJ...816...17W}. Subsequent analysis found no narrow-band radio signals \citep{2015arXiv151101606H} and no periodic pulsed optical signals \citep{2015arXiv151202388S,2016arXiv160200987A}. The infrared flux is equally unremarkable \citep{2015ApJ...815L..27L, 2015ApJ...814L..15M, 2015arXiv151203693T}.

\section{Method}
\label{sec:trend}
\subsection{Selecting benchmark stars}
We select comparison stars using three different, complementary approaches. The first tests a wide range of $g$-band magnitudes ($9<g<14$) to explore a potential correlation between brightness and long-term slope and uses the KIC calibration of the DASCH data. The results are presented in section~\ref{sec:widecompare}.

Afterwards, we focus on a benchmark which is more closely matched to our example case of KIC8462852. This time, we use the APASS calibration and select the stars most similar to KIC8462852, with Kepler magnitude $11.5<K_P<12.2$, surface gravity $3.8<$log~$g<4.2$ (dwarfs) and temperature $6700<T_{eff}<6900$ (F-stars) to match the general properties of KIC8462852. We also checked the stars are constant during the 4.25 years of Kepler observations, with amplitudes (e.g. due to rotation) $<1$\% and trends $<0.1$\%. We obtained APASS colors of these stars, and selected only those with $11.5<B<12.5$ and $0.4<B-V<0.7$. 

Finally, we examine the Landolt standard stars \citep{1992AJ....104..340L,2013AJ....146..131L} to assure testing a sample of well-known constant stars. The results are presented in section~\ref{sec:landoltcompare}.

\subsection{Retrieval of scanned photometry}
\label{subsec:scanned}
The DASCH data are available online\footnote{http://dasch.rc.fas.harvard.edu/lightcurve.php, retrieved on 01-Feb 2016} in multiple versions, each calibrated against a different set of absolute photometric standards.  There are versions calibrated against the Kepler Input Catalog (KIC, \citet{2011AJ....142..112B}), the Guide Star Catalog (GSC, \citet{1990AJ.....99.2019L, 2008AJ....136..735L}), and the AAVSO Photometric All-Sky Survey (APASS, \citet{2011SASS...30..121S, 2012JAVSO..40..430H}). We downloaded the data files based on each of the three calibrations, and following the procedures described by S16, selected all observations, excluding data from Yellow and Red plates.  
  
Following \S§2.1 in S16 we make the following additional cuts:

\begin{itemize}
\item[a)] Identify and remove all values with quality flag indicator ``AFLAGS''$>9000$. AFLAGS are pipeline flags to denote possible problems such as blends, possible plate defects (24 separate criteria) where quality is potentially compromised.\footnote{http://dasch.rc.fas.harvard.edu/database.php, retrieved on 01-Feb 2016}

For a specific removal of selected AFLAG criteria, the AFLAG word has to be unpacked to reveal which bits are set. A tool will be provided in the upcoming DR5 data release for users to do this.

\item[b)] Remove all data values with one-sigma error bars $>0.33$mag.
\item[c)] Remove all data values with the flux within 0.2 magnitudes of the quoted plate limit, with the limiting magnitude taken from the DASCH data as defined in \citet{2013PASP..125..857T}.
\end{itemize}

Following these criteria, we show the number of data points for the different steps in Table~\ref{tab:cleaning}. Again, we note that the quality flag indicator ``AFLAGS''$>9000$ is ill-defined, but required to reproduce the cleaning in S16.

Since the light curves using the APASS calibration yields the largest number of usable observations in almost all cases, we will use these light curves for the analysis below unless noted otherwise.

We followed previous papers (e.g., \citet{2016arXiv160103256S,2013PASP..125..793T}) in opting for a conservative cut on AFLAGS, although this means sacrificing data points and thus increasing errors on least-squares fits which then have fewer data points that typically have $\sim0.12$mag errors.

\subsection{Test for normality}
We test the DASCH light curves for normality \citep{Shapiro}, before and after the cleaning steps listed in section~\ref{subsec:scanned}. As an example, we present the results for KIC8462852. As is well known, error estimates in linear least-squares regressions are only valid for Gaussian distributions. Consequently, it is advised to use regression analytics that account for such a distribution, if this is relevant in a particular science case.

\subsection{Linear trends and flux discontinuities}
In order to test the stability of the stars in question, we fit linear regressions and estimate the uncertainties. We do this for large populations (hundreds of stars) and generate probability distributions (section~\ref{sec:kiccompare}). In the next section, we will continue with the selection an of adequate regression method. We will also test the hypothesis of a sudden change in the flux zero point (a flux discontinuity); the background of this will be explained in section~\ref{sec:break_statistics}.

\section{Results}

\subsection{Regression diagnostics}
\label{subsec:regressions}
As an example case, we test the normality of KIC8462852 using the methods mentioned in the previous section. The test for normality rejects a Gaussian distribution at very high significance. Consequently, we perform different linear regression methods and compare the results.

For the original light curve (with 1953 observations, Table~\ref{tab:cleaning}), we fit a linear regression and use normal error estimates. Without binning, we obtain a slope of $+0.12\pm0.02$ mag per century.

After cleaning the light curve, we fit a linear regression and get a formal slope of $+0.14\pm0.016$ mag per century. As noted, these errors are incorrect due to non-normality.

To estimate errors in the non-Gaussian data, we can use a robust regression (e.g., \citet{McKean}), which gives $+0.13\pm0.018$ mag per century.

Finally, we can test the effect of 5-year bins as used in S16. With that modification, we find a slope of  $+0.12\pm0.02$ mag per century.

All regressions present significant slopes similar to that reported by S16. For the following tests of hundreds of stars, we will employ simple linear regressions. Compared to robust regressions, and as seen above, this under-estimates the errors typically by $\sim10\%$; we judge this error as irrelevant given the systematic uncertainties presented in the following sections.

\begin{table*}
\center
\caption{Selection and cleaning of DASCH data for KIC8462852 \label{tab:cleaning}}
\begin{tabular}{lrrr}
\tableline
Calibration & APASS & GSC & KIC \\
\tableline
Nonzero flux values &  1953 & 1809 & 1234 \\
After removing points with bad data flags & 1609 & 993 & 976 \\
After removing points with large errors & 1470 & 694 & 708 \\
\tableline
Average B-magnitude of remaining points & 12.37 & 12.26 & 12.06\\
\tableline
(After removing all AFLAGS and all negative BFLAGS, section~\ref{sec:strict}) & 689 & 337 & 343 \\
\tableline

\end{tabular}
\end{table*}

\begin{figure*}
\includegraphics[width=0.5\textwidth]{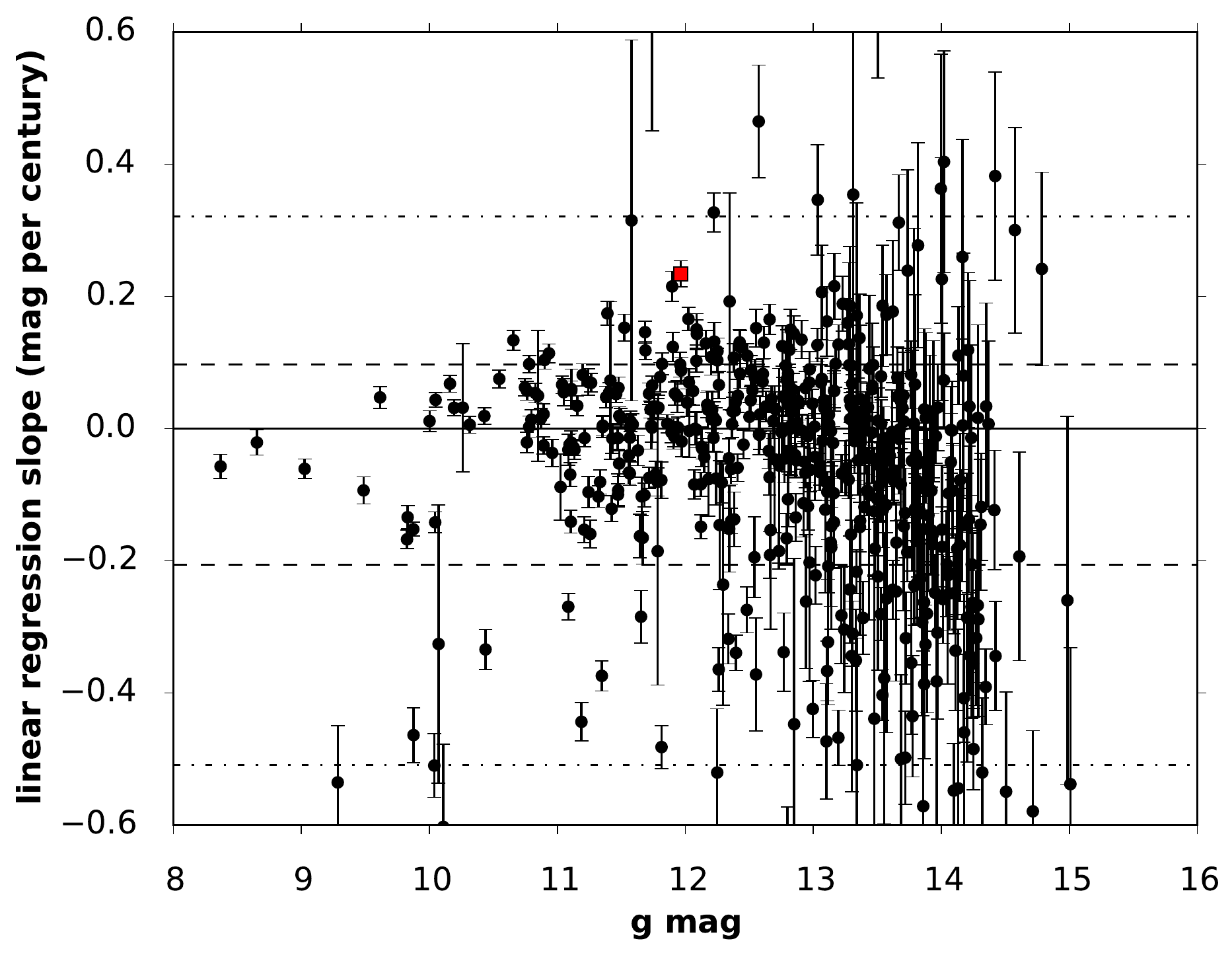}
\includegraphics[width=0.5\textwidth]{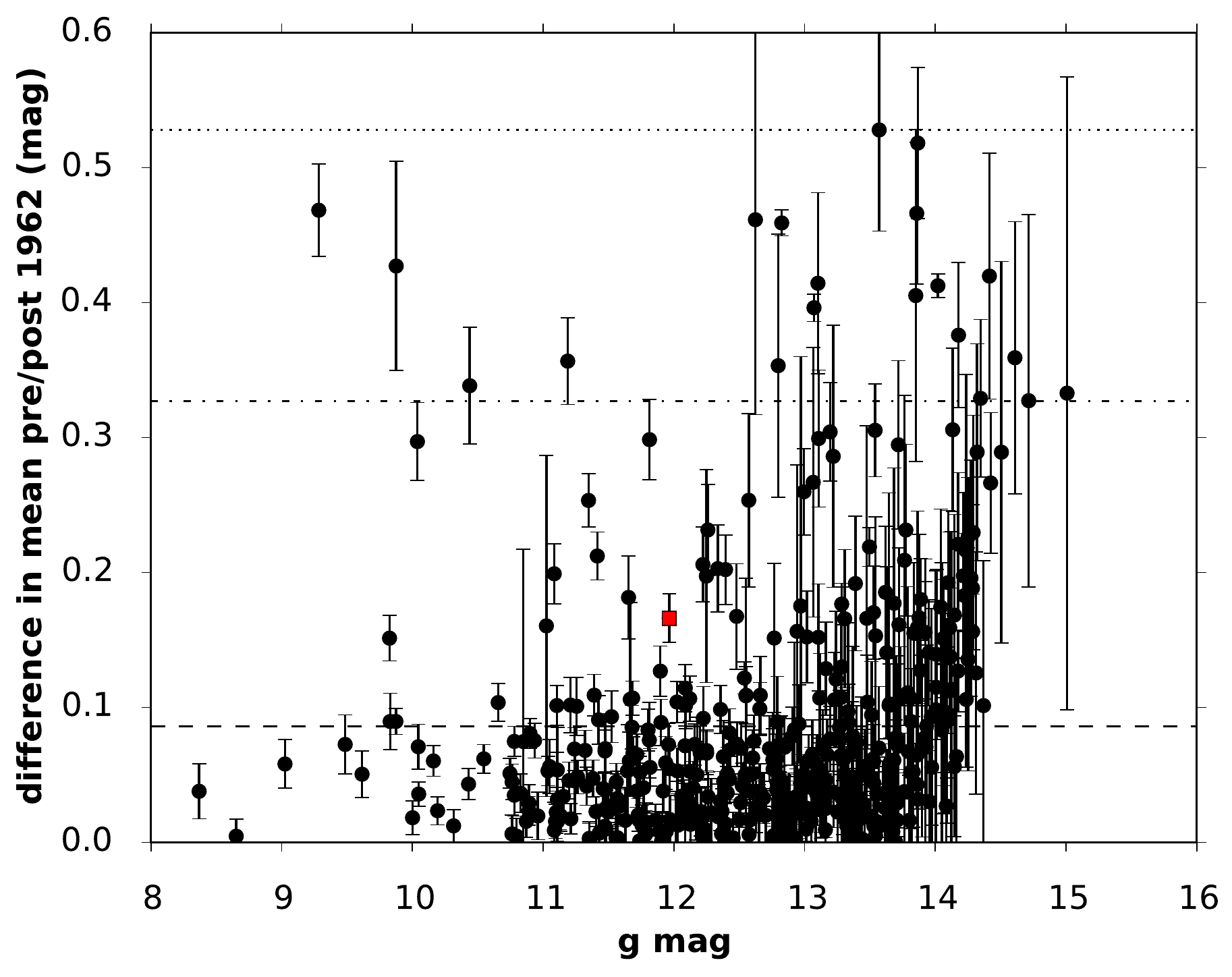}
\caption{\label{fig:lund}DASCH KIC calibrated data. Left: Linear regression slopes of 513 DASCH stars (black circles) and ``Tabby's star'' from the same data (red square). The horizonzal lines indicating the maximum slopes for 68\% ($1\sigma$, dashed) and 95\% ($2\sigma$, dash-dotted) of all stars. Right: Absolute value of the difference in means before and after 1962, for the same dataset. Horizontal lines show the $1\sigma$, $2\sigma$, and $3\sigma$ regions of maximum absolute differences. The average number of data points per lightcurve was $n=832\pm161$}.
\end{figure*}

\subsection{Benchmarking a wide range of $g$-band magnitudes}
\label{sec:widecompare}
We now test a wide range of $g$-band magnitudes ($9<g<14$) to explore a potential correlation between brightness and long-term slope and uses the KIC calibration of the DASCH data. From the Kepler Input Catalog, we select stars similar to KIC8462852 by requiring the temperature to be within 100K ($6484<T_{eff}<6684$), the stellar radius within 5\% ($1.614<R_{\odot}<1.784$), and log~$g$ within 10\% ($3.712<$log~$g$$<4.536$). We performed the usual data cleansing described in S16. We removed a few stars that have less than 10 DASCH observations to avoid very large error bars. The result in Figure~\ref{fig:lund} (left panel) shows the slopes and discontinuities (right panel). The results are virtually identical with, or without removal of red and yellow plates.

\begin{figure*}
\includegraphics[width=0.5\textwidth]{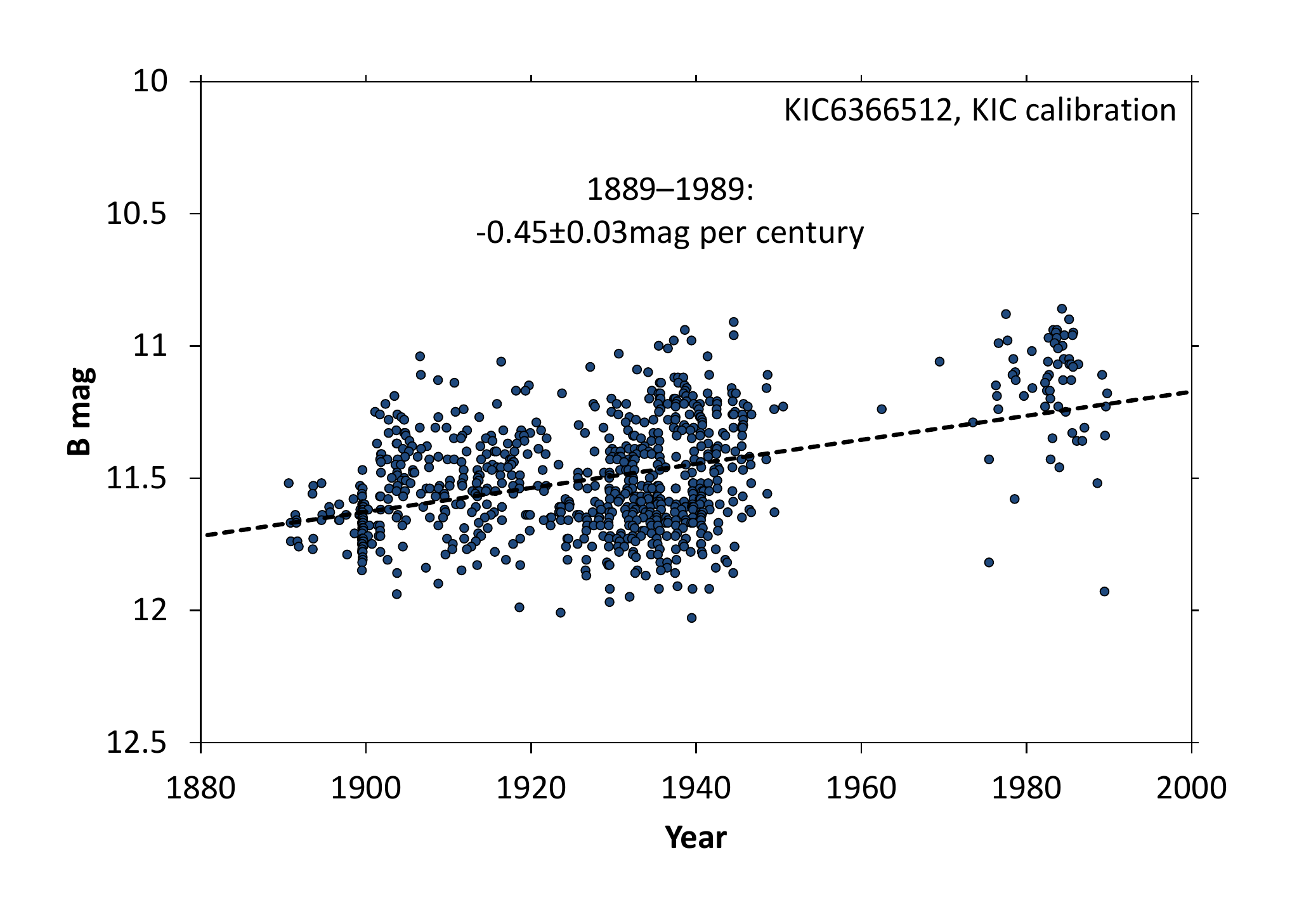}
\includegraphics[width=0.5\textwidth]{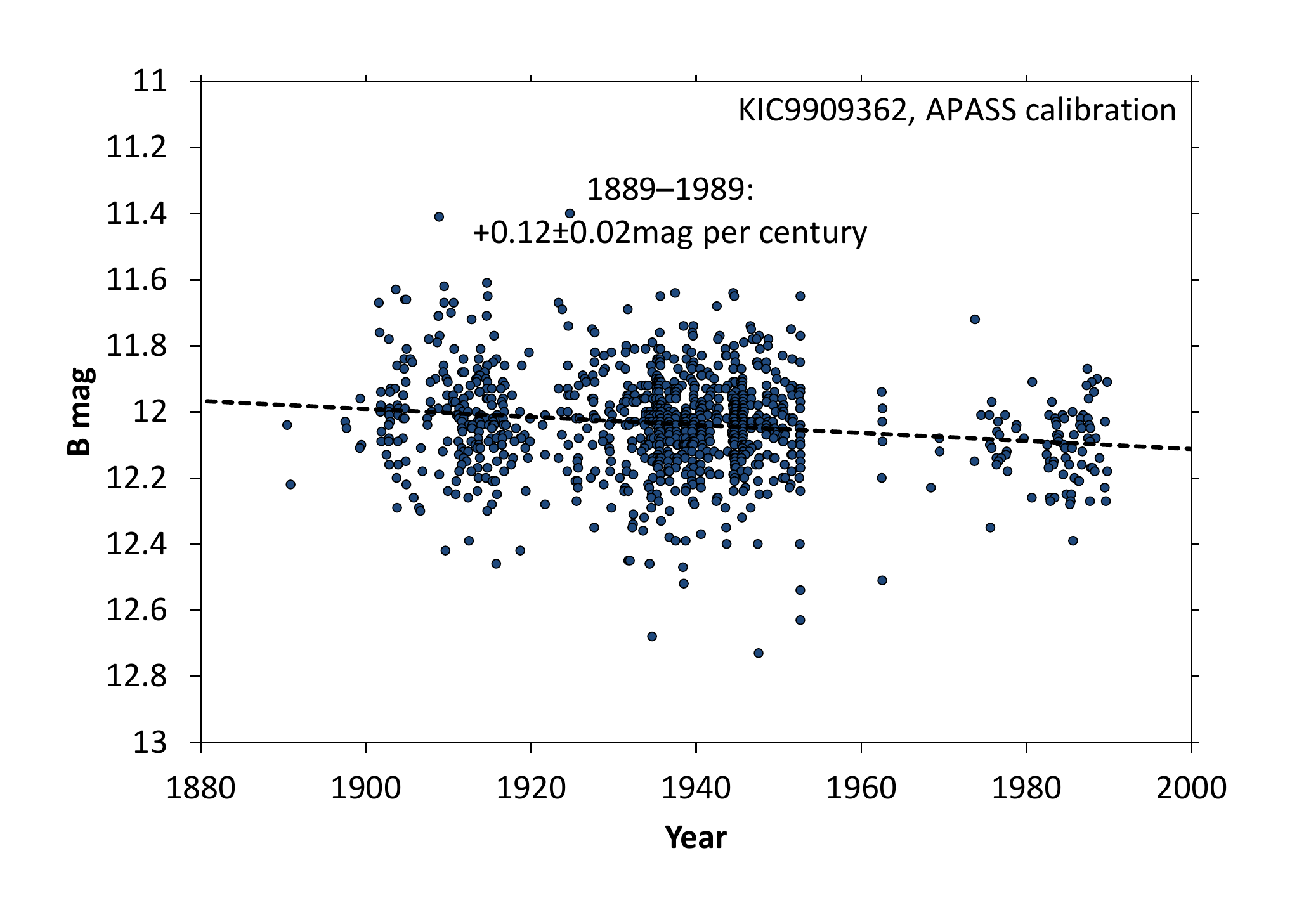}
\caption{\label{fig:KIC6366512}DASCH KIC calibrated data for KIC6366512 (left, $n=832$) and APASS calibrated data for KIC9909362 (right, $n=1181$) with data cleansing as performed by S16. The trends are highly significant, although flux discontinuities are statistically preferred.}
\end{figure*}

\begin{figure*}
\includegraphics[width=0.5\textwidth]{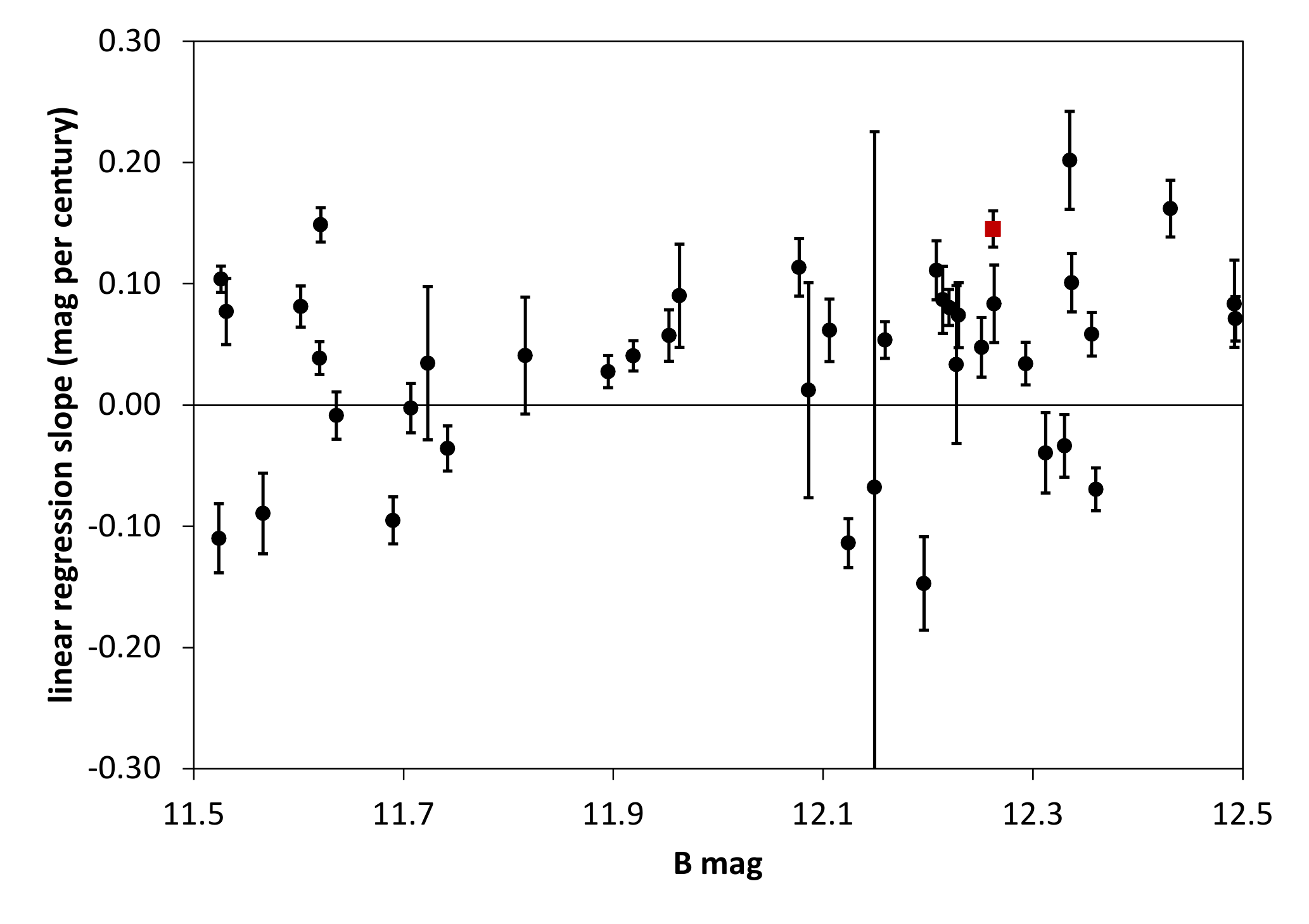}
\includegraphics[width=0.5\textwidth]{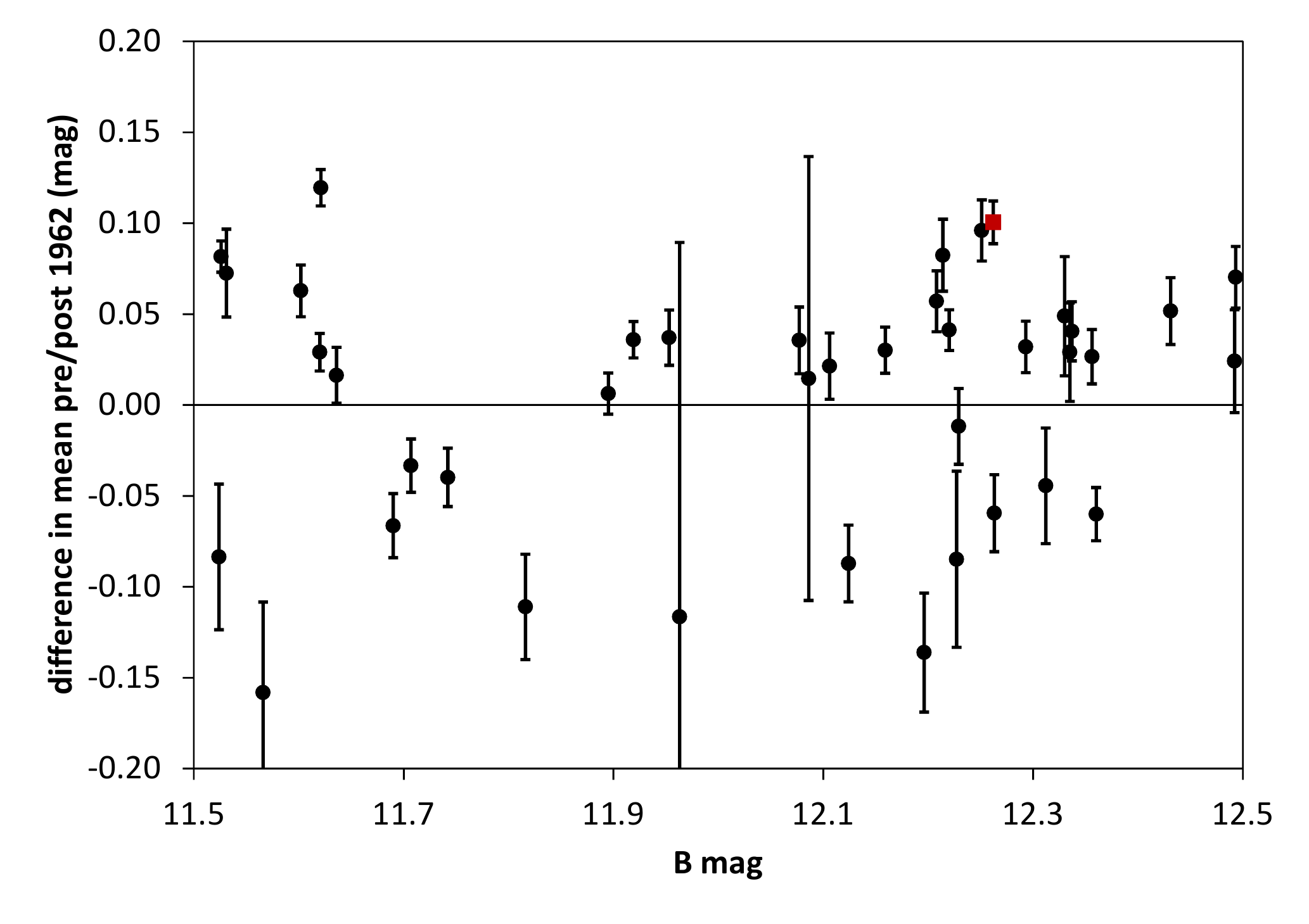}

\includegraphics[width=0.5\textwidth]{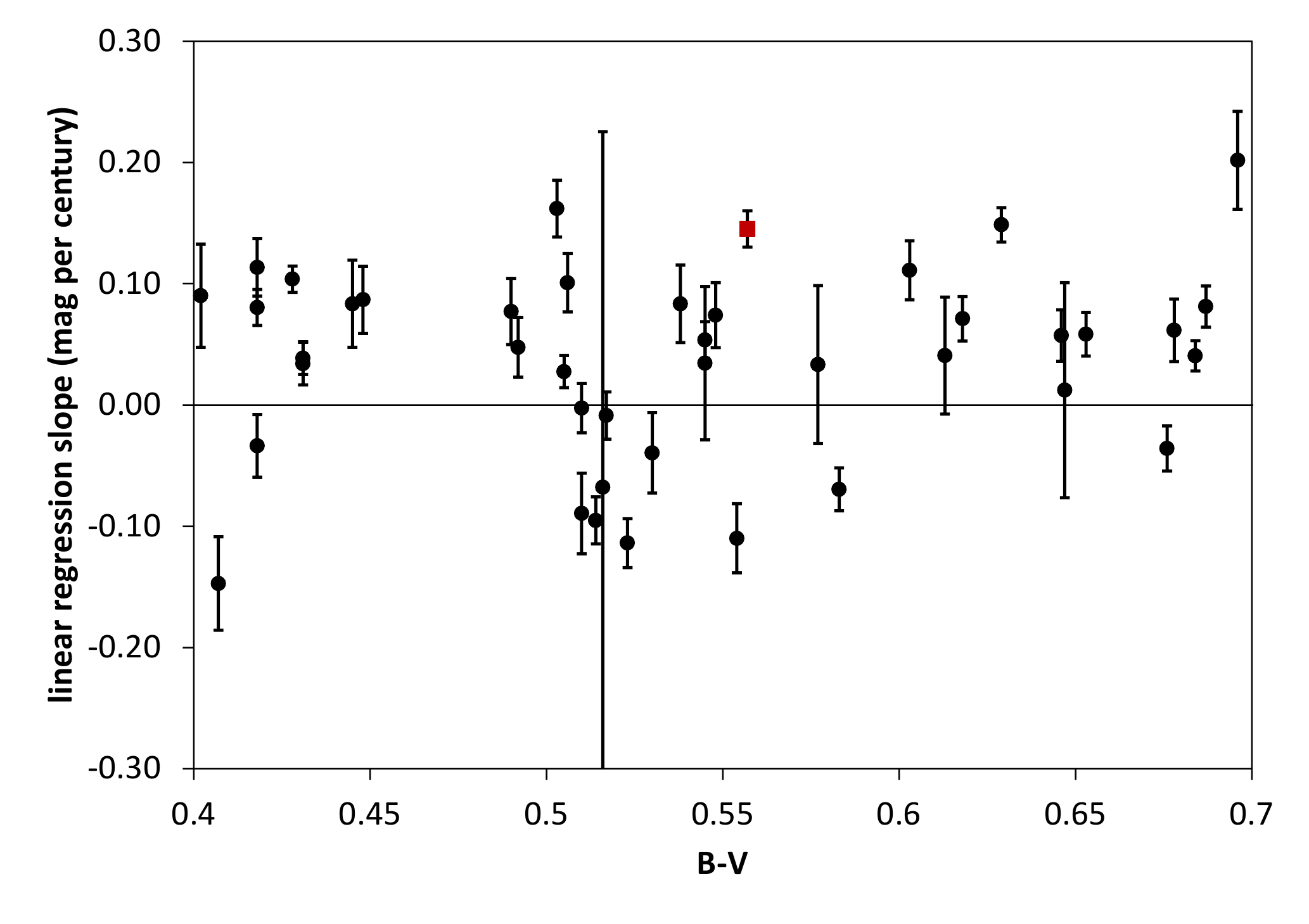}
\includegraphics[width=0.5\textwidth]{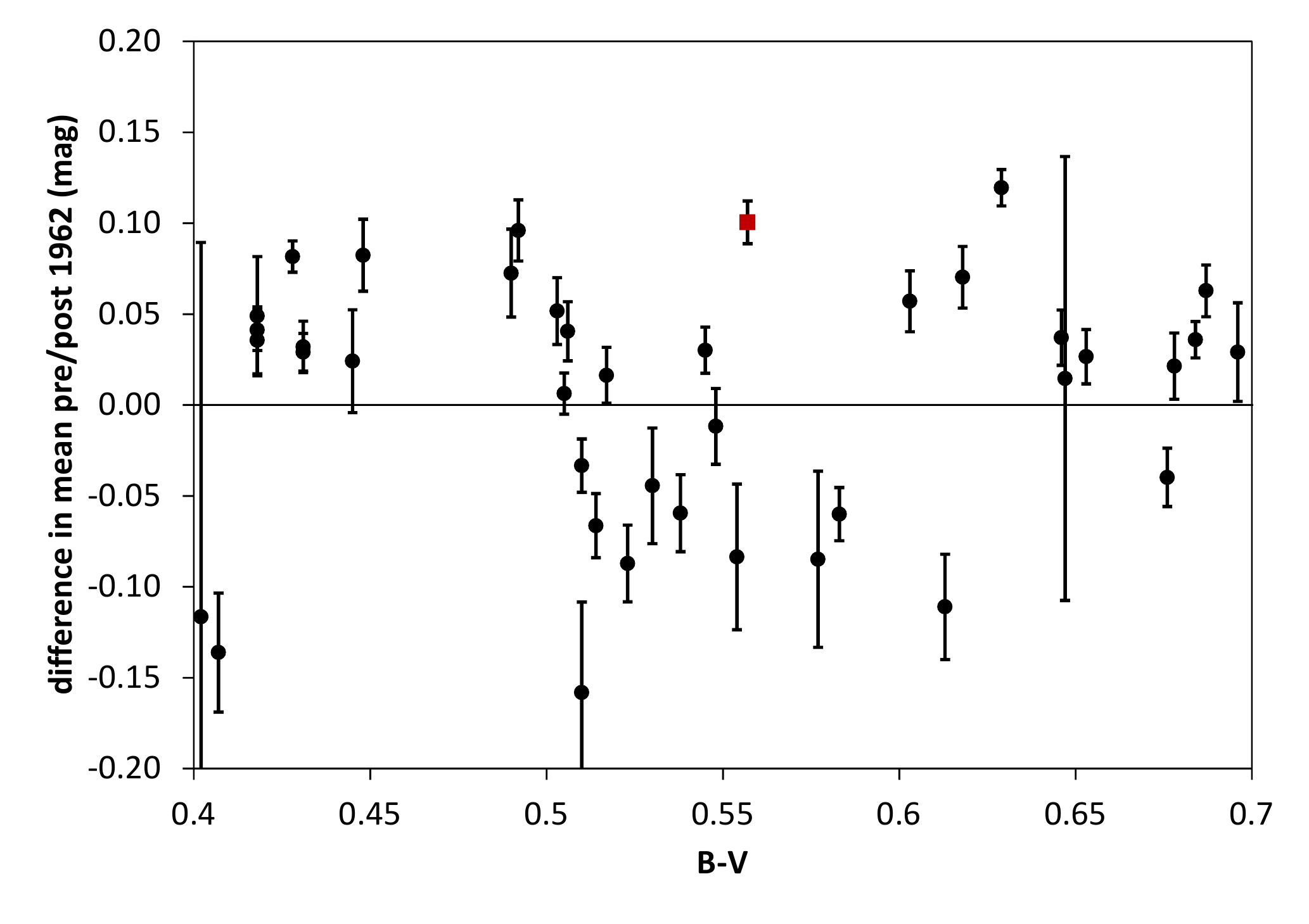}

\includegraphics[width=0.5\textwidth]{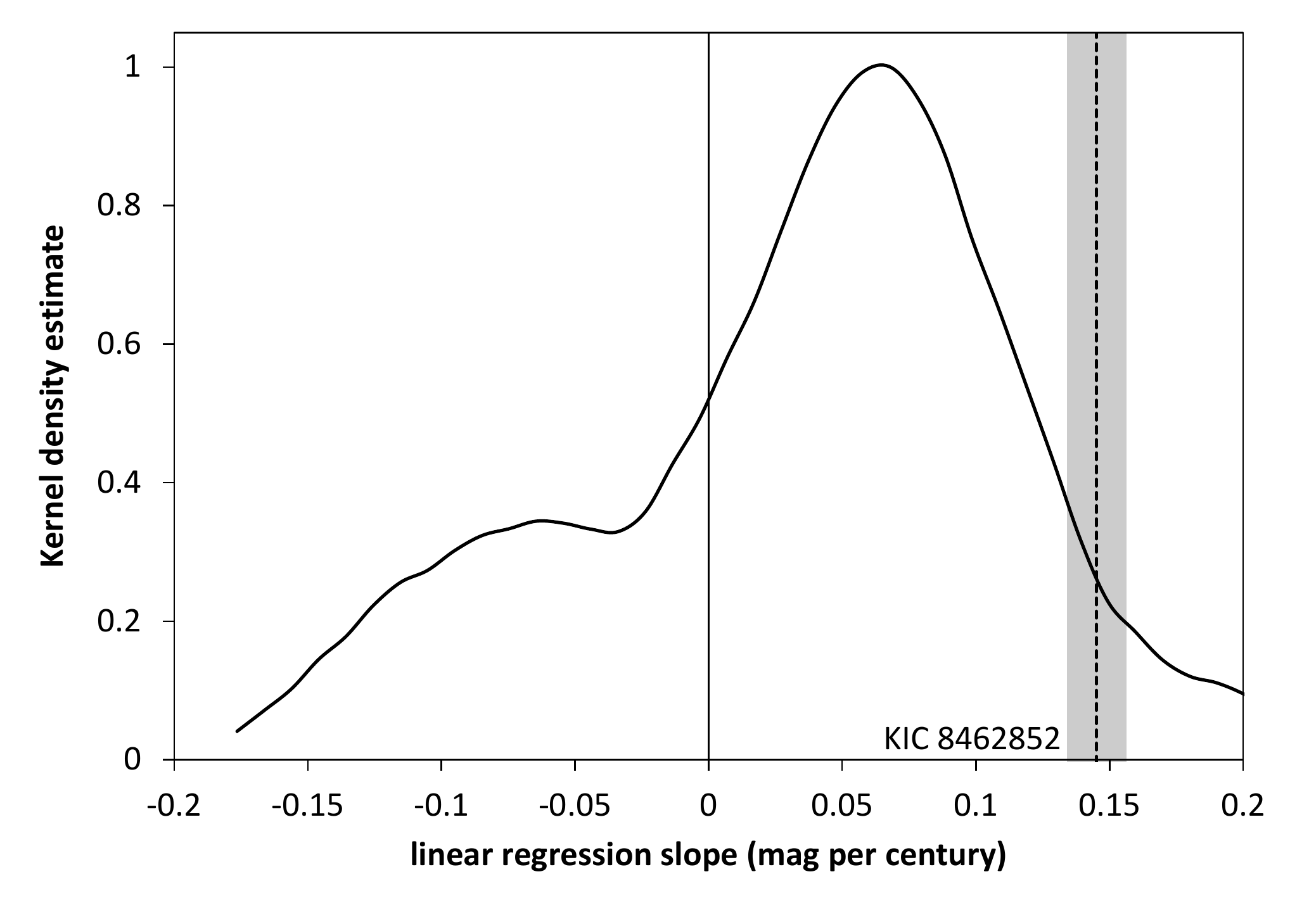}
\includegraphics[width=0.5\textwidth]{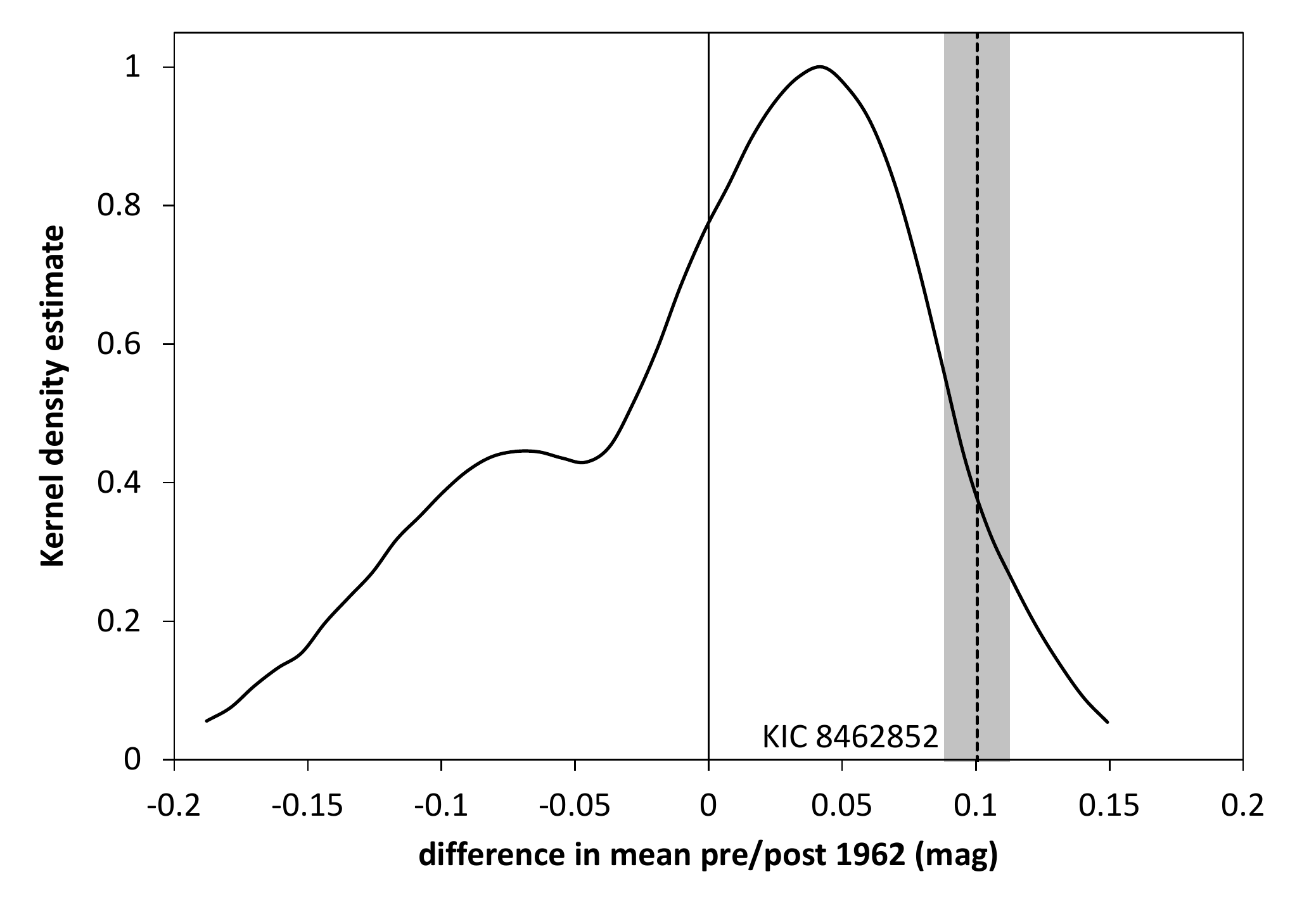}
\caption{\label{fig:dist}DASCH APASS calibrated data. Left side: Linear regression slopes of 41 stars (black circles) and ``Tabby's star'' (red square) versus $B$ magnitude (top) and $B-V$ (middle). The bottom panel shows the probability density. We use the Kernel density \citep{Rosenblatt1956,Parzen1962} with a non-parametric Gaussian Kernel to estimate the probability density function of the distributions, which has the advantage of smoothness (continuity), in contrast to the conventional histogram, which would also require bin size decisions. Right side: The same for differences in mean before and after 1962. The average number of data points per lightcurve was $n=721\pm123$. As noted in the text, the large errors come from over-rejection and too few data points for a meaningful least-squares slope.}
\end{figure*}

\subsection{Benchmarking stars similar to KIC8462852}
\label{sec:kiccompare}
We also tested stars similar to KIC8462852. We selected from the Kepler Input Catalog stars within three degrees of KIC8462852, with Kepler magnitude $11.5<K_P<12.2$, surface gravity $3.8<$log~$g<4.2$ (dwarfs) and temperature $6700<T_{eff}<6900$ (F-stars) to match the general properties of KIC8462852. We also checked the stars are constant during the 4.25 years of Kepler observations, with amplitudes (e.g. due to rotation) $<1$\% and trends $<0.1$\%. We obtained APASS colors of these stars, and selected only those with $11.5<B<12.5$ and $0.4<B-V<0.7$. 

We remove red and yellow plates, and apply the data quality cuts listed in section~\ref{subsec:scanned}. We show two examples of stars with significant slopes in Figure~\ref{fig:KIC6366512}. The left panel presents KIC6366512 (B$=11.8$mag, $B-V=0.3$). This star is constant within 0.2\% rms during the 4.25 years of Kepler observations. In the DASCH data, it shows a brightening trend of $-0.45\pm0.03$mag per century (or a flux discontinuity at very high significance). 

A second example is shown in the right panel of Figure~\ref{fig:KIC6366512} for KIC9909362, this time in the APASS calibration. KIC9909362 is a constant F-star with B$=12.2$mag, $B-V=0.7$, very similar to KIC8462852. We performed the same data cleansing and find a dimming trend with slope $0.12\pm0.02$mag per century.

The slopes for all 41 stars in this selection is shown in Figure~\ref{fig:dist} (left). We also split the light curves for each star at the year 1962 and calculate the mean for each segment (Figure~\ref{fig:dist}, right). The error bars are the added uncertainties of both segments. The plots show KIC8462852 is placed towards the end of the distribution (i.e. its slope/break is relatively high), with comparably low error bars. Still, 4 of 41 comparison stars have higher absolute slopes, and 5 of 41 have higher absolute flux discontinuities. Therefore, KIC8462852 does not appear to be extraordinary by these measures. Instead, it is within the range of linear trends and flux discontinuities for normal Harvard DASCH photometry.

The probability distributions in Figure~\ref{fig:dist} appear, at first glance, asymmetric. We believe that the underlying calibration is, however, Gaussian. The effect is likely to come from crowding and/or blending, which will increase systematic errors. Therefore, a 3-sigma criterion for discovery of variability is not adequate. 

The evidence is strong for a systematic step at the year 1962 in the case of ``Tabby's star''; it is not yet clear how general this effect occurs in the data. The Kepler field is significantly more crowded than most of the Landolt standard star fields. Grindlay and Los 2016 (in prep) shall also show the importance of this, and will also deal with the possible causes of the (sometimes) ``step'' in derived magnitudes post-1962, which may contribute to the ``tail'' and ``second bump'' in the probability density distribution shown in the left side of Figure~\ref{fig:dist} (bottom panel) of this paper.

\begin{figure*}
\includegraphics[width=0.5\textwidth]{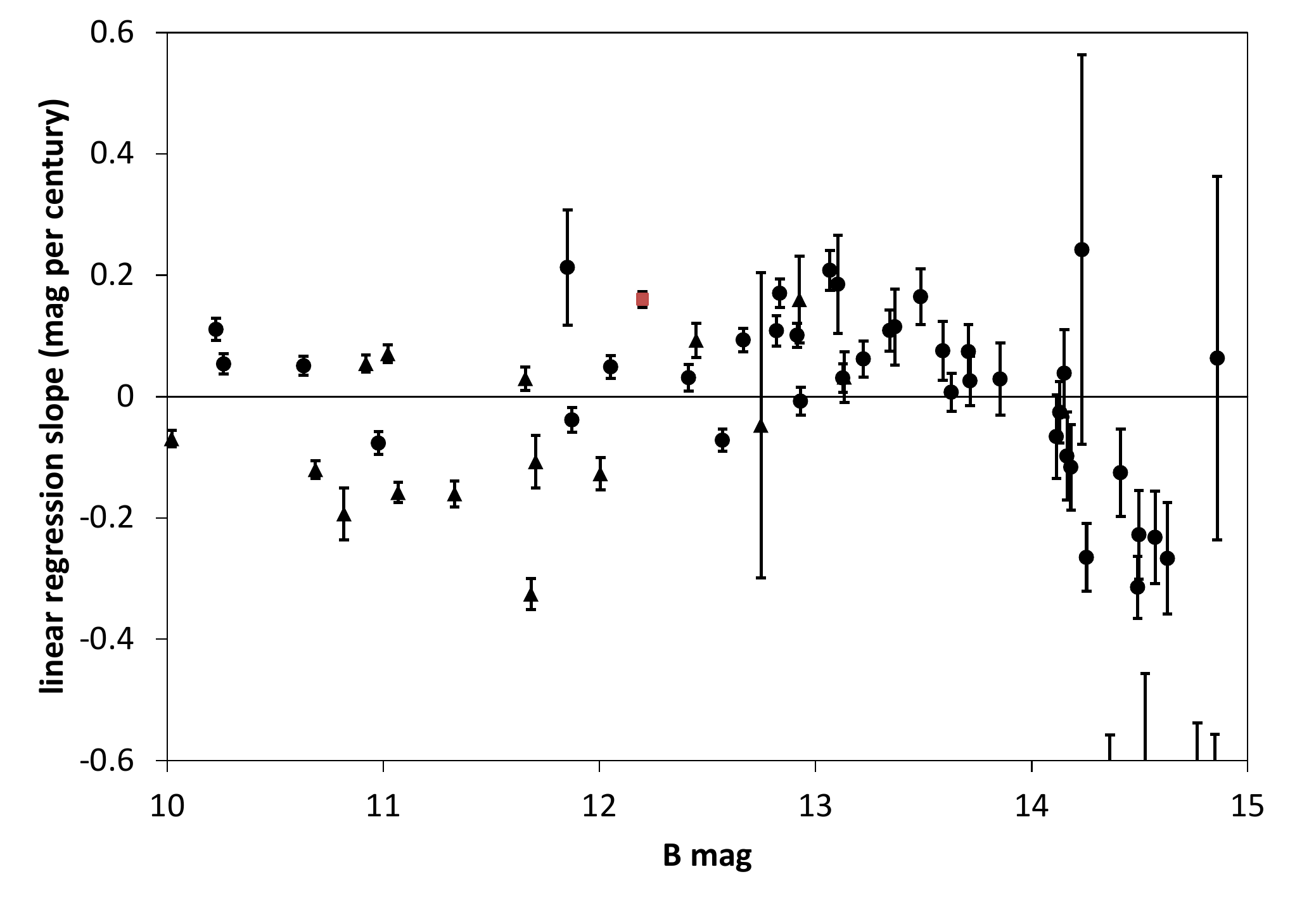}
\includegraphics[width=0.5\textwidth]{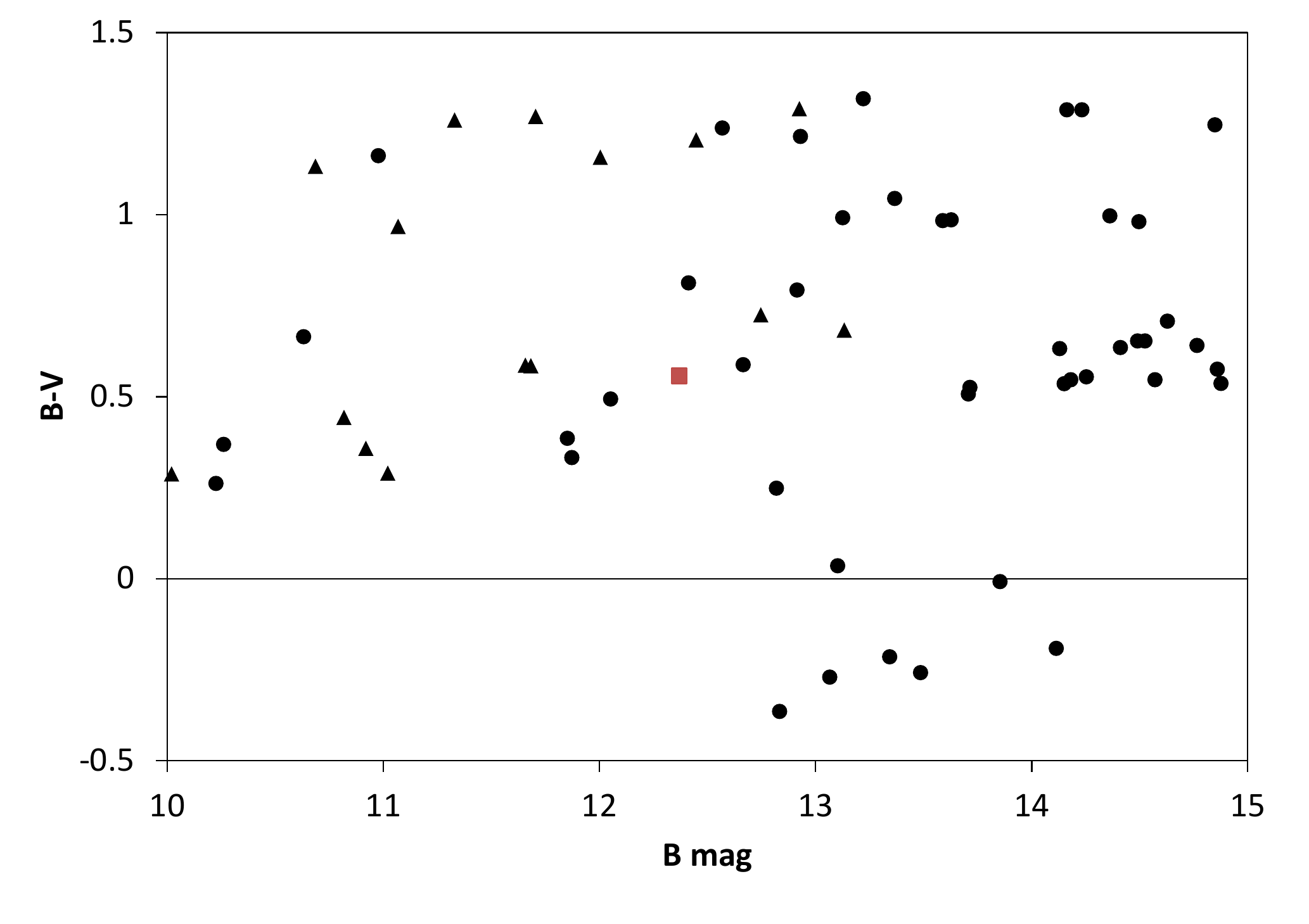}
\caption{\label{fig:landolt}Left: $B$ Magnitude versus linear trend for the Landolt standard stars from \citet{1992AJ....104..340L} (circle symbols) and \citet{2013AJ....146..131L} (triangles). All data values with any flags have been removed before the fitting. KIC8462852 (square) does not appear extraordinary among these benchmark stars. Right: $B$ magnitude versus color ($B-V$) with the same symbol coding. The parameter space is well sampled by these stars. The average number of data points per lightcurve was $n=687\pm144$. As noted in the text, the large errors come from over-rejection and too few data points for a meaningful least-squares slope.}
\end{figure*}

\subsection{Landolt standard stars}
\label{sec:landoltcompare}
A common photometric standard system was established by \citet{1992AJ....104..340L} in the Johnson-Kron-Cousins filters for 526 stars centered on the celestial equator. This extensive work was referenced in $\sim4000$ publications; its dataset provides 78 constant benchmark stars for which DASCH photometry is available. We have also added the stars from the follow-up work in \citet{2013AJ....146..131L}, which contains another 16 stars with DASCH photometry. In the data cleansing, this time we used a very strict approach and removed all data values with \textit{any} AFLAG and all warning BFLAG and kept only the good data points. After data cleansing, there were 72 stars left that could be used for fitting linear trends. We show the result in Figure~\ref{fig:landolt}. It is clear that most Landolt stars show linear trends, as was the case for other comparison stars in the previous section. The parameter space of $B$ magnitude and color $B-V$ is well-sampled, and again KIC8462852 does not appear extraordinary among these benchmark stars.

We note some possible improvements that may be used in future analyses. In particular, we note that the large deviations in Figure~\ref{fig:landolt} (left) are likely due to over-rejecting data points and having so few (down to 100) points that least-sqare fit errors are large, given the typical photometry errors of $\sim0.12$mag per point. This is almost certainly why the faint stars ($B\sim14-15$), with far fewer points in their lightcurves and the predominance of the shallow ac series plates for the Kepler field, have large deviations -- e.g. the 4 points with slopes $<-0.6$.

\begin{figure*}
\includegraphics[width=0.5\textwidth]{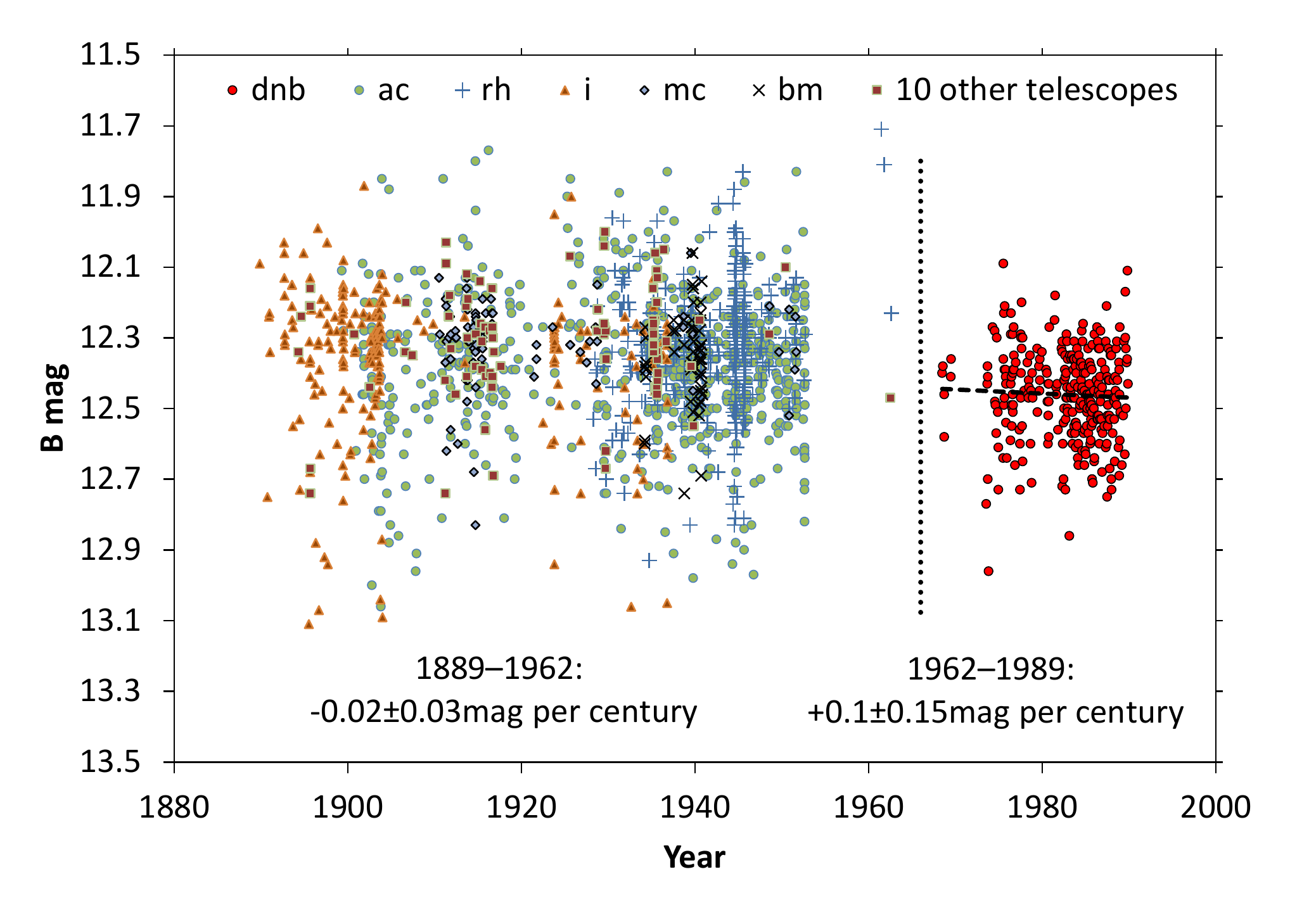}
\includegraphics[width=0.5\textwidth]{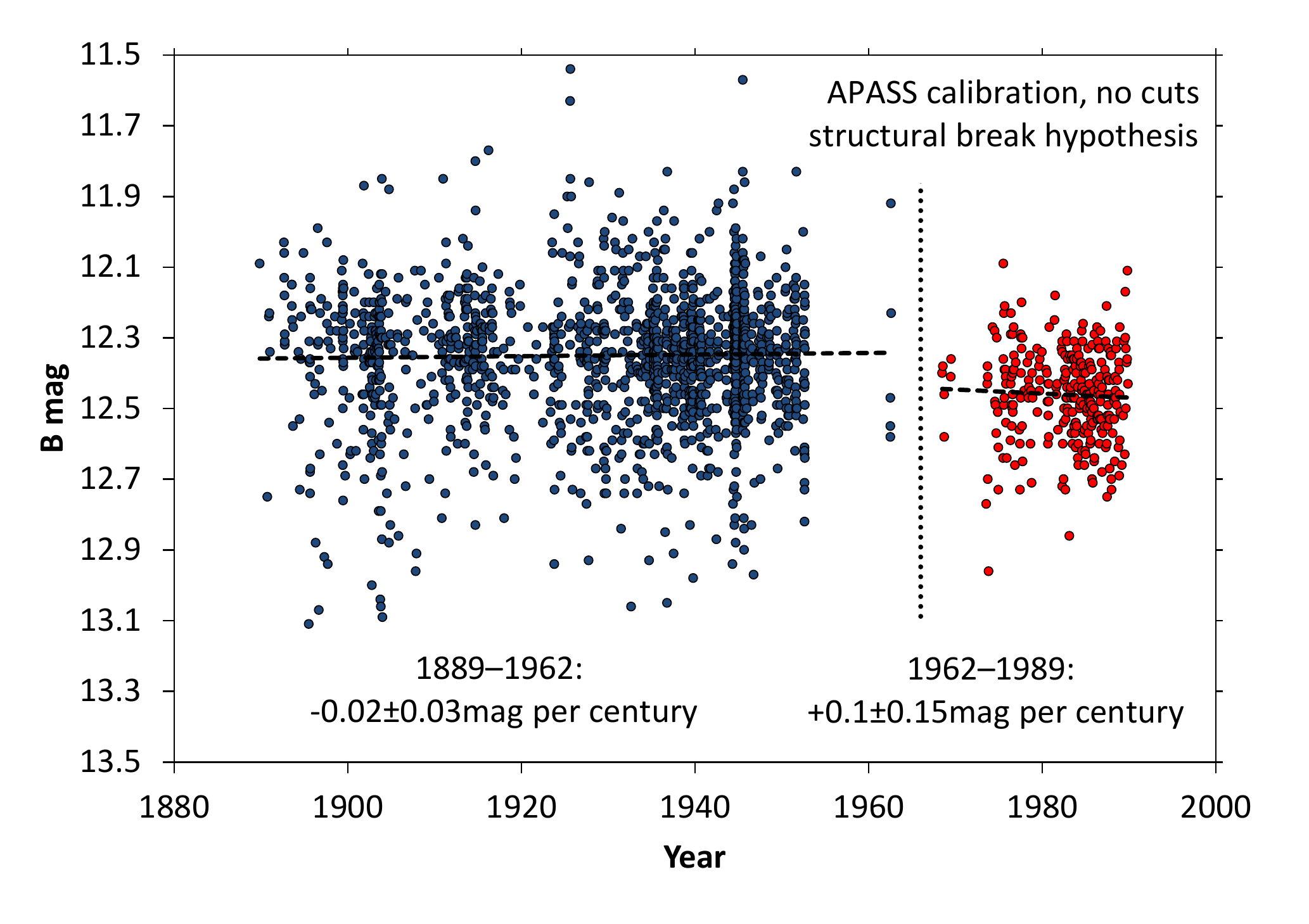}
\caption{\label{fig_break}Hypothesis of a flux discontinuity. Left: The data before 1962 comes from 16 different telescopes, while the data after 1962 (red symbols) comes from only one telescope and shows an offset. Right: Linear regressions for both segments separately indicate constant luminosity within the errors. We hypothesize that the flux discontinuity is due to a different technology used after 1962 in ``dnb'' data, e.g. due to a different emulsion. Figure shows $n=1470$ data points.}
\end{figure*}

\subsection{Linear trend of flux discontinuity?}

\begin{table*}
\center
\caption{Sources of DASCH observations of KIC8462852 \\ (APASS calibration, all plates)\label{tab:telescopes}}
\begin{tabular}{ccccc}
\tableline
Series & Telescope & From & To & Number of Plates \\
\tableline
ac & 1.5-inch Cooke Lenses&1899.3&1952.7&613 \\
rh&3-inch Ross Fecker&1928.3&1962.6&410 \\
\underline{dnb}&\underline{Damons North Blue}&\underline{1962.5}&\underline{1989.9}&300 \\
i&8-inch Draper Doublet&1889.8&1936.8&233 \\
mc&16-inch Metcalf Doublet&1910.5&1951.6&79 \\
bm&3-inch Ross&1934.2&1940.8&54 \\
ay&2.6-inch Zeiss-Tessar&1924.3&1927.9&44 \\
md&4-inch Cooke Lens&1911.2&1940.5&36 \\
ca&2.5 inch Cooke Lens&1935.4&1936.7&15 \\
ir&8-inch Ross Lundin&1935.2&1962.5&14 \\
a&24-inch Bruce Doublet&1894.3&1906.7&10 \\
mb&4-inch Cooke&1929.5&1929.8&10 \\
ax&3 inch Ross-Tessar Lens&1923.2&1923.8&6 \\
am&1-inch, 1.5-inch Cooke Lenses&1903.5&1904.7&2 \\
me&1.5-inch Cooke Lenses&1911.2&1911.3&2 \\
ma&12-inch Metcalf Doublet&1907.4&1907.4&1 \\
mf&10-inch Metcalf Triplet&1917.6&1917.6&1 \\
\tableline
\multicolumn{5}{l}{Dates represent the first and last observation of KIC8462852 for the respective} \\
\multicolumn{5}{l}{telescope. The underlined series ``dnb'', is a possible source for structural} \\
\multicolumn{5}{l}{breaks in the light curve of KIC8462852.} \\
\end{tabular}
\end{table*}

After visual inspection of many light curves, we hypothesize that rather than a linear trend, there is actually a flux discontinuity in the data. We show the statistical evidence for that explanation in section~\ref{sec:break_statistics} and discuss it in section~\ref{subsec:whatif}.

We suggest that this happened at the ``Menzel Gap'', a time of missing data in the 1960s \citep{2012IAUS..285..243G}. The underlying cause for such a break could be that the time series is not a single, perfectly calibrated data stream. Instead, many telescopes contributed measurements to the DASCH archive. In our example case of KIC8462852, a total of 17 telescopes were involved. As can be seen in Table~\ref{tab:telescopes} and in the left panel of Figure~\ref{fig_break}, 16 of the 17 telescopes were active between 1889 and 1962. After 1962, all data come from the ``Damons North Blue'' telescope. We therefore hypothesize the underlying root cause of a flux discontinuity would be the use of different optical setup (camera, lens, coatings, plates, emulsions; geography, light-pollution, airmass, etc.) after 1962. Such differences might have canceled out from combining 16 different telescopes in the time before 1962. Also, there is a considerable difference in limiting magnitude.  According to DASCH\footnote{http://dasch.rc.fas.harvard.edu/series.php, retrieved 15-Feb 2016}, ``Damons North Blue'' used a 4.2 cm lens, with a limiting magnitude of 14.18.  All other telescopes used for the KIC8462852 observations had an average limiting magnitude of 13.57.  The different limiting magnitudes affect the quality cuts.

The only overlap of telescopes between the two potential segments occurs in the year 1962. Unfortunately, only three observations from the ``3-inch Ross Fecker'' and four observations from ``Damons North Blue'' overlap, all with large scatter. Formally, for that segment of time, we find $12.22\pm0.17$ for average magnitude of KIC8462852 from ``Damons North Blue'' and $11.91\pm0.13$ from the ``3-inch Ross Fecker'', so that the $1\sigma$ error bars nearly overlap. 

We can employ statistical tests to check the confidence (if any) of a flux discontinuity at this point in time.

\subsection{Structural break: Statistical results}
\label{sec:break_statistics}
We use the test from \citet{Chow} to test the likelihood of a statistical break in the data.  The Chow test splits the data in two segments and compares the compatibility of linear regressions in both. 

For KIC8462852, the Chow test prefers the year 1962 for a split, coincident with the ``Menzel gap''. When splitting the data this way, a flux discontinuity is significant at 12$\sigma$ confidence, and removes any linear trends in both parts ($p$=0.78 for the first part, $p$=0.31 for the second part). This is also visually evident from Figure~\ref{fig_break}.

\subsection{Partial data analysis}
\label{subsec:whatif}
The Chow test raises a question about the observer's perspective. What if we performed the test at an earlier time, and only part of the dataset were available?  The test should yield a consistent result, albeit with larger error bars. Using this method, we can ask the question: Would an observer in the year X also have found a linear trend with the then-available data? What are required to produce a trend?

To investigate this question, we start with the complete light curve, and then repeatedly delete the newest observation, until the dimming becomes insignificant.  We find that to find a significant dimming, one needs the data from 1889 through 1976.42 (at the 5\% significance level), or 1889 through 1978.84 (at the 1\% level). The light curve ends in 1989.89. In other words, for a researcher with only the data from 1889 to 1976 (or 1978) at hand, the star would have appeared constant within the errors. This is in contrast to a continuous linear trend, which should be significant even for a third of the dataset. Therefore, we argue it is more likely that a sudden jump in apparent luminosity occurred, rather than a linear trend.

\begin{figure}
\includegraphics[width=.5\textwidth]{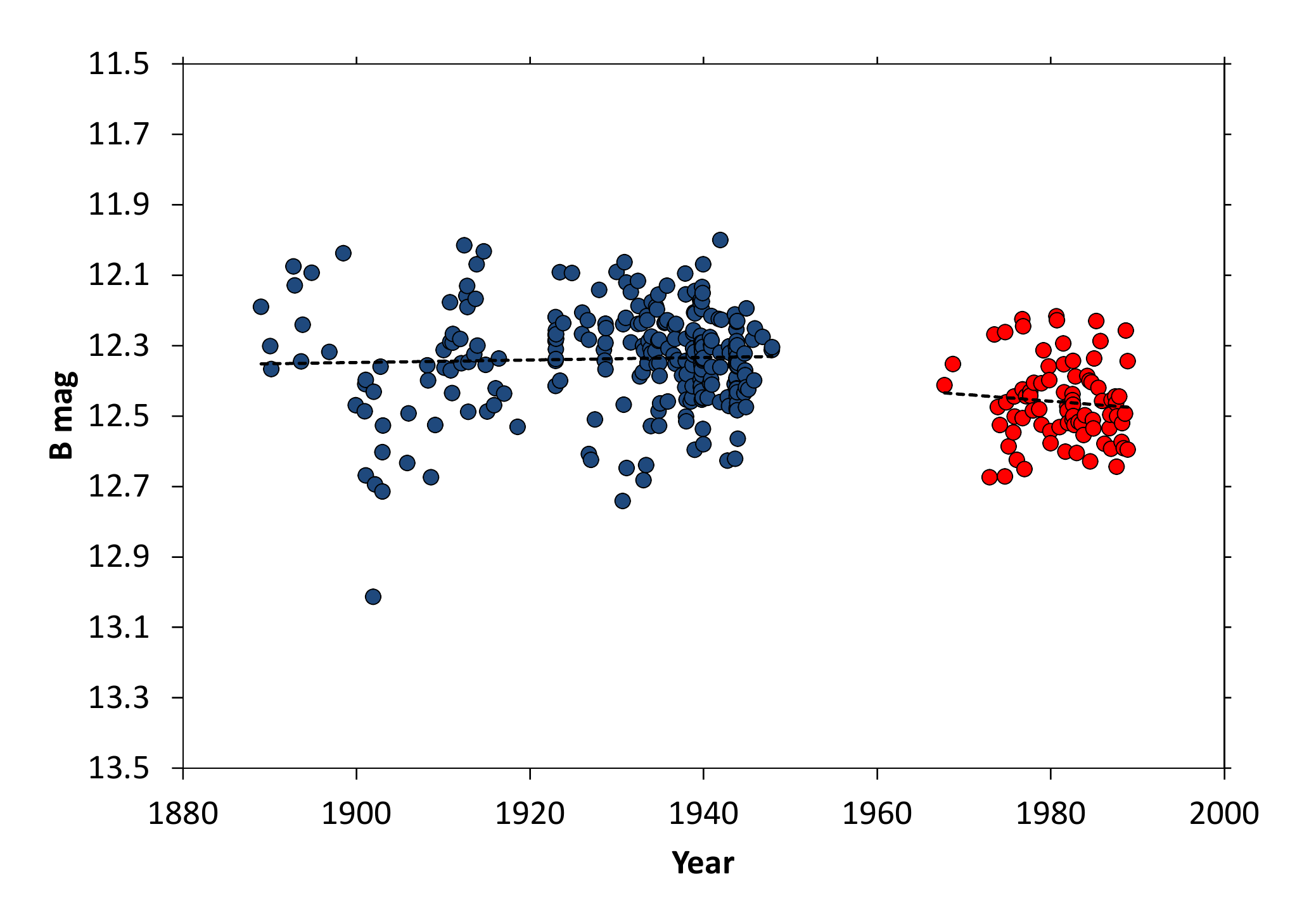}
\caption{\label{fig:fullclean}KIC8462852 photometry with all AFLAG and all warning BLFAG datapoints removed; $n=343$ data points.}
\end{figure}

\subsection{Testing very strict data cleansing}
\label{sec:strict}
For the Landolt stars benchmark (section~\ref{sec:landoltcompare}), we have employed a very strict data cleansing by removing all data values with \textit{any} AFLAG and all warning BFLAG and keeping only the good data points. We have repeated this exercise for KIC8462852 and show the result in Figure~\ref{fig:fullclean}. The visual result is virtually identical to Figure~\ref{fig_break} despite the lower amount of data points. The formal slopes are, within the errors, also identical. Consequently, even the most restrictive data cleansing cannot remove the existing low-magnitude trends.

\section{Literature comparison of long-term accuracy of DASCH photometry}
\label{sec:longterm}

The long-term photometric calibration is described by the DASCH team to have an accuracy of $\sim\pm$0.1mag per century. A best case example is shown on the DASCH website\footnote{http://dasch.rc.fas.harvard.edu/photometry.php}, together with the description: ``For quality control purposes we are also interested in stars that do \textbf{NOT} vary. Such constant-brightness stars enable sensitive determination of various systematic effects and provide a completely independent measure of uncertainties. At left is the lightcurve of such a star demonstrating about +/-0.1mag photometry over 600 plates, that span 100 years and 19 different plate-series.'' (their emphasis). A very similar example, with the same accuracy over a century, is given in the original DASCH calibration paper (\citet{2010AJ....140.1062L}, their Figure 15 and caption).

This judgment regarding the level of long-term accuracy is similar to the one found by \citet{2013PASP..125..857T} (0.1mag) and \citet{2013PASP..125..793T} who checked 997 Kepler planet host stars and found that ``Our typical photometric uncertainty is ̃0.1--0.15 mag''. In their Figure 3 in \citet{2013PASP..125..793T}, they present the uncertainty as a function of magnitude, indicating an rms of 0.2mag at 2$\sigma$ for K$_P$=12mag. In their Figure 1 in the same paper, a decade-long bump is apparent for the Kepler planet host star KIC8191672, although it is unclear whether this trend is of an instrumental or astrophysical nature. 

Lastly, there is a paper by the DASCH core team, analyzing ``KU Cyg, a 5 year accretion event in 1900'' \citep{2011ApJ...738....7T}. They state that, for the star KU Cyg, ``There  seems  to  be  a  slight  trend  of  0.1--0.2mag brightening from 1910 to 1990; however, given our systematic uncertainty over 100 years of $\sim$0.1 mag (S. Tang et al. 2011, in preparation), it is not convincing.''

A trend on the 0.1mag level is therefore not extraordinary, and can be said to be within the normal fluctuation of post-calibrated Harvard plate data.

\begin{figure}
\includegraphics[width=0.5\textwidth]{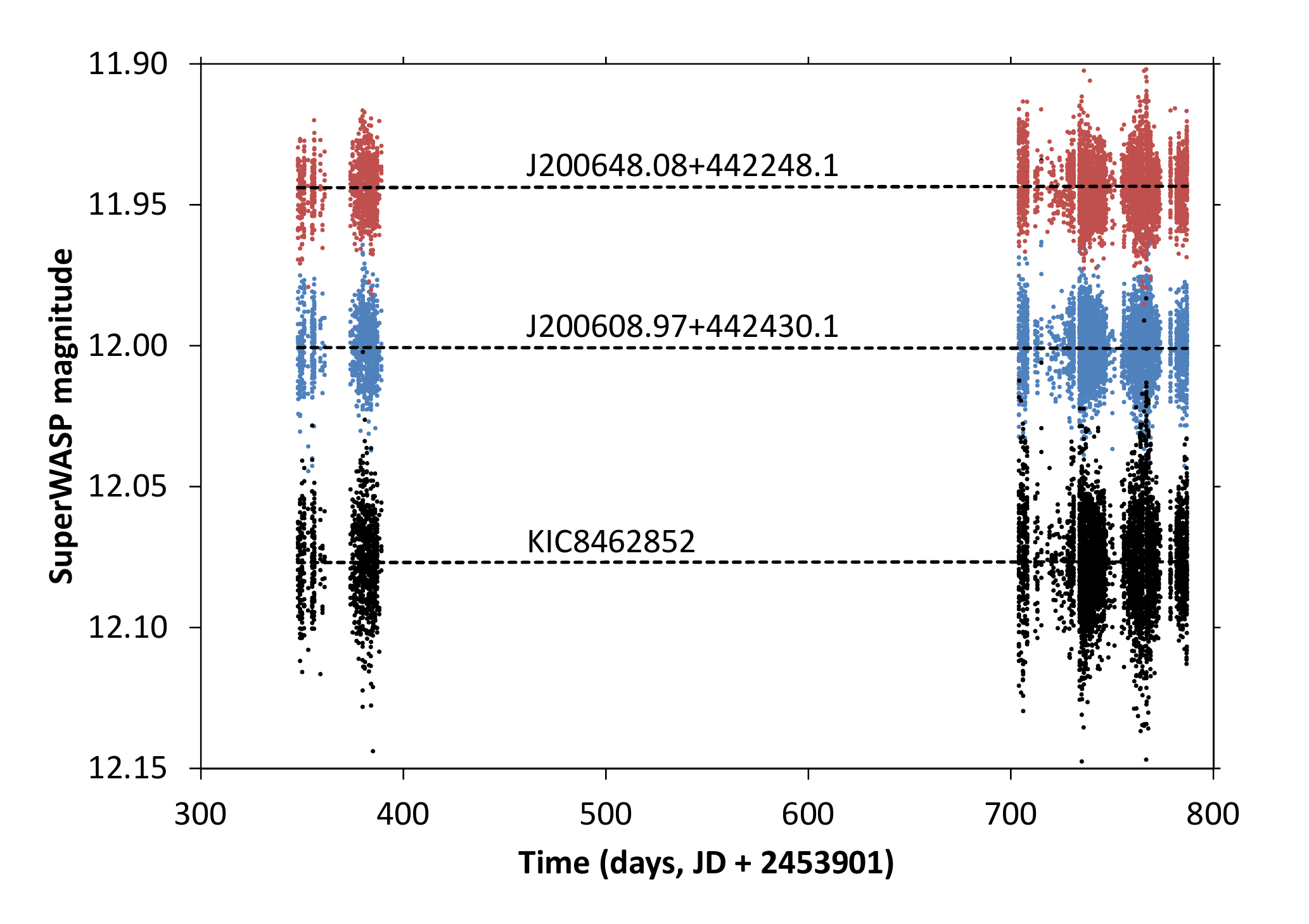}
\caption{\label{fig_superwasp}Time versus flux for SuperWASP data. Red and blue are the two check stars used by \citet{2016arXiv160103256S}, black is KIC8462852 shifted by 0.5mag for visibility. Dashed lines are linear regressions; all with insignificant slopes. The first 22 (of 5351) data values have been removed due to a 0.2mag offset.}
\end{figure}

\section{Literature comparison for KIC8462852}
\label{sec:lit}
S16 claims to have found a ``highly significant and highly confident secular dimming at an average rate of 0.165$\pm$0.013 magnitudes per century'', which is described as ``completely unprecedented for any F-type main sequence star''. We have attempted to independently check the photometry described by S16, and to evaluate the claim of the dimming trend (section~\ref{sec:trend}), as well as the long-term accuracy of the data at hand (section~\ref{sec:longterm}). Afterwards, we compare our results to the literature (section~\ref{sec:lit}).

\subsection{Linear trend in S16}
S16 claims a dimming of $0.165\pm0.013$ magnitudes per century for KIC8462852. In this section, we compare those claims with the results of our analysis.

S16 does not mention which calibration was used, or which search radius for the allowed astrometric deviation between scanned source position and catalogue position, but does mention that after performing the data cleansing, ``I have 1232 magnitudes from DASCH''. As described in section~\ref{subsec:scanned}, we find 699 values (KIC calibration), 1386 (APASS,) or 679 (GSC).  The origin of the differences between our DASCH light curve for KIC8462852 and the one used by S16 is unclear. One possible origin is the selection of which astrometric deviation between the measured (scanned) source and nominal (catalog) position is considered acceptable. The DASCH server allows for values between 1 and 60 arc seconds, with a default value of 5 arc seconds. Changing this value changes the number of resulting observations.

When performing a linear regression with the 1386 cleaned APASS values, we find a formal $+0.14\pm0.02$mag per century (section~\ref{subsec:regressions}), consistent with S16. When accounting for non-normality, this changes only slightly.  We have not attempted to try further binning choices, such as 1-year bins, 2.5-years bins, 5-year bins etc. as we do not believe this adds any value to the analysis.

\subsection{By-eye measurements in S16}
S16 performed manual (``by-eye'') brightness estimates using a method described in \citet{1981PASP...93..253S} with a 10$\times$ loupe, for 131 of 1581 data points. 

These ``by-eye'' measurements have neither been publicly released nor been shown in the study, so that we cannot compare the results. The results seem to be different, because a linear fit ``yields a slope of $+0.310\pm0.029$ magnitudes per century'', which is described as ``formally different from the slope that I get from DASCH''. In fact, this trend has almost twice (1.88$\times$, $5\sigma$ confidence of differing) the slope of the scanned data.

\subsection{Comparing the check stars in S16}
The dimming trend mentioned in S16 was compared to check stars. The author ``used the same procedures and selections to produce DASCH light curves for five nearby stars with similar magnitudes'' and found ``that check stars have constant light curves to a level of 0.03 mag over a full hundred years.'' The paper reveals, however, only the identities of two of the check stars. We re-examined these.

For TYC 3162-879-1 (KIC8462775), we pulled the data from DASCH for the KIC and the APASS calibration. We found no trend, and no flux discontinuity, within the errors of 0.02mag per century.

For TYC 3162-1001-1 (KIC8398290) we tried the same. The KIC calibration, uncleaned and unbinned, results in a brightening trend of -0.07$\pm$0.01mag per century. After data cleansing as described in S16, we obtain 1002 data points and a formal slope of -0.05$\pm$0.01mag per century. This is slightly larger than the 0.03 mag found in S16, and formally highly significant. As discussed in section~\ref{sec:longterm}, we believe that all these formally significant trends are overshadowed by long-term systematics.

Also, the slope of TYC 3162-1001-1 is $\sim$50\% of the one we find for KIC8462852 ($+0.1\pm0.02$mag per century, section~\ref{subsec:regressions}). This means that the check stars revealed in S16 do indeed fluctuate less than KIC8398290, but one of the two also shows evidence for systematics in the data.

\subsection{Cross-checking SuperWASP data}
As noted by \citet{2016MNRAS.457.3988B}, KIC8462852 was observed by SuperWASP for 3 seasons. The first season, which includes only 22 observations, shows a 0.2mag offset, which is also seen for the check stars in S16. Therefore, we discard the first season (dropping 22 of 5351 total data points) and examine only seasons 2 and 3. As can be seen in Figure~\ref{fig_superwasp}, KIC8462852 as well as the two check stars show constant luminosity within the errors. Due to the large number of observations, we can determine precise (nominal) values for the average of both seasons separately. For KIC8462852, we get $12.65701\pm0.00054$mag for the first season, and $12.65663\pm0.00026$mag for the second season, which is a brightening of $-0.00038\pm0.00054$mag. In other words, constant luminosity within the errors. The same is true for the two check stars. While the standard deviations support this assessment, there might be systematic errors in the SuperWASP data. A detailed analysis is beyond the scope of this paper, so that we advice caution about this result.

If a linear dimming trend in the data was present, we would expect a luminosity decrease. As the two seasons are separated by an average of 377 days, luminosity would have to been decreased by 0.00170mag (based on the purported dimming trend of 0.165$\pm$0.013 magnitudes per century in S16). As our systematic errors are larger than the theoretical dimming for this period, we cannot exclude (or confirm) a dimming trend.

\section{Discussion and next steps}

\subsection{Issues in defining quality cuts}
There is the problem of human bias that \textit{does} affect the selection of acceptable criteria for a comparison. Assume that some researcher analyzed photometry of KIC8462852, and found a slope of 0.165 mag per century. One could now define criteria for quality cuts, so that the trend in KIC8462852 persists, but trends in most other stars vanish. Clearly, quality cuts (if any!) must be defined completely independently. Then, a large number (thousands) of constant stars must be processed consistently and automatically. Afterwards, the distribution of slopes can be used to assess the significance of a slope of 0.165 mag per century. A similar approach is used for the removal of instrumental systematics from the Kepler light curves, dubbed ``Cotrending Basis Vectors'' \citep{2012PASP..124.1000S}.

The use (or rejection) of red and yellow plates is relevant. A literature review shows that all publications accessible for us (except S16) do use these data (e.g., \citet{2010AJ....140.1062L, 2013PASP..125..857T, 2013PASP..125..793T, 2015OEJV..173....1L}). Indeed, the DASCH team itself does this in their publications, and simply mentions the plates ``are mostly blue sensitive'' \citep{2011ApJ...738....7T}. It is preferable to \textit{keep} all available data (that are not amongst a few multi-$\sigma$ outliers), and propagate their large(r) error bars accordingly. Indeed, these values are also classified as ``Johnson B magnitude'' data and have been calibrated by the DASCH team.

Plate selection aside, other quality cuts are done equally arbitrary in the literature. In \citet{2013PASP..125..793T}, for example, the authors analyzed 997 Kepler stars and included all plates (also the red and yellow ones), but defined their own series of quality cuts. These include blended images, ``measurements within 0.75 mag of the limiting magnitude'' (in contrast to S16, who set a limit of 0.2mag), ``images within the outer border of the plates whose width is 10\%'' (which was accepted in S16), and more. Unfortunately, these criteria are also not precisely defined, e.g. the rejection of ``Stars with strong correlation between magnitude measurements and plate limiting magnitudes'', so that the results are also not reproducible.

\section{Conclusion}
We have re-analyzed time-series photometry from the Harvard plates and find a photometric sensitivity limit of $\sim0.1$mag per century (in agreement with Grindlay \& Los (2016, in prep.)), which is an extraordinary achievement for a historical archive like this, and confirms the number given in other DASCH studies (e.g., \citet{2010AJ....140.1062L, 2011ApJ...738....7T, 2013PASP..125..857T, 2013PASP..125..793T}).

For our example case of KIC8462852, we therefore find that the slope (or flux discontinuity) is not notably greater ($<1.7\sigma$ among all samples) when compared to $>500$ twin stars. Assuming no long-term dimming is present, the puzzling day-long dips in KIC8462852 might indeed be the result of a family of large comets \citep{2015arXiv151108821B}.

\acknowledgments
{{\it Acknowledgments.} We thank Bradley E. Schaefer for his feedback on possible issues with quality cuts and plate selection. We also thank Prabal Saxena for indicating that some comparison stars are unsuitable, and Rene Hudec and Peter Kroll for advice on photographic plates.

We express our gratitude to the DASCH project which does invaluable work for the community.

\bibliography{hippke-2016-03-30}
\bibliographystyle{aasjournal}

\appendix

\begin{figure*}
\includegraphics[width=.5\linewidth]{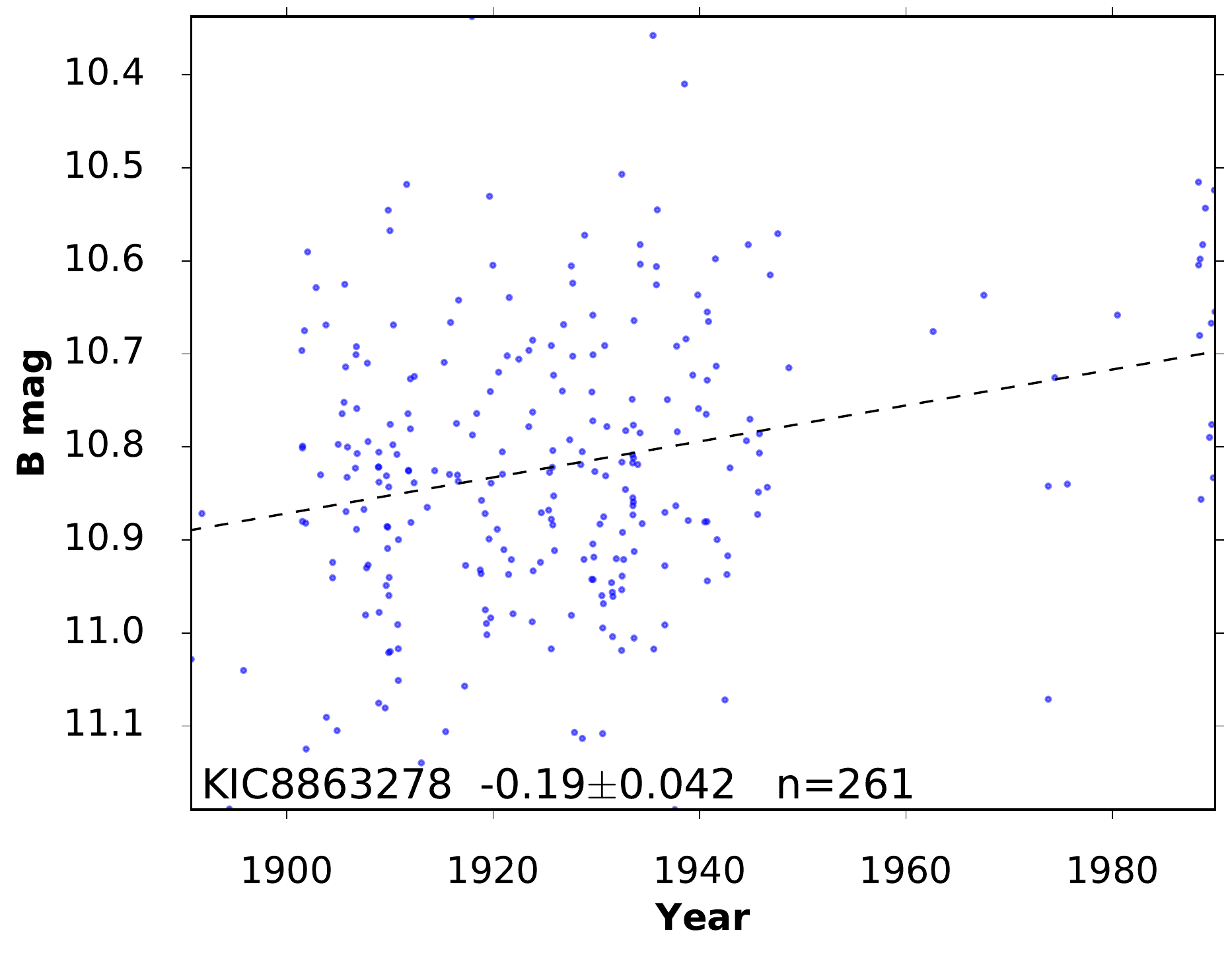}
\includegraphics[width=.5\linewidth]{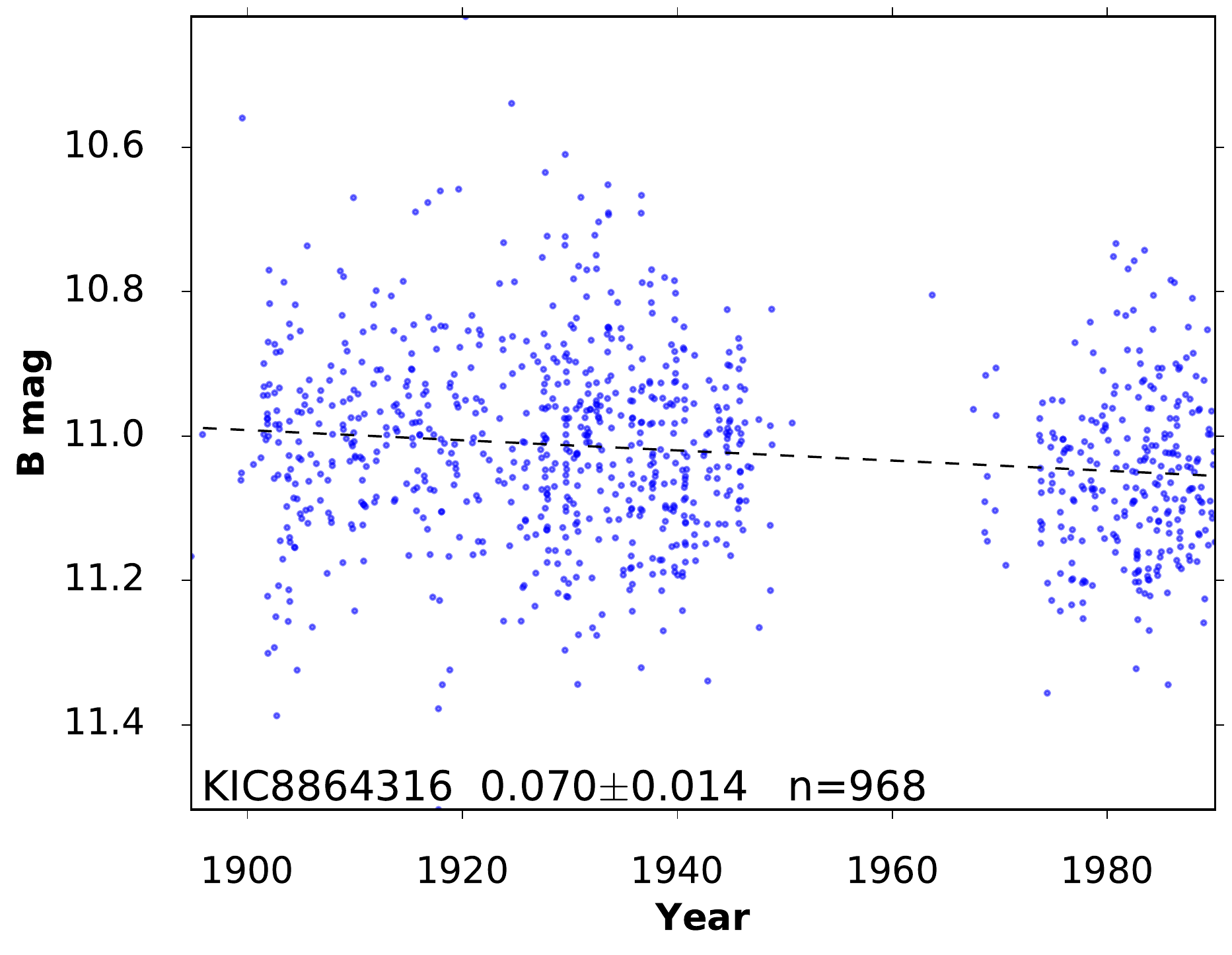}

\includegraphics[width=.5\linewidth]{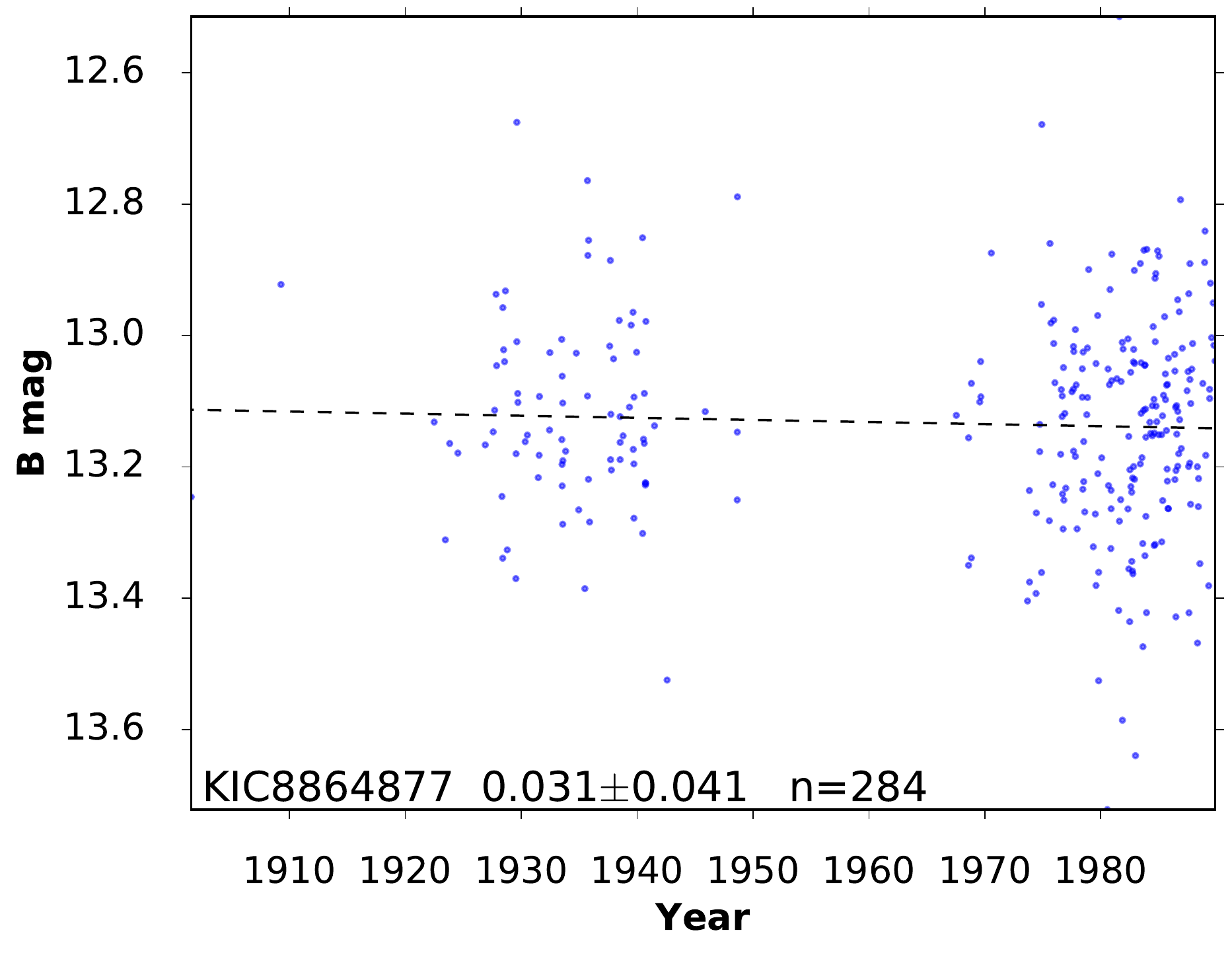}
\includegraphics[width=.5\linewidth]{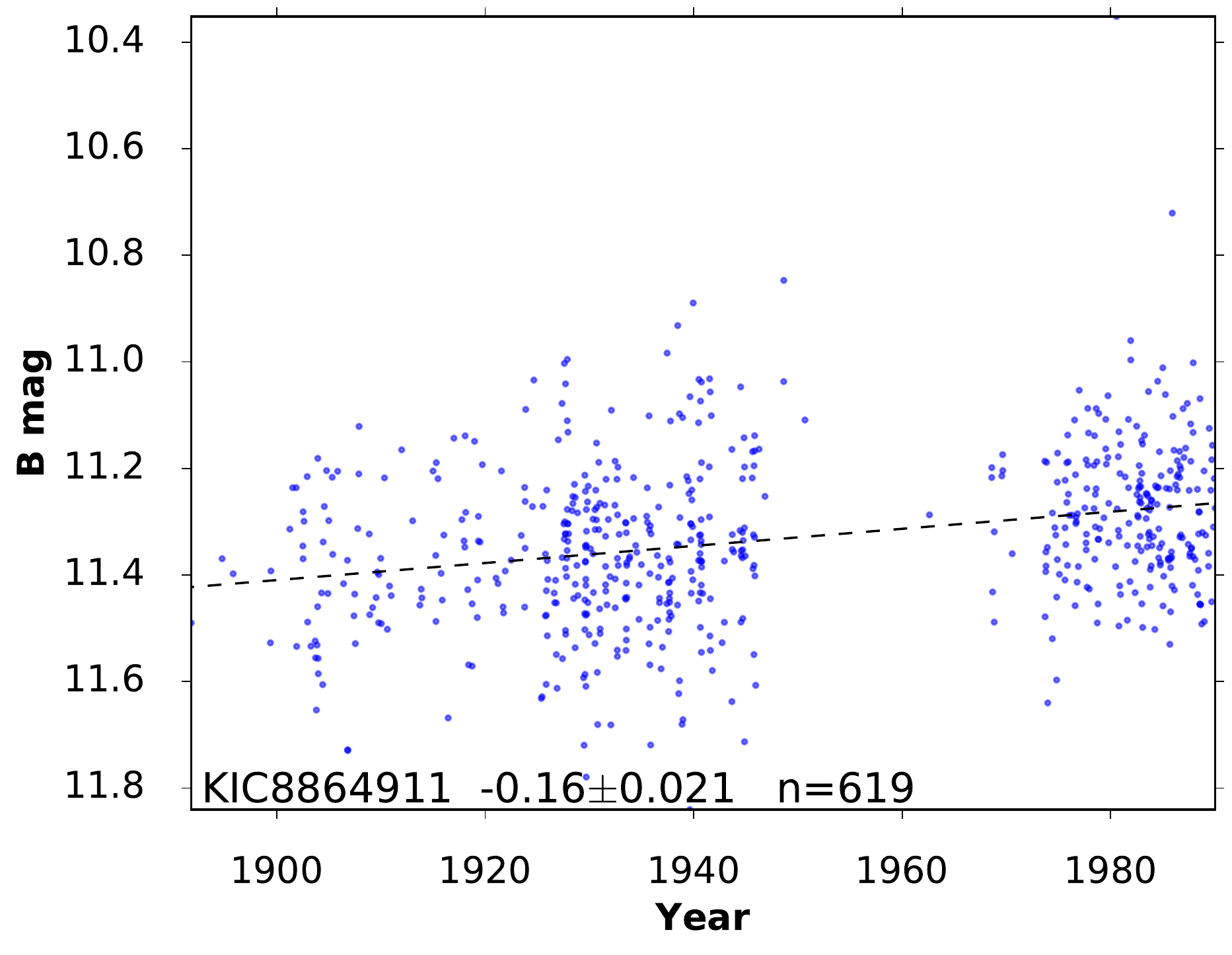}

\includegraphics[width=.5\linewidth]{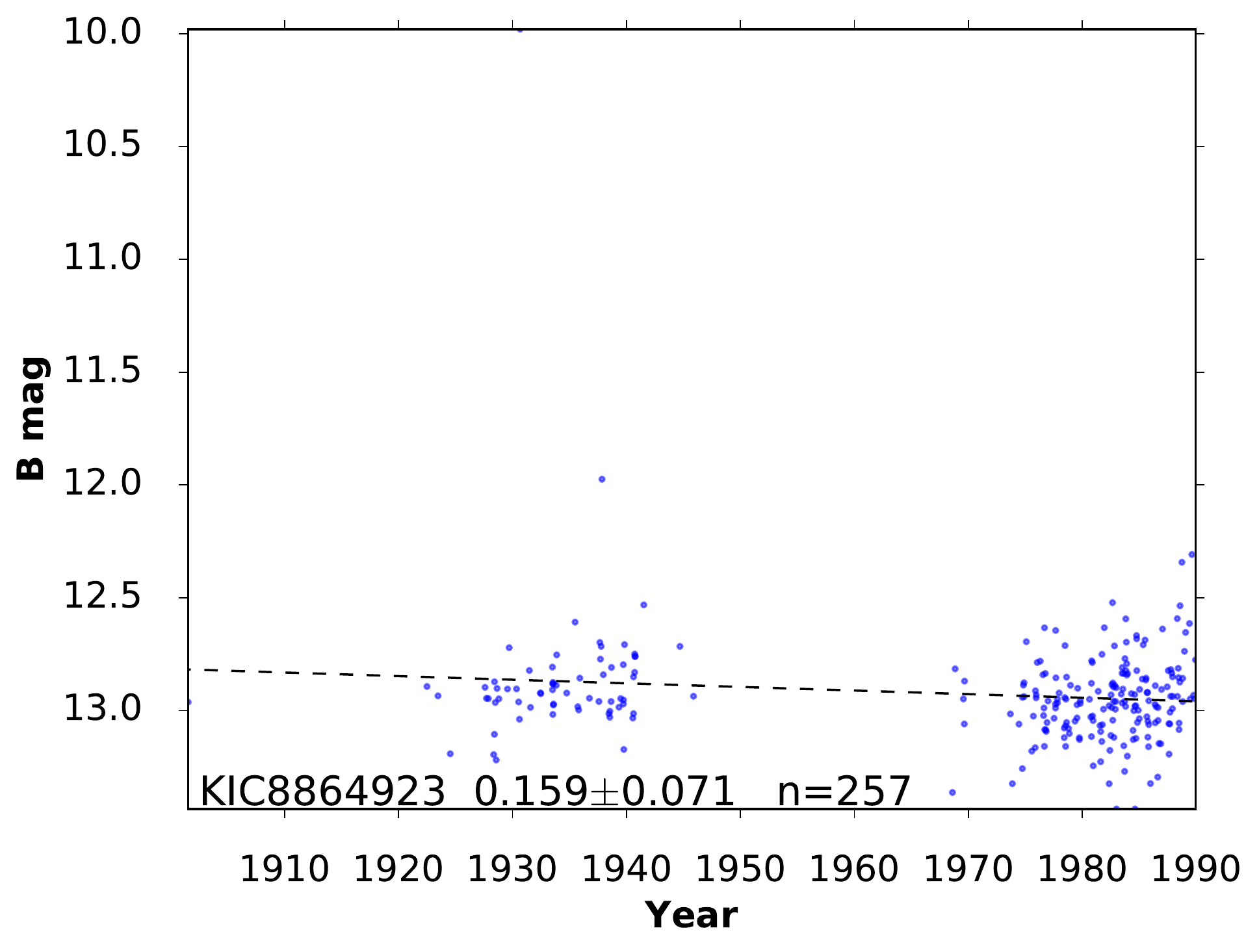}
\includegraphics[width=.5\linewidth]{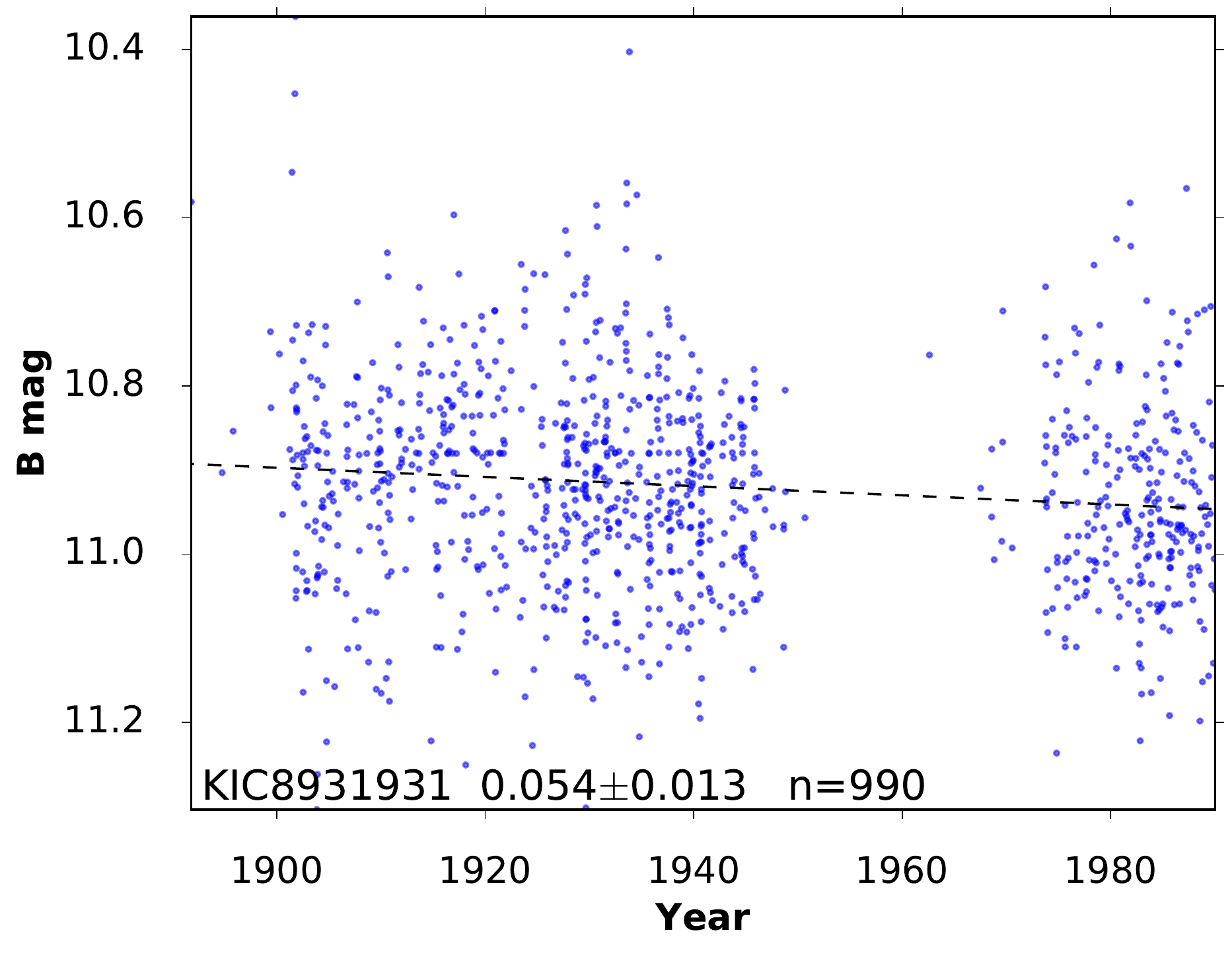}
\caption{\label{fig:land1}Photometry of Landolt standard stars with $n>100$ and all AFLAGS removed.}
\end{figure*}

\begin{figure*}
\includegraphics[width=.5\linewidth]{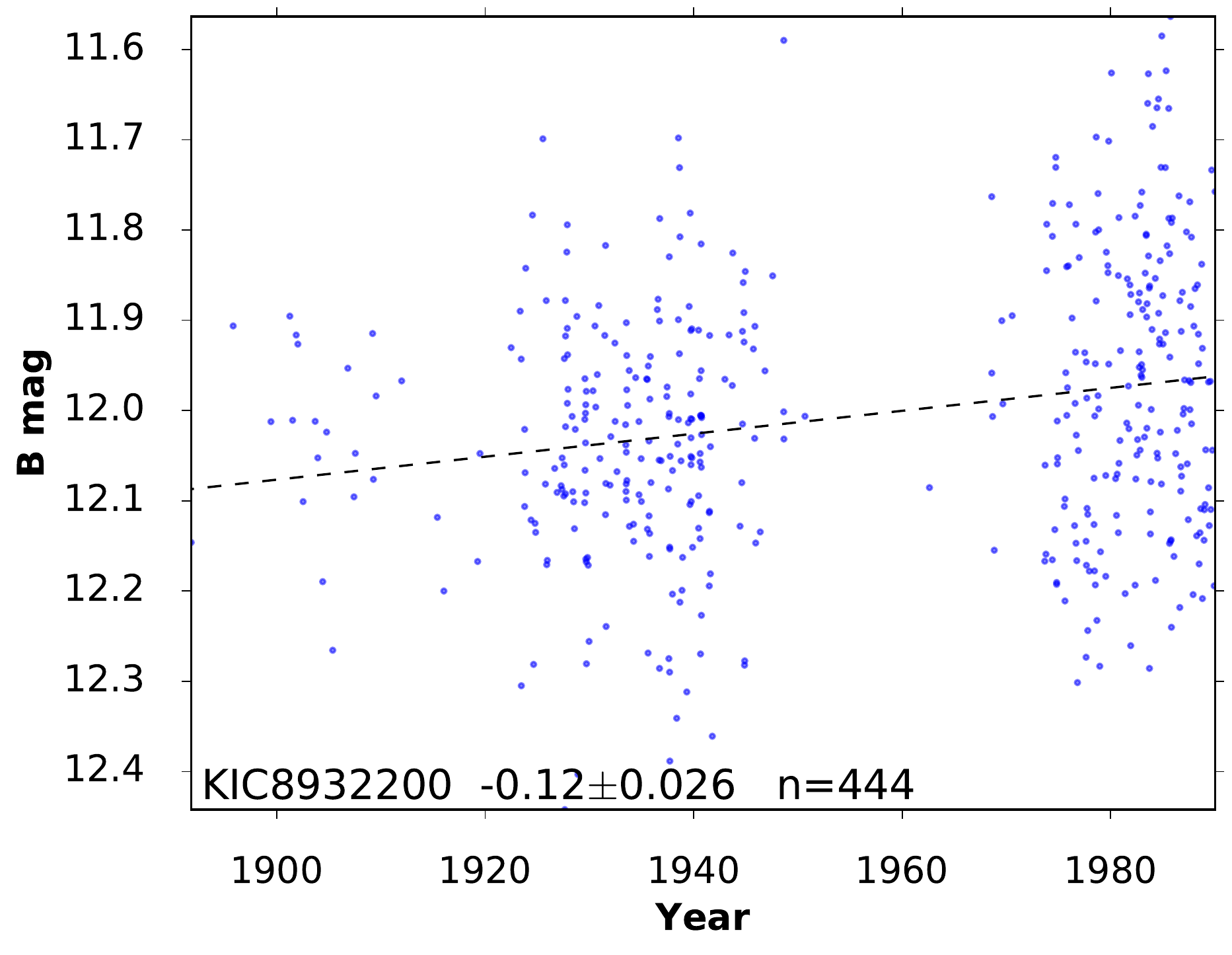}
\includegraphics[width=.5\linewidth]{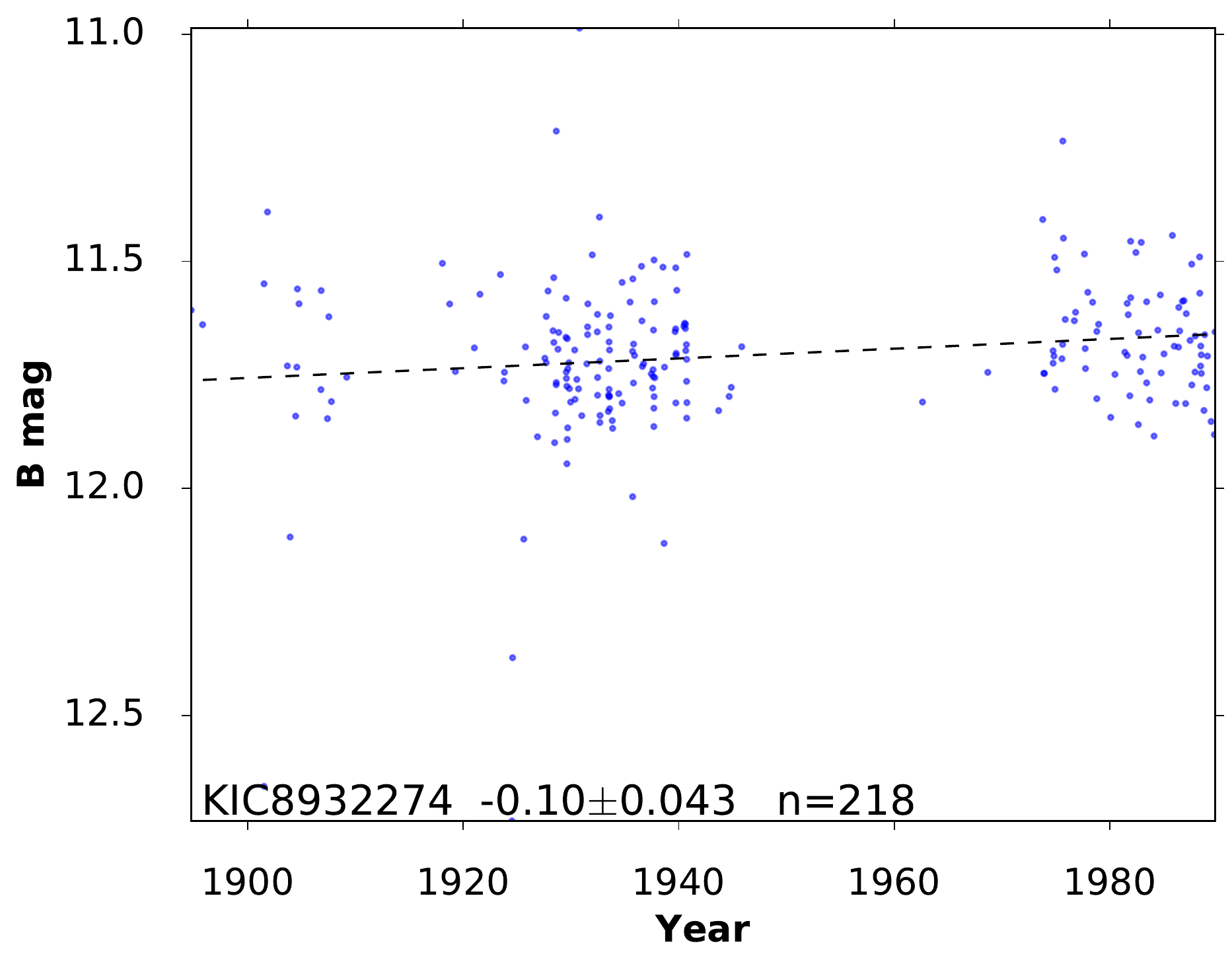}

\includegraphics[width=.5\linewidth]{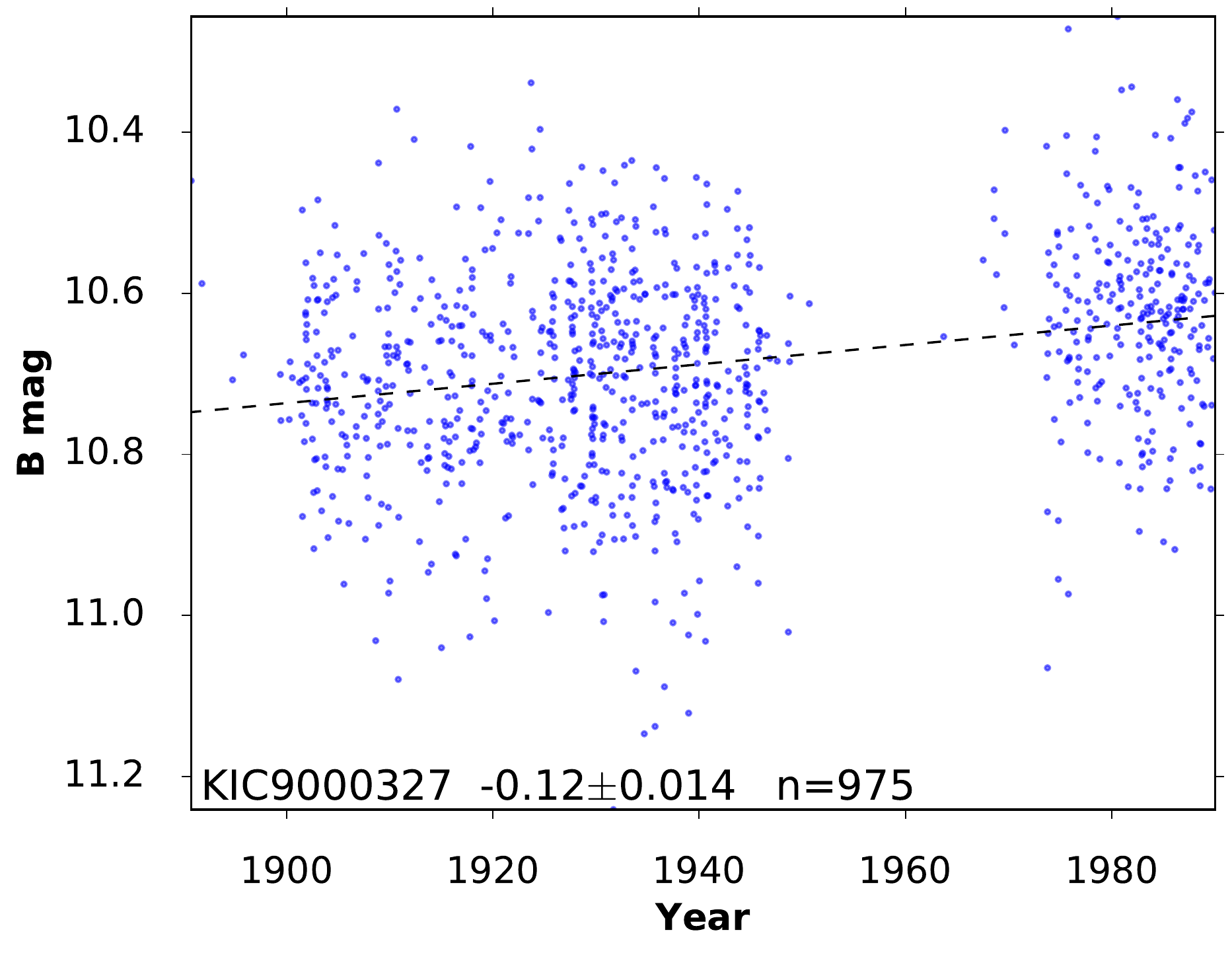}
\includegraphics[width=.5\linewidth]{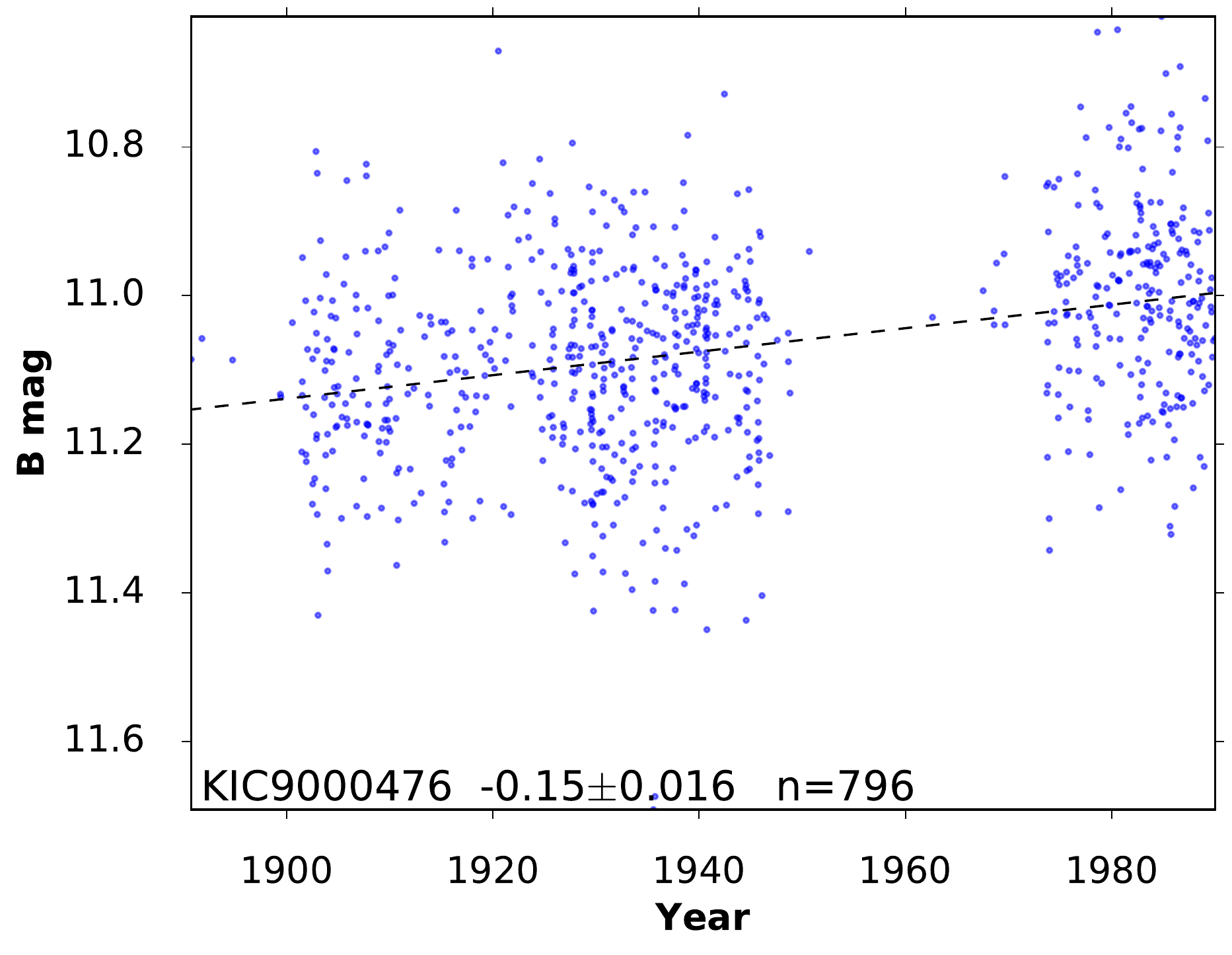}

\includegraphics[width=.5\linewidth]{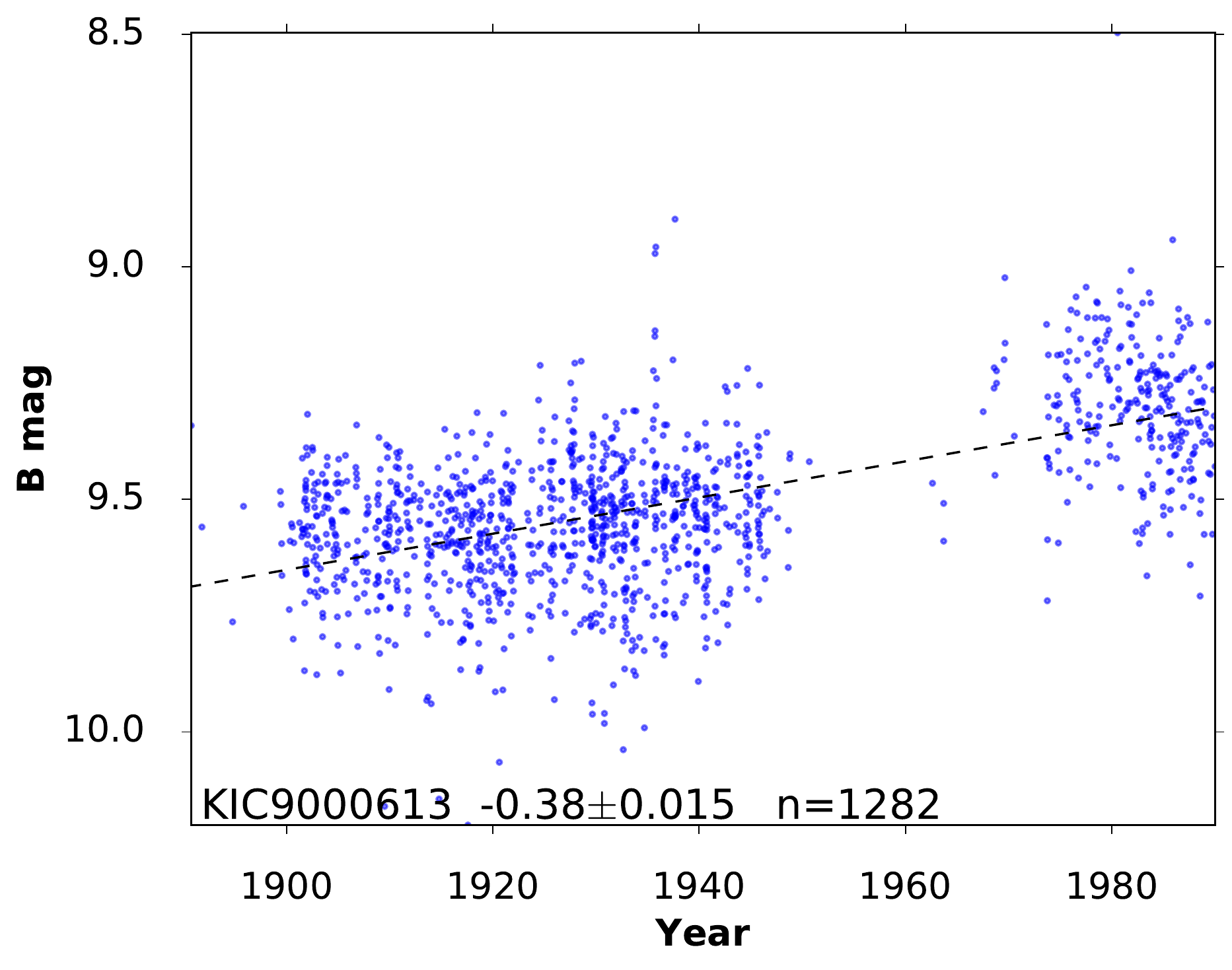}
\includegraphics[width=.5\linewidth]{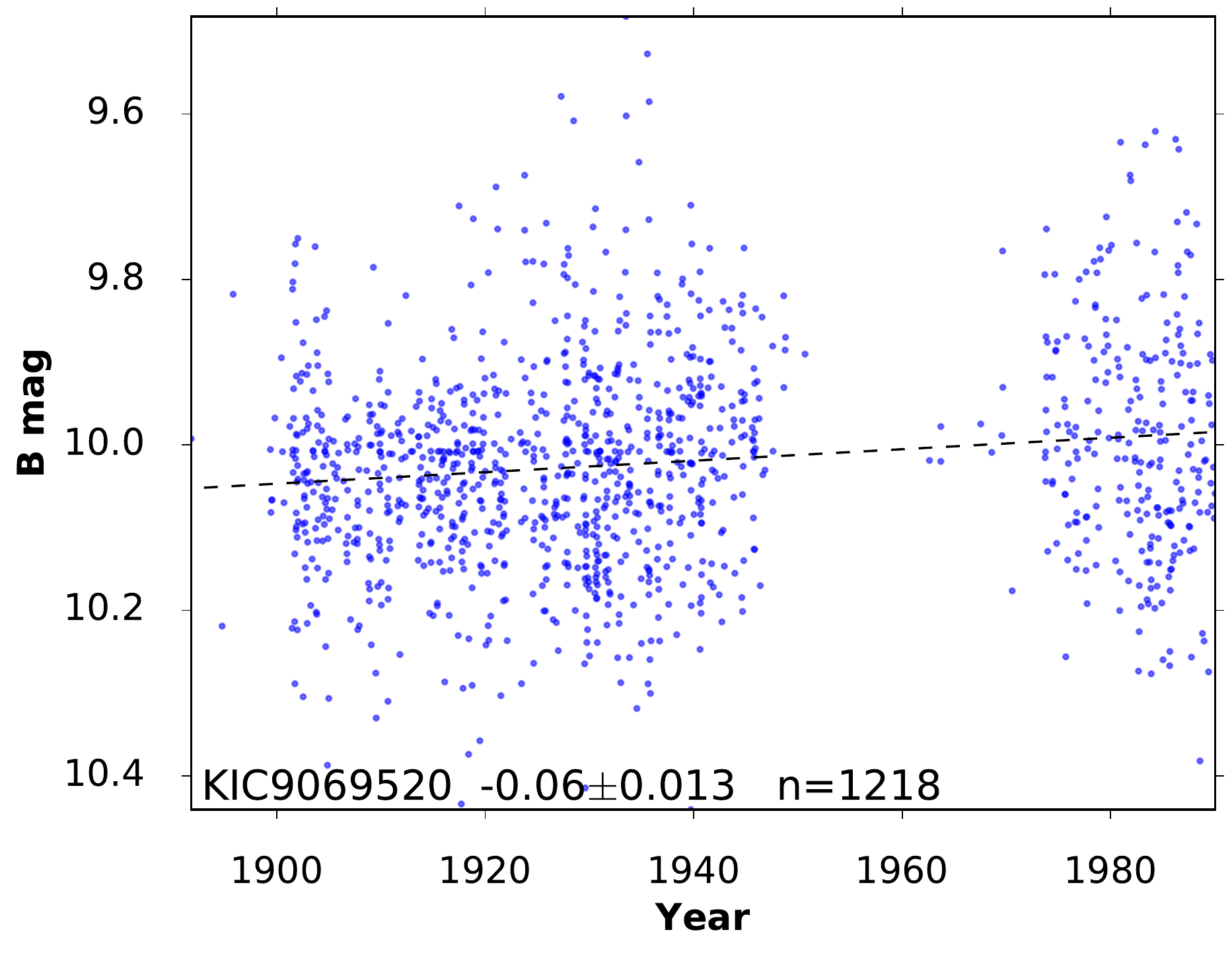}
\caption{\label{fig:land3}Photometry of Landolt standard stars with $n>100$ and all AFLAGS removed (continued).}
\end{figure*}

\begin{figure*}
\includegraphics[width=.5\linewidth]{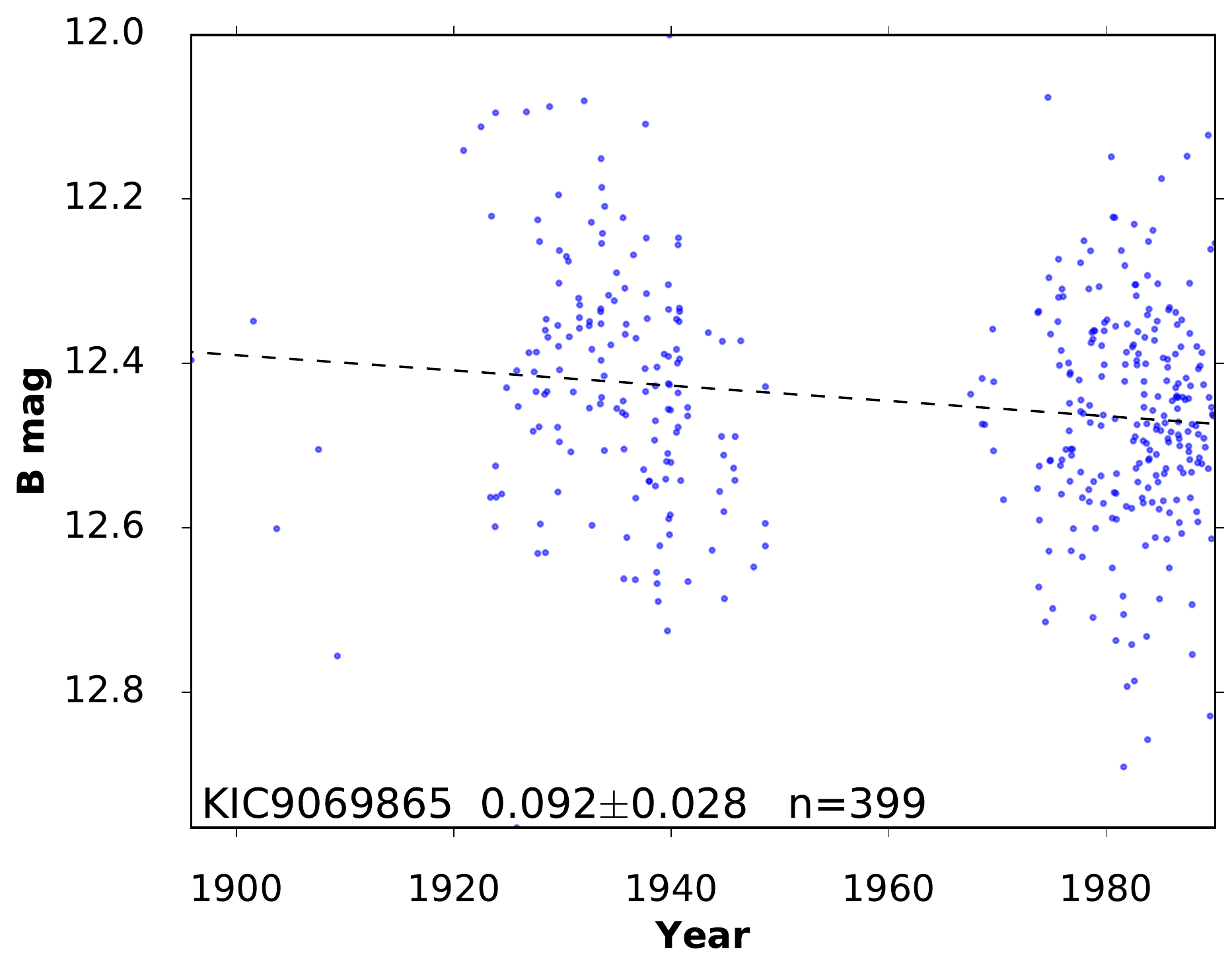}
\includegraphics[width=.5\linewidth]{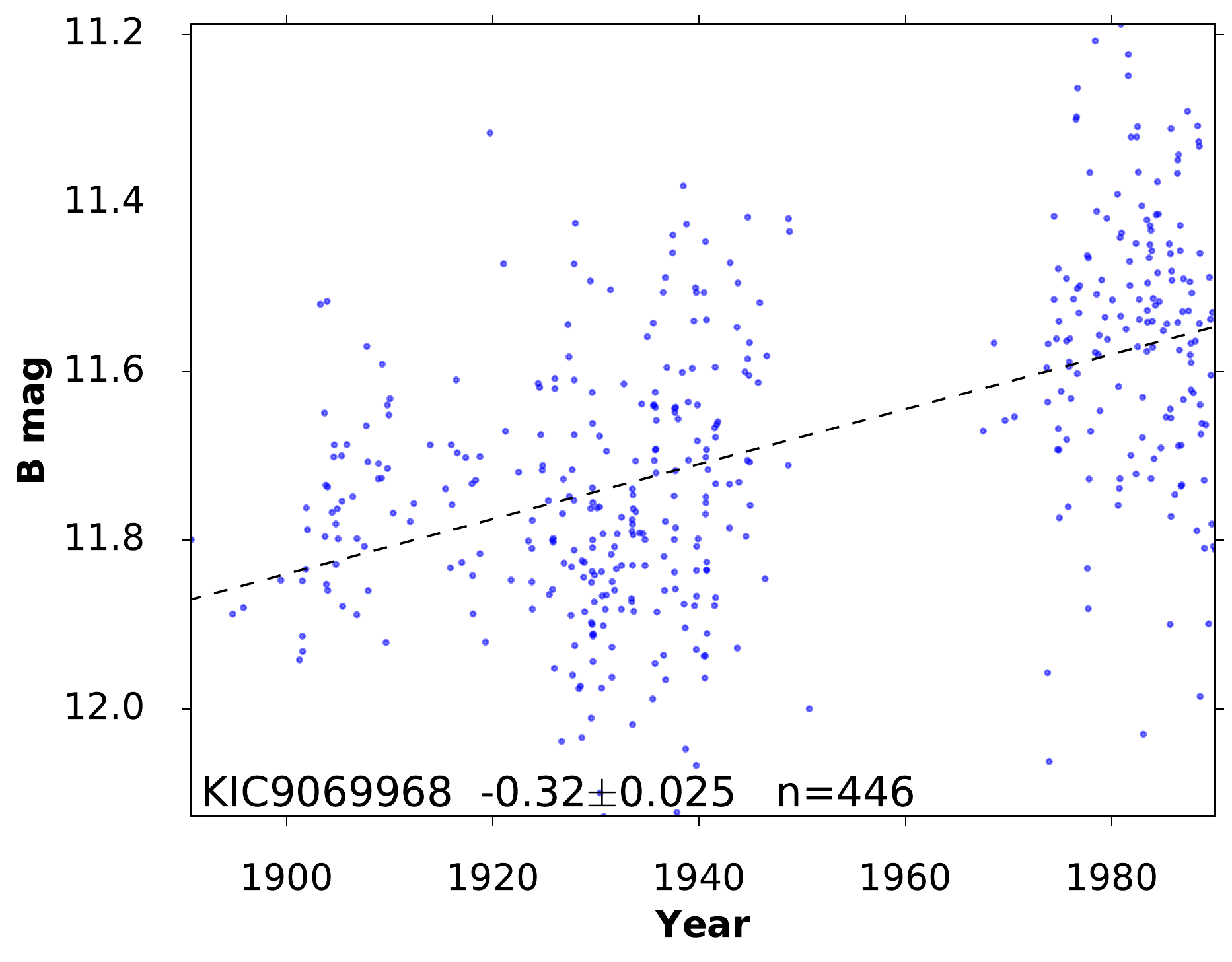}

\includegraphics[width=.5\linewidth]{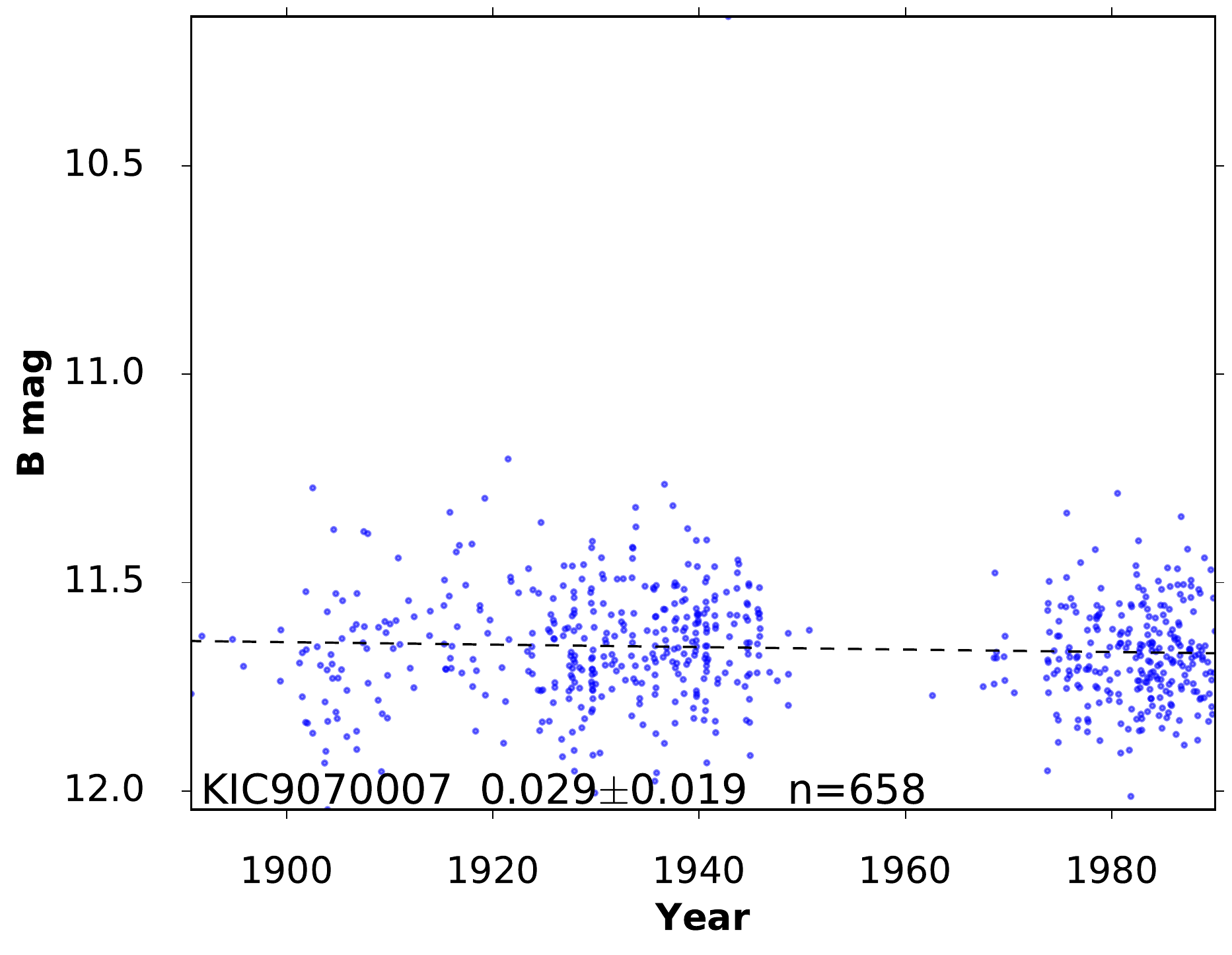}
\includegraphics[width=.5\linewidth]{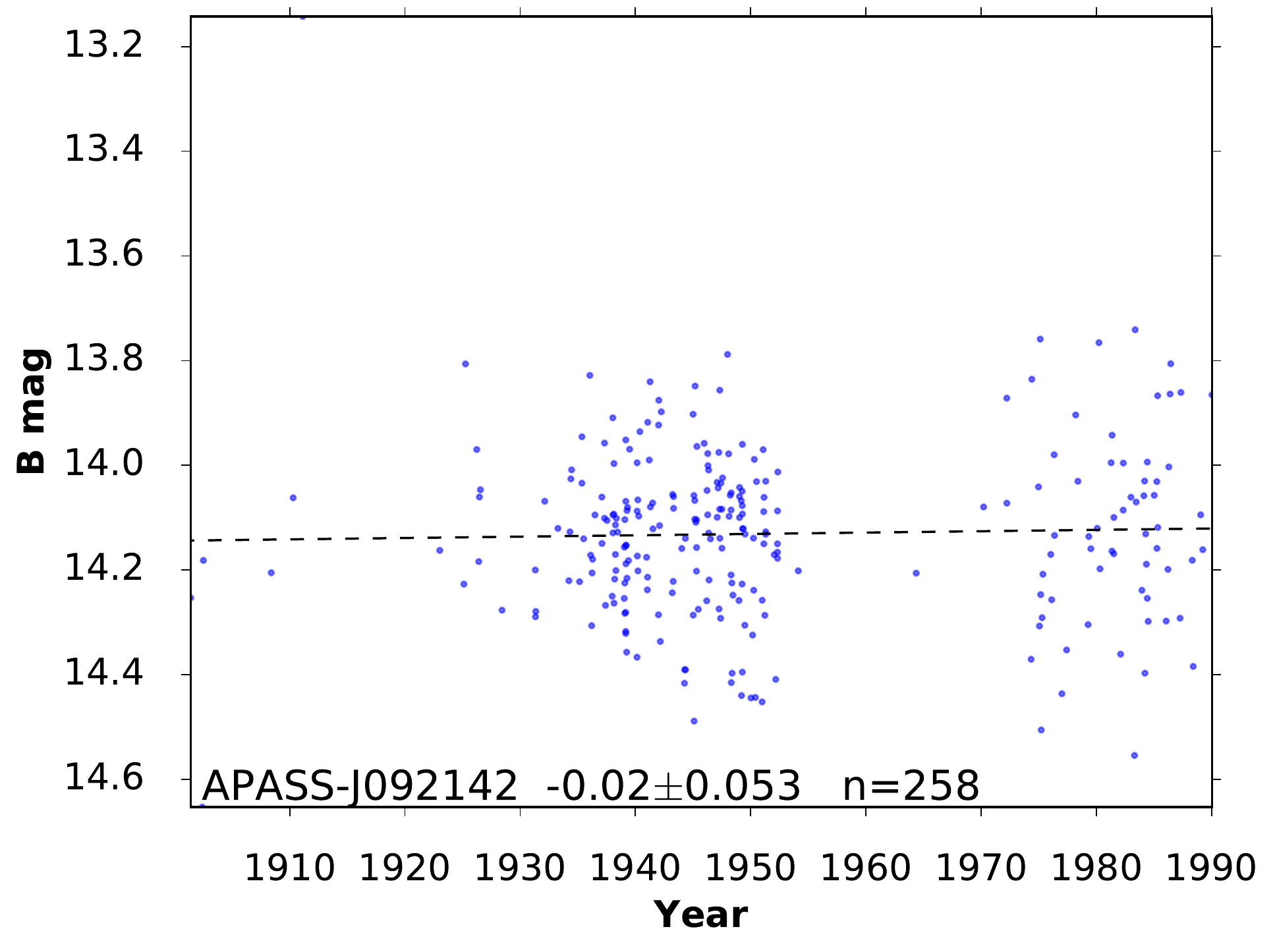}

\includegraphics[width=.5\linewidth]{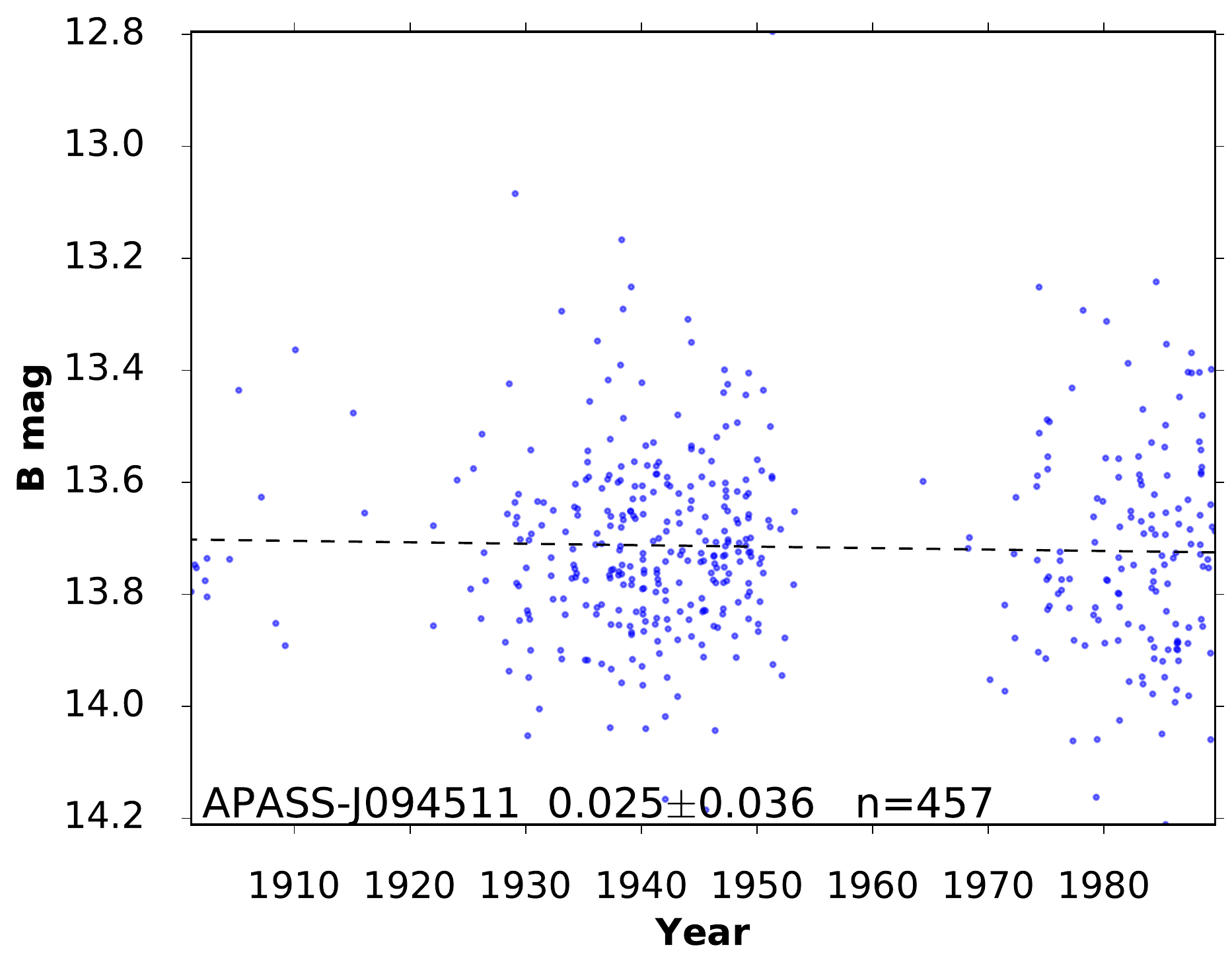}
\includegraphics[width=.5\linewidth]{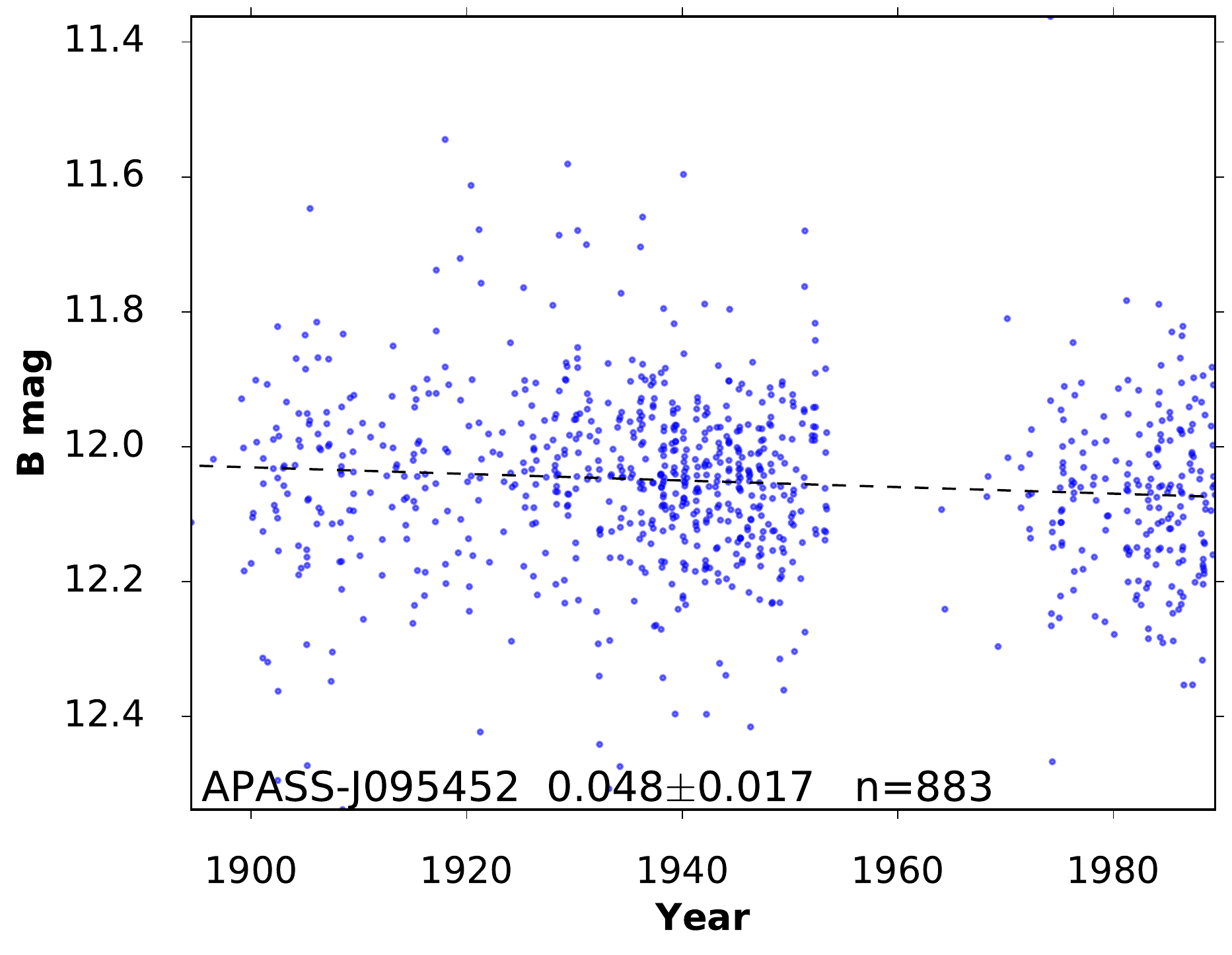}
\caption{\label{fig:land4}Photometry of Landolt standard stars with $n>100$ and all AFLAGS removed (continued).}
\end{figure*}

\begin{figure*}
\includegraphics[width=.5\linewidth]{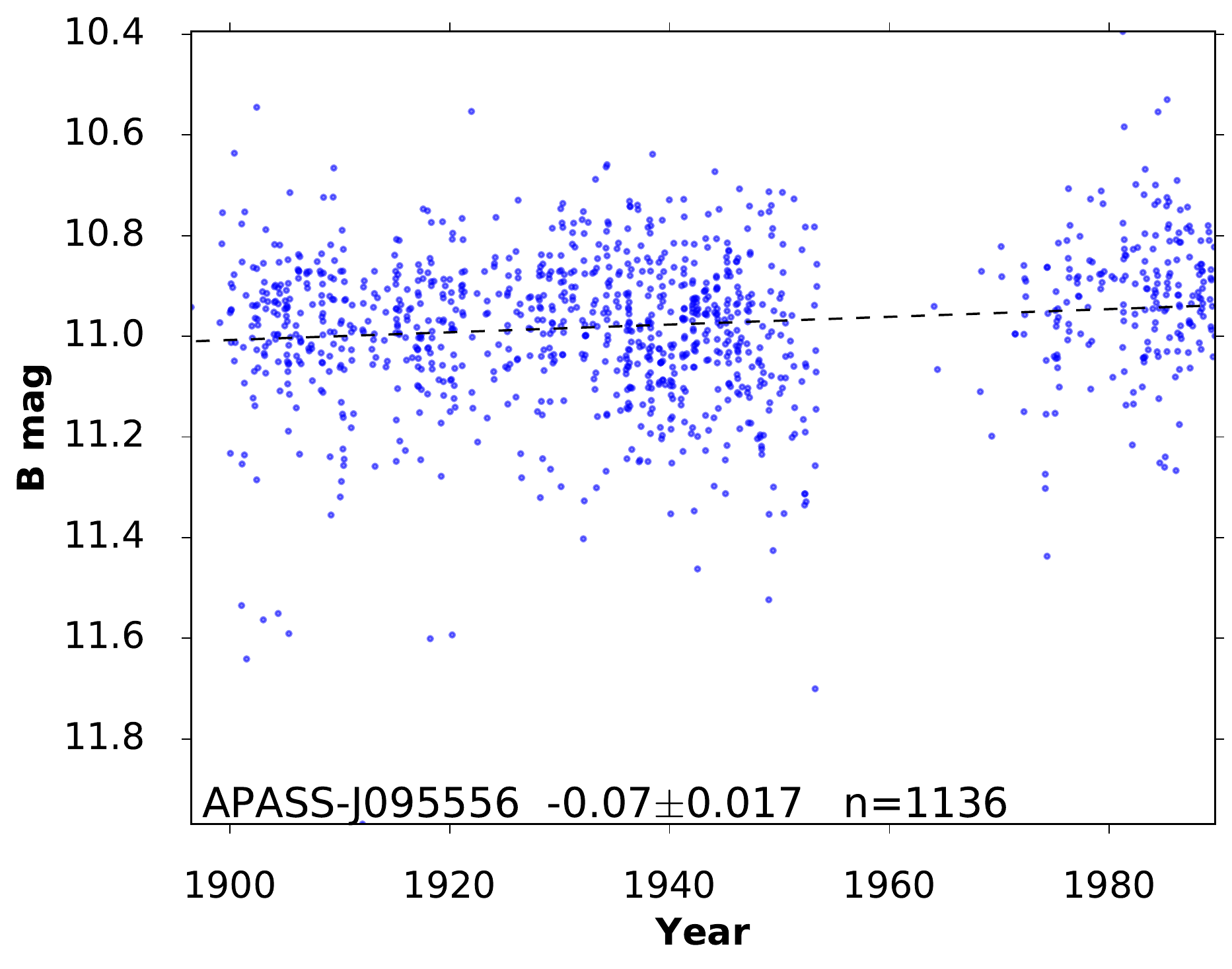}
\includegraphics[width=.5\linewidth]{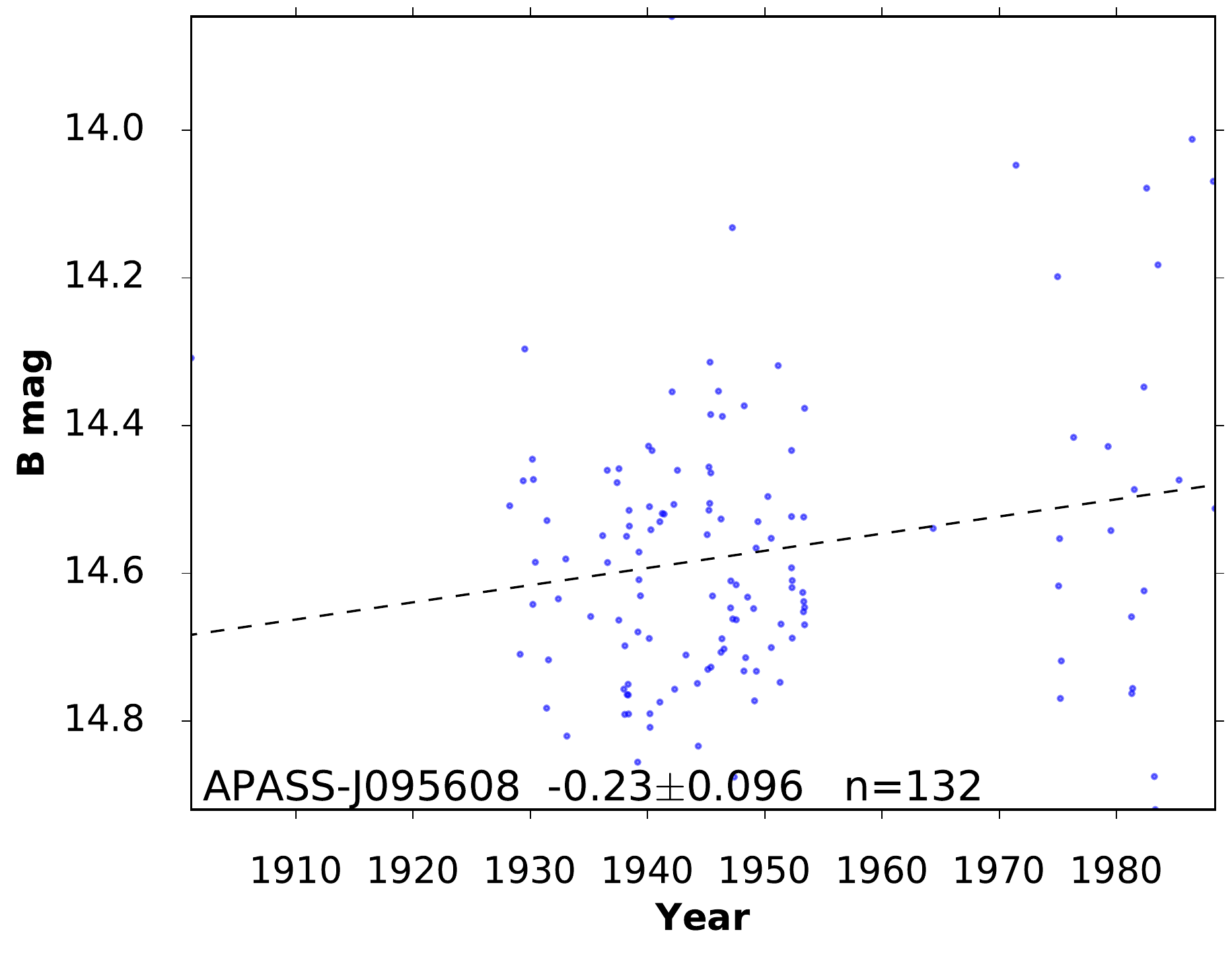}

\includegraphics[width=.5\linewidth]{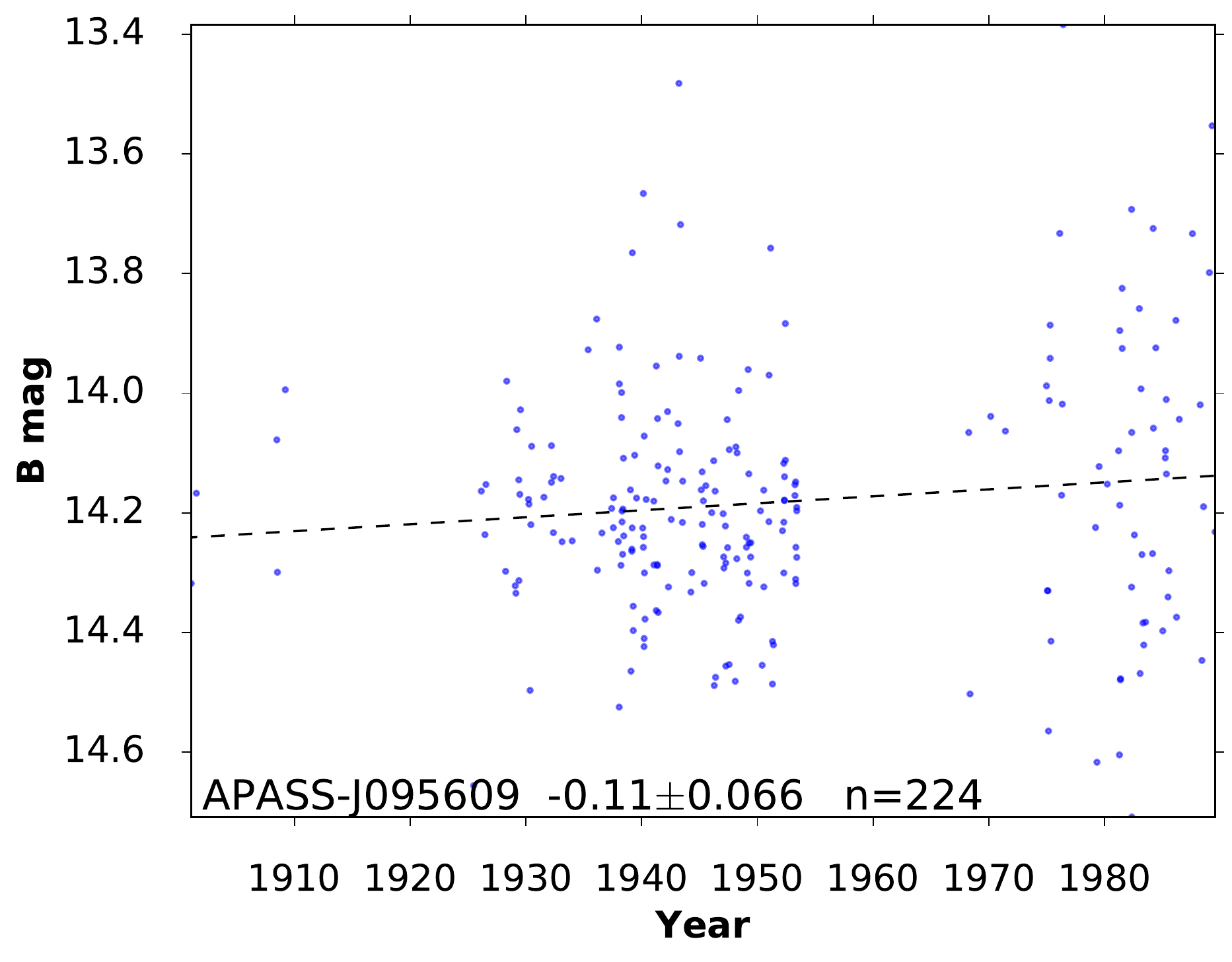}
\includegraphics[width=.5\linewidth]{fig_092142}

\includegraphics[width=.5\linewidth]{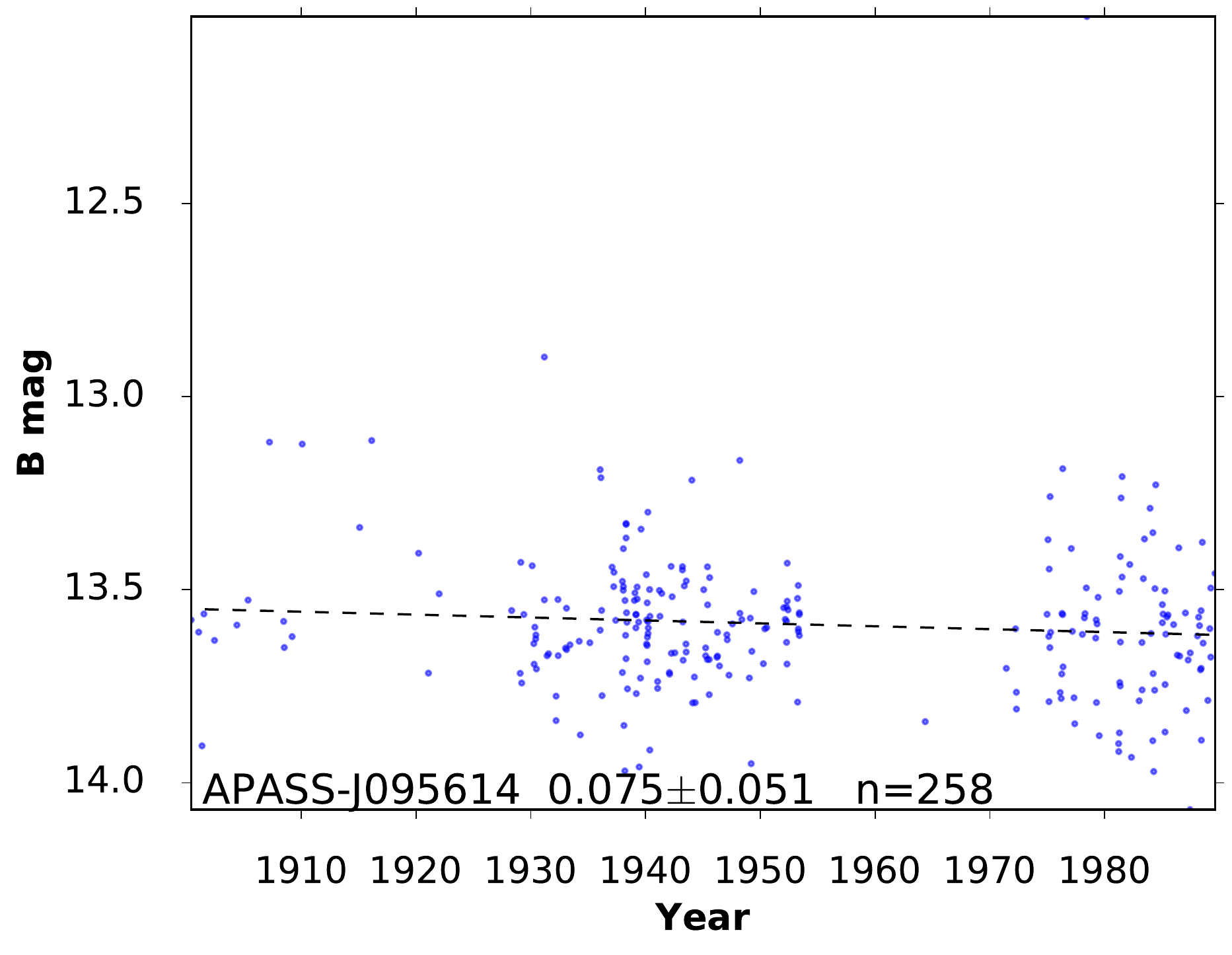}
\includegraphics[width=.5\linewidth]{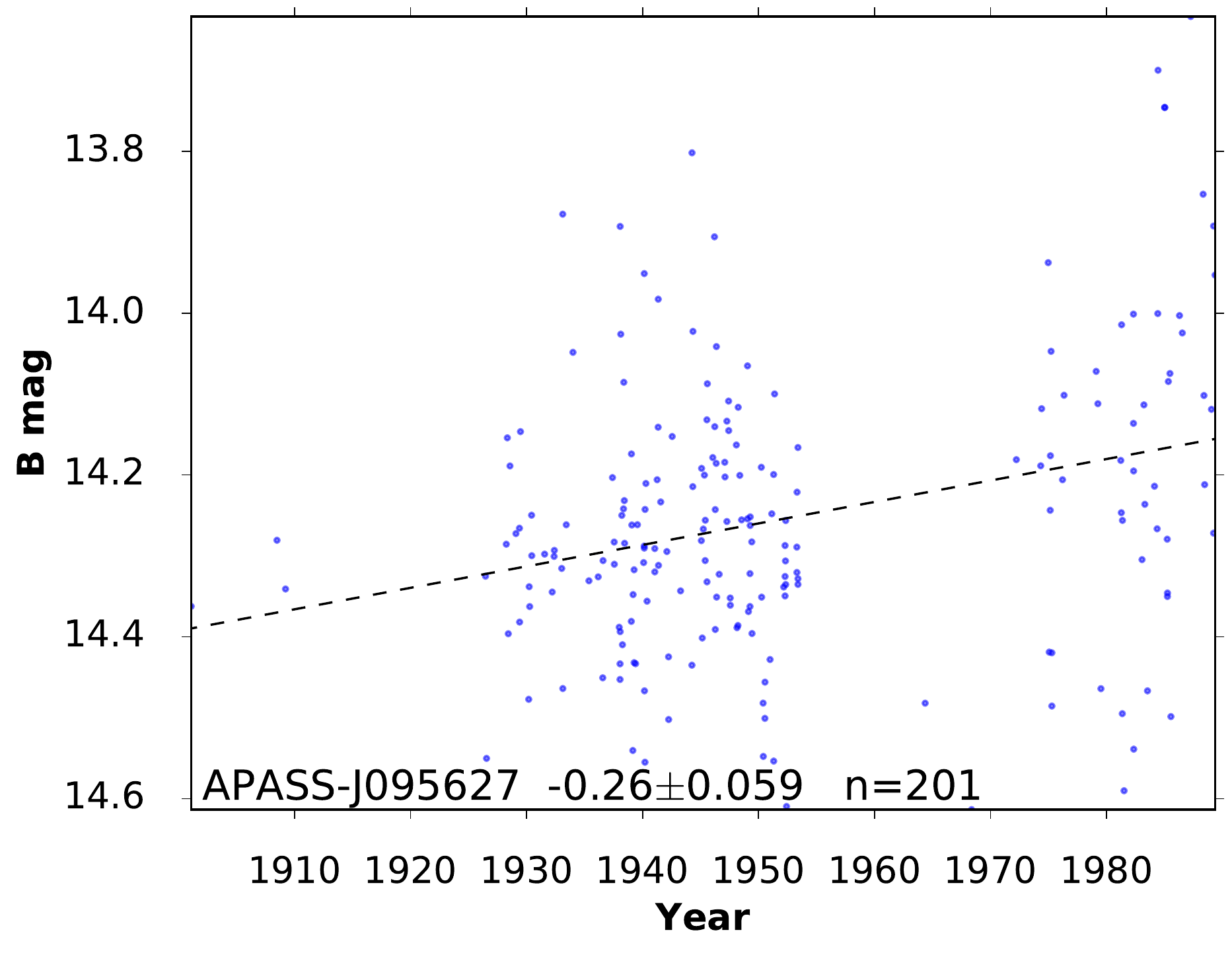}
\caption{\label{fig:land5}Photometry of Landolt standard stars with $n>100$ and all AFLAGS removed (continued).}
\end{figure*}

\begin{figure*}
\includegraphics[width=.5\linewidth]{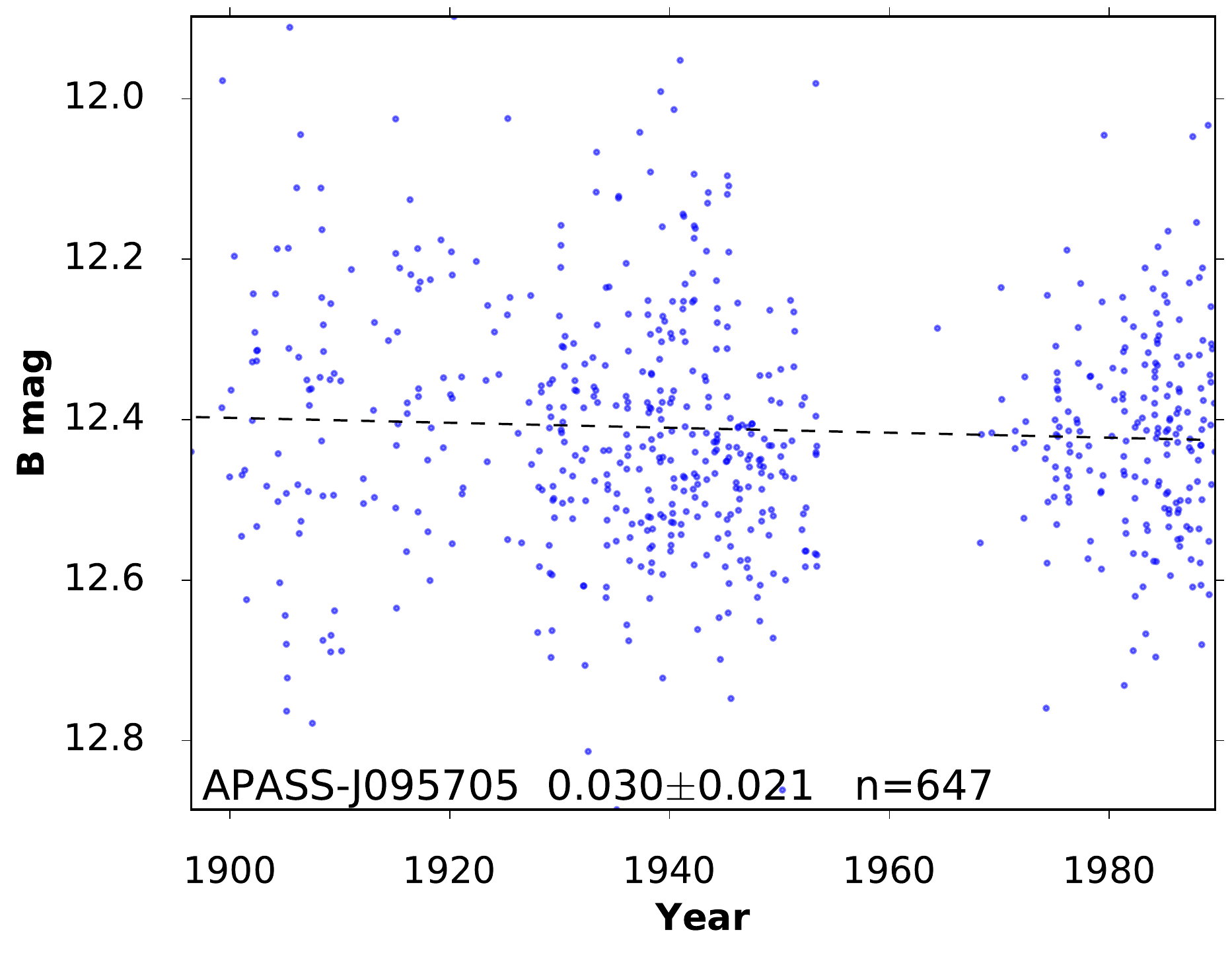}
\includegraphics[width=.5\linewidth]{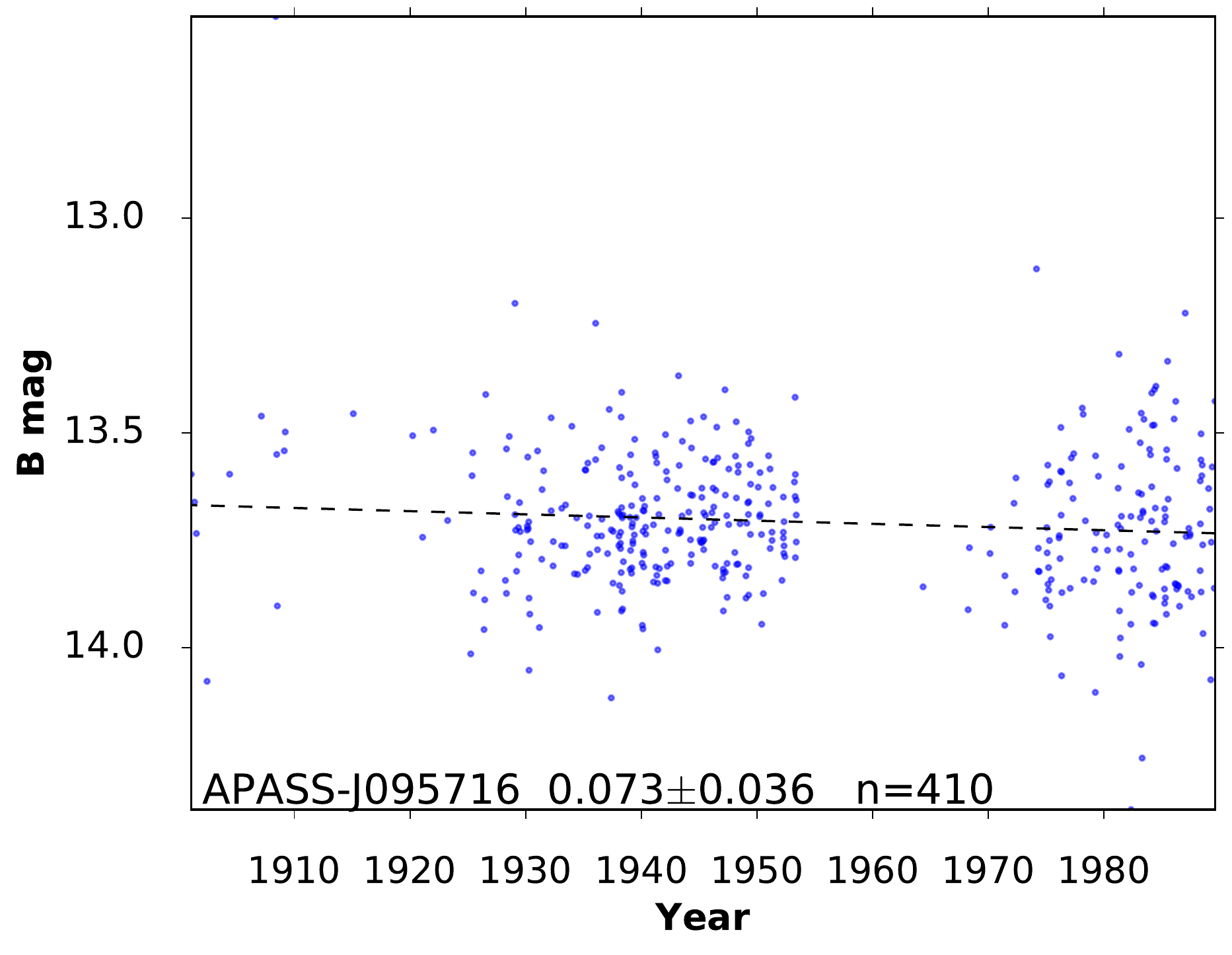}

\includegraphics[width=.5\linewidth]{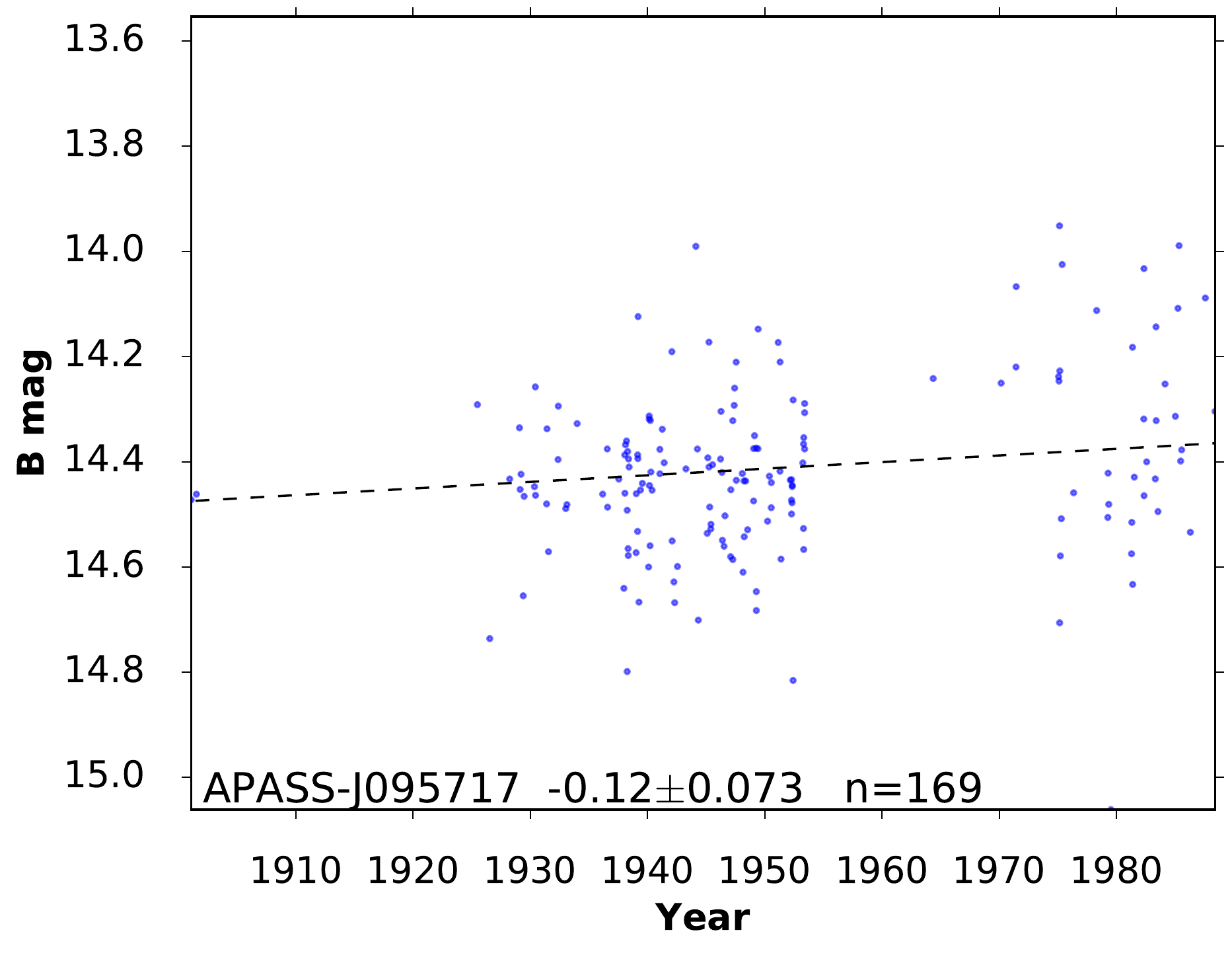}
\includegraphics[width=.5\linewidth]{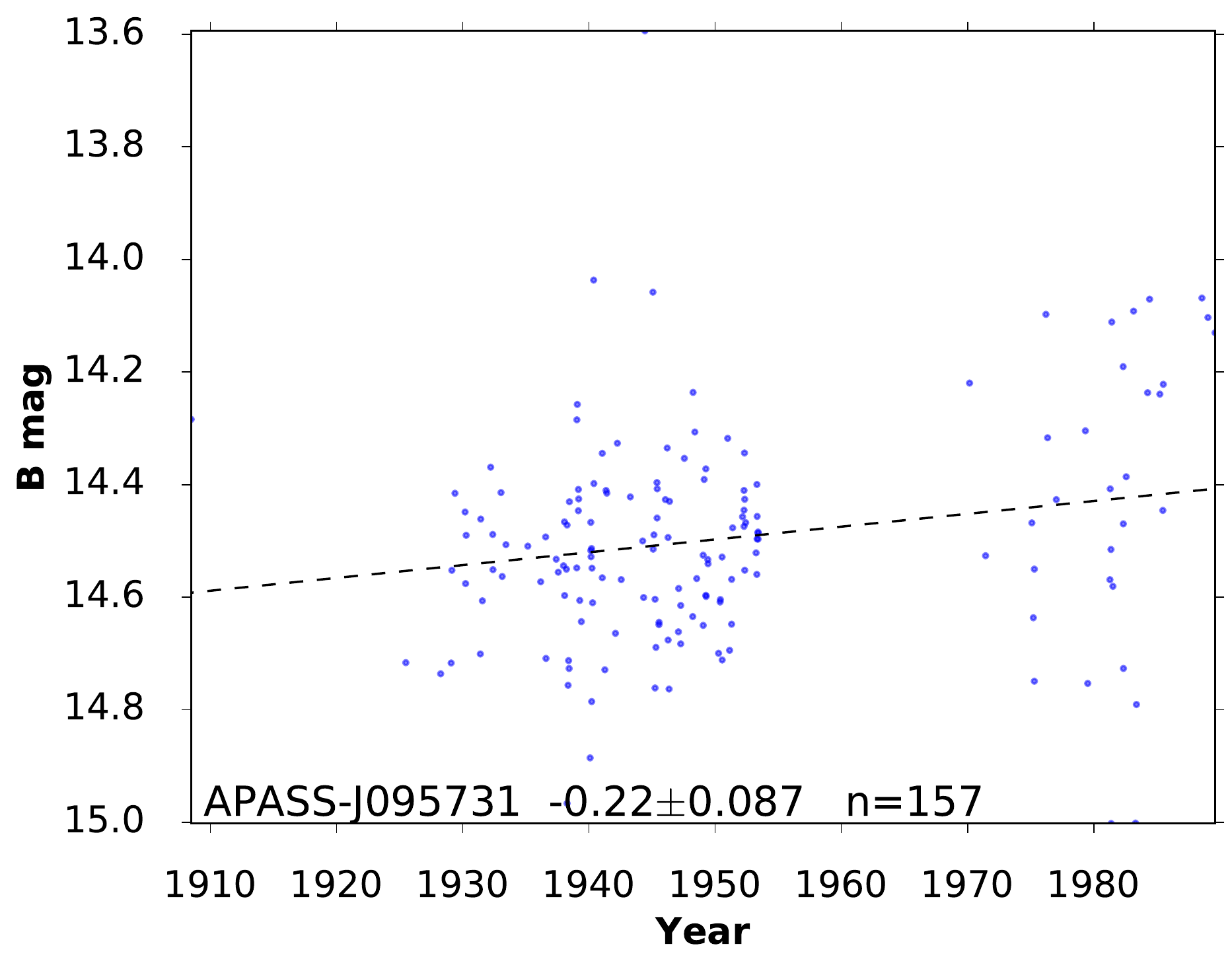}

\includegraphics[width=.5\linewidth]{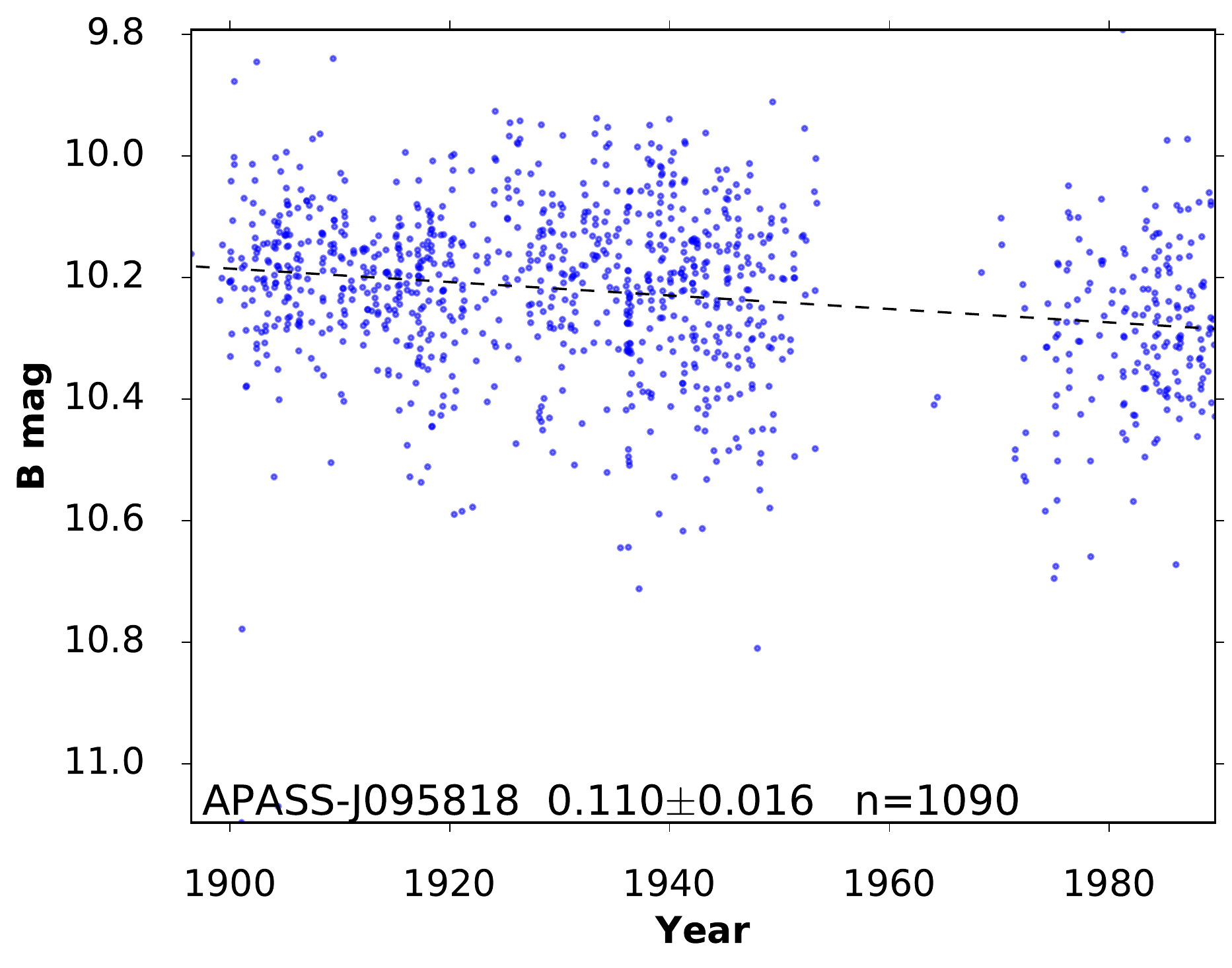}
\includegraphics[width=.5\linewidth]{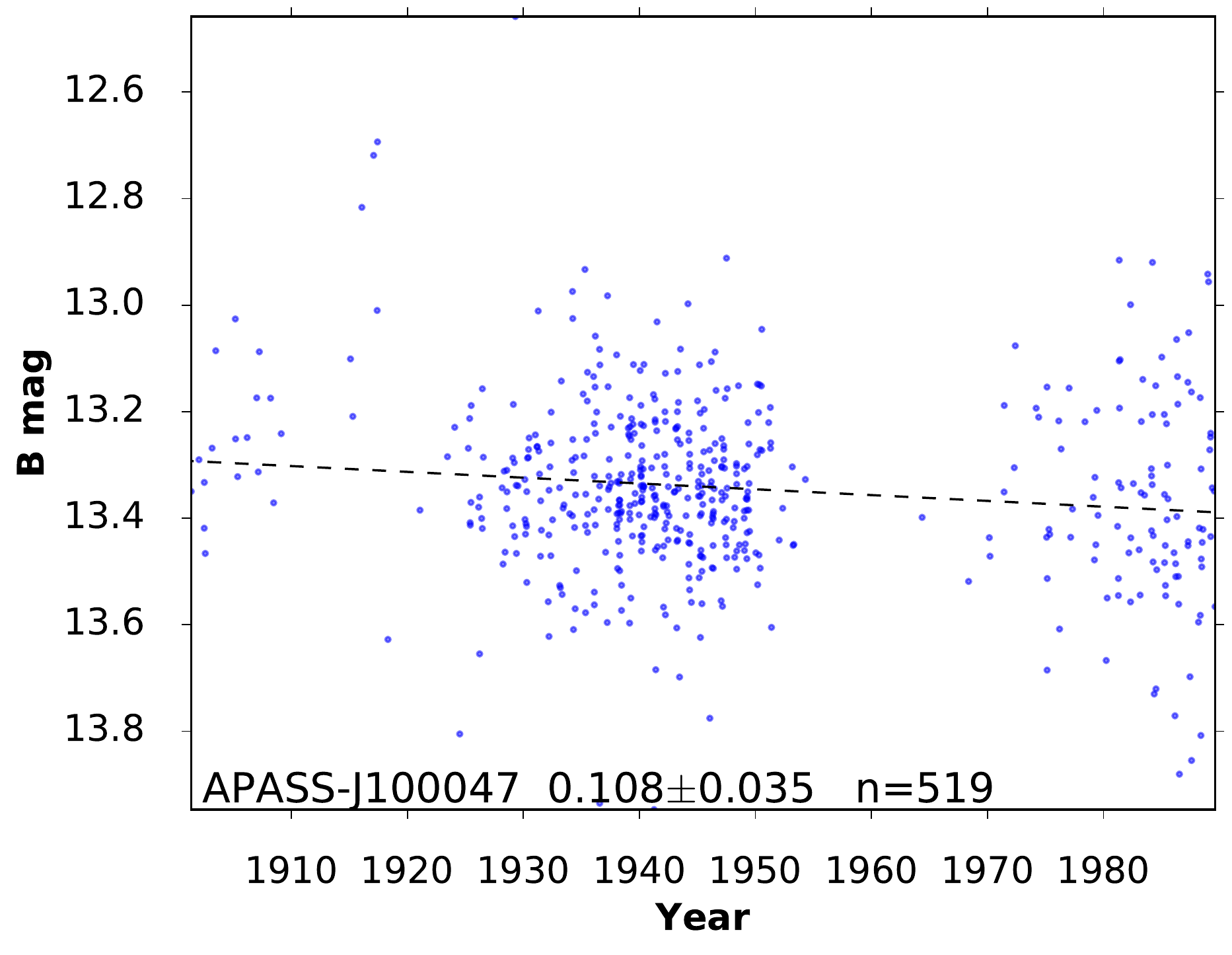}
\caption{\label{fig:land6}Photometry of Landolt standard stars with $n>100$ and all AFLAGS removed (continued).}
\end{figure*}

\begin{figure*}
\includegraphics[width=.5\linewidth]{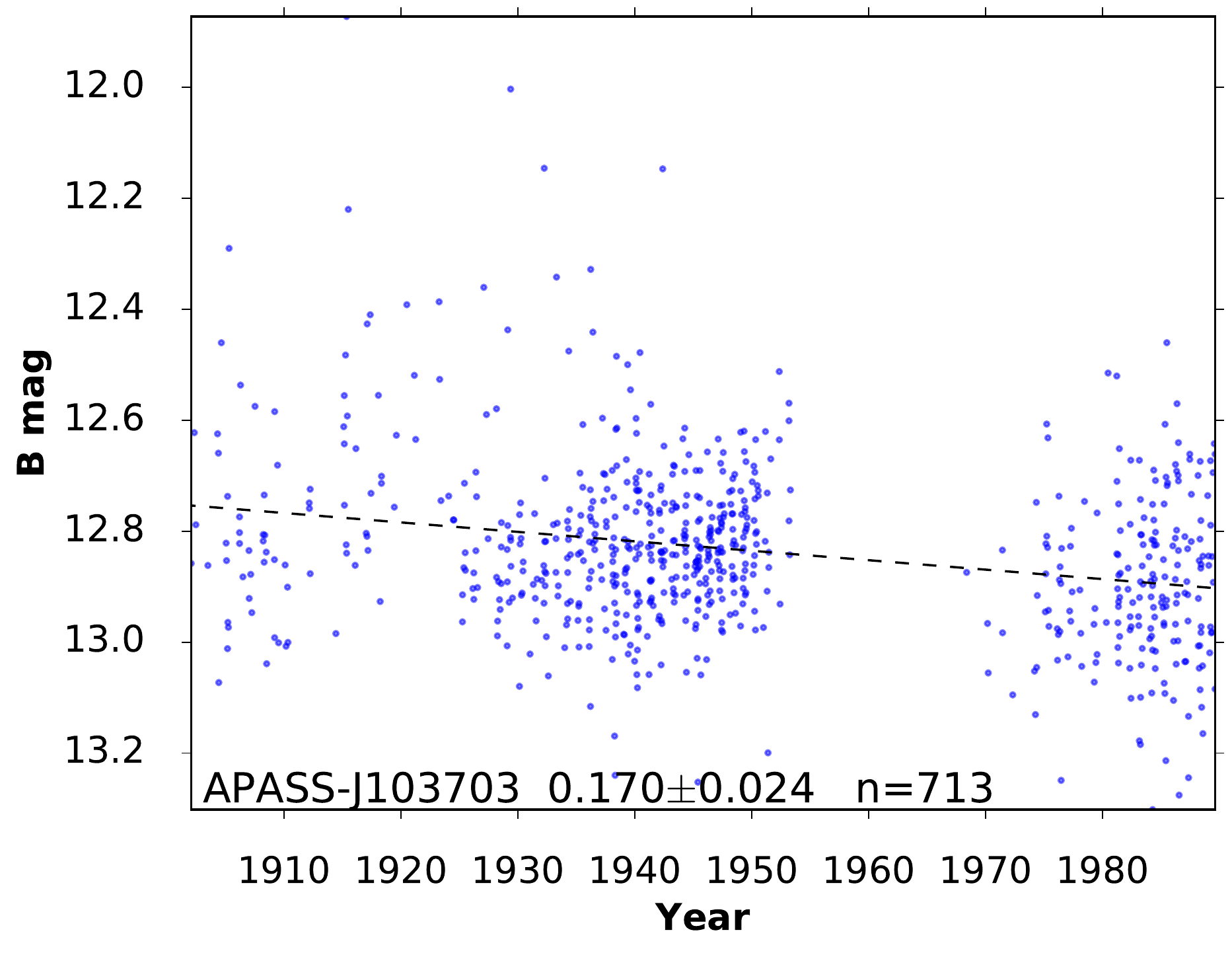}
\includegraphics[width=.5\linewidth]{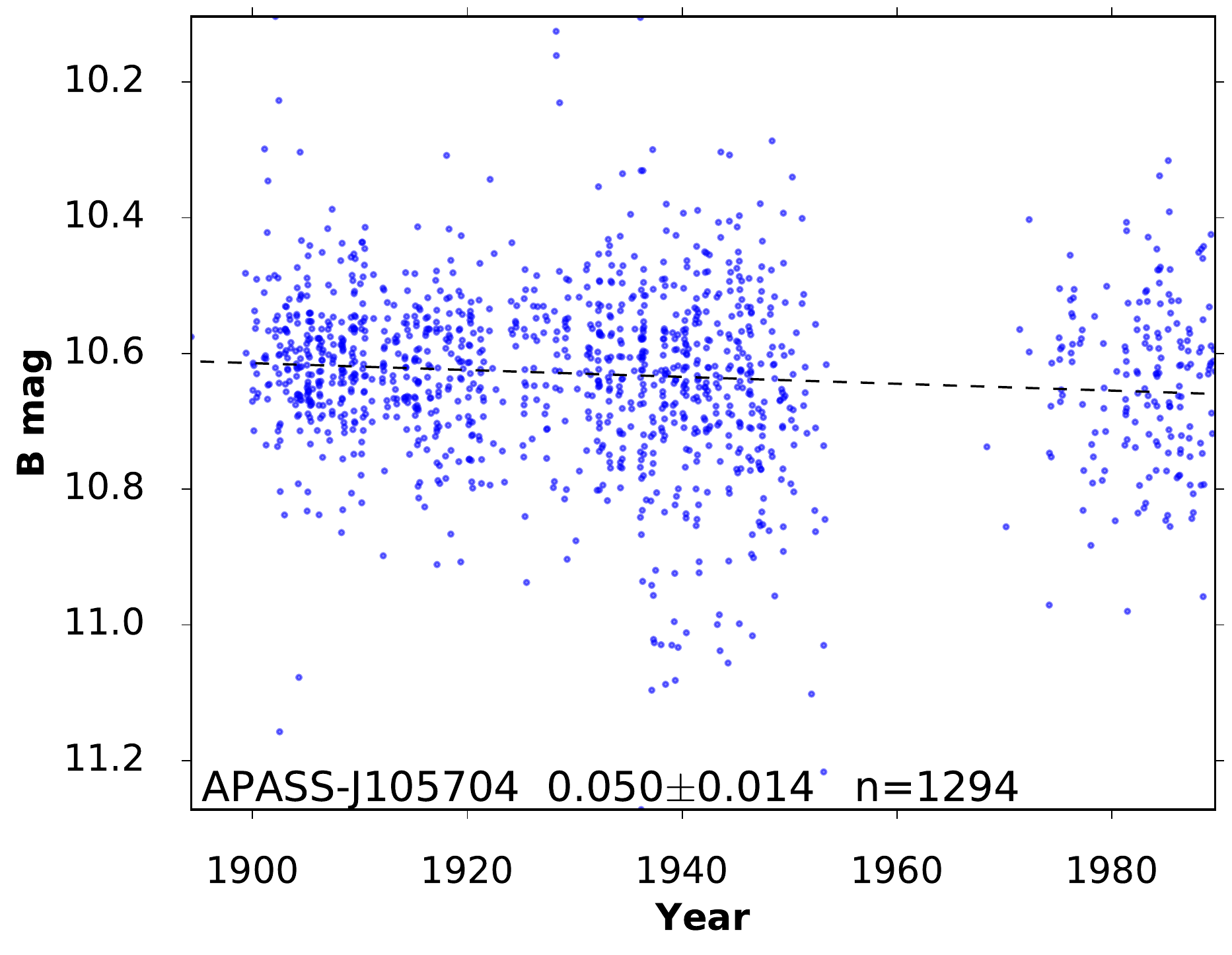}

\includegraphics[width=.5\linewidth]{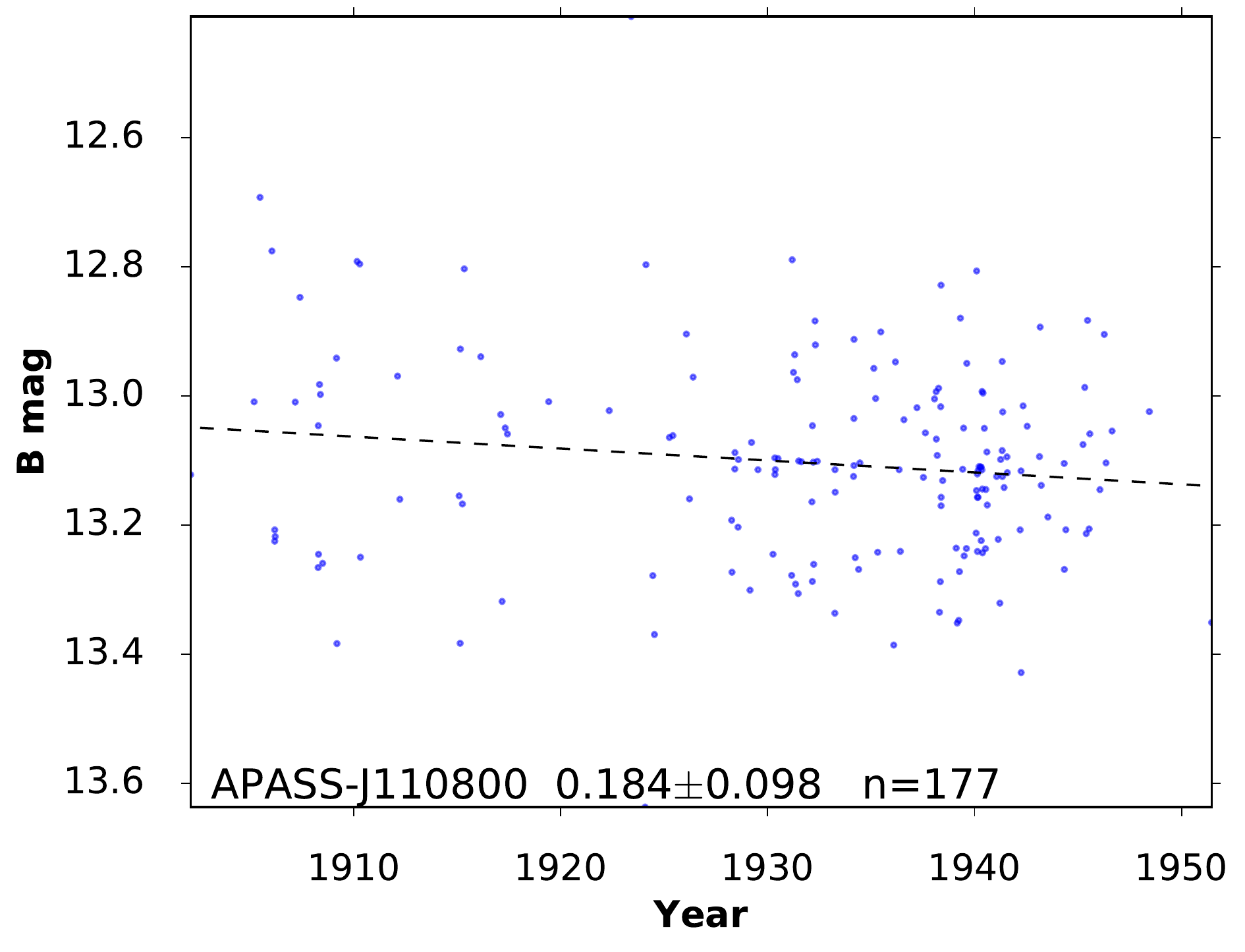}
\includegraphics[width=.5\linewidth]{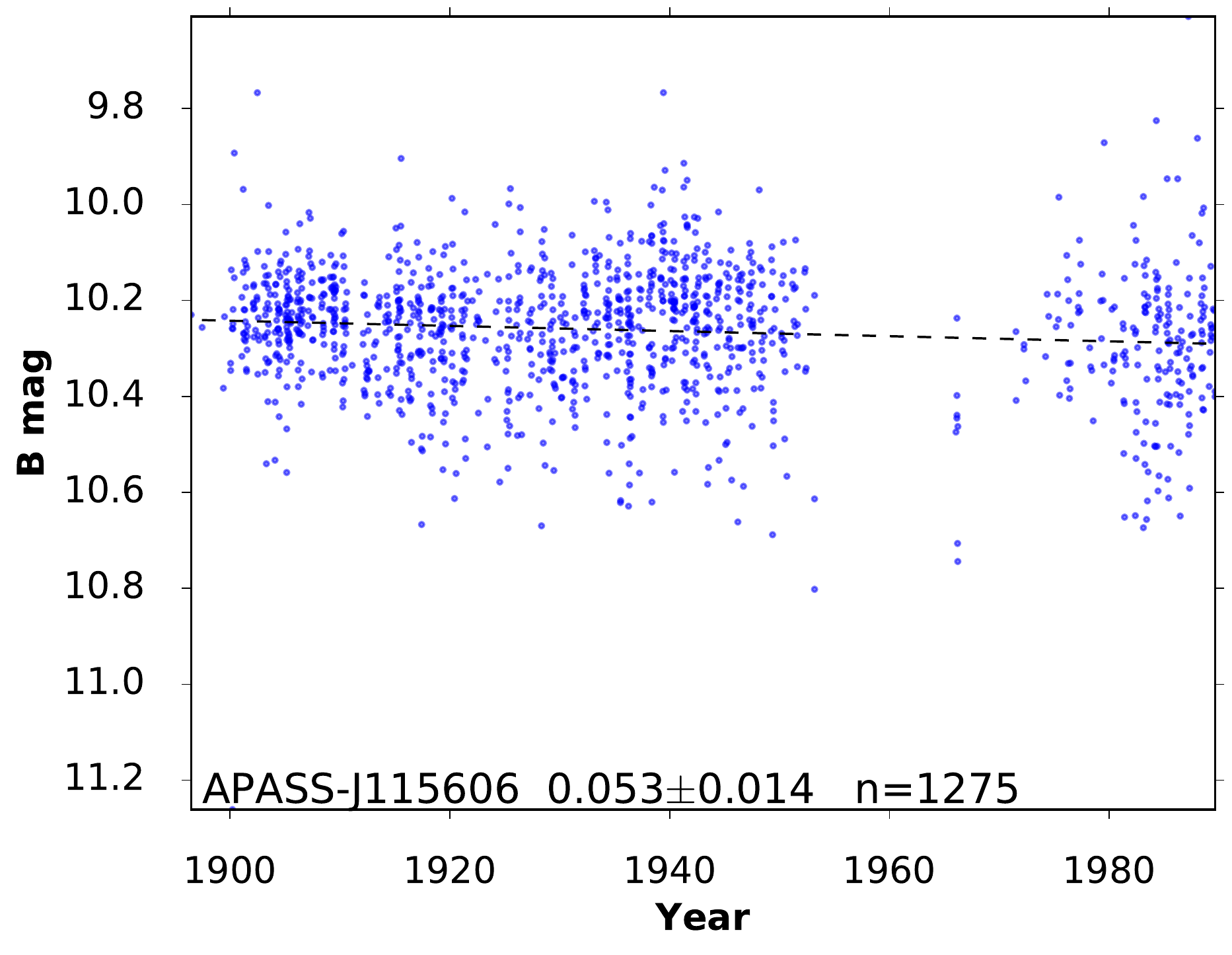}

\includegraphics[width=.5\linewidth]{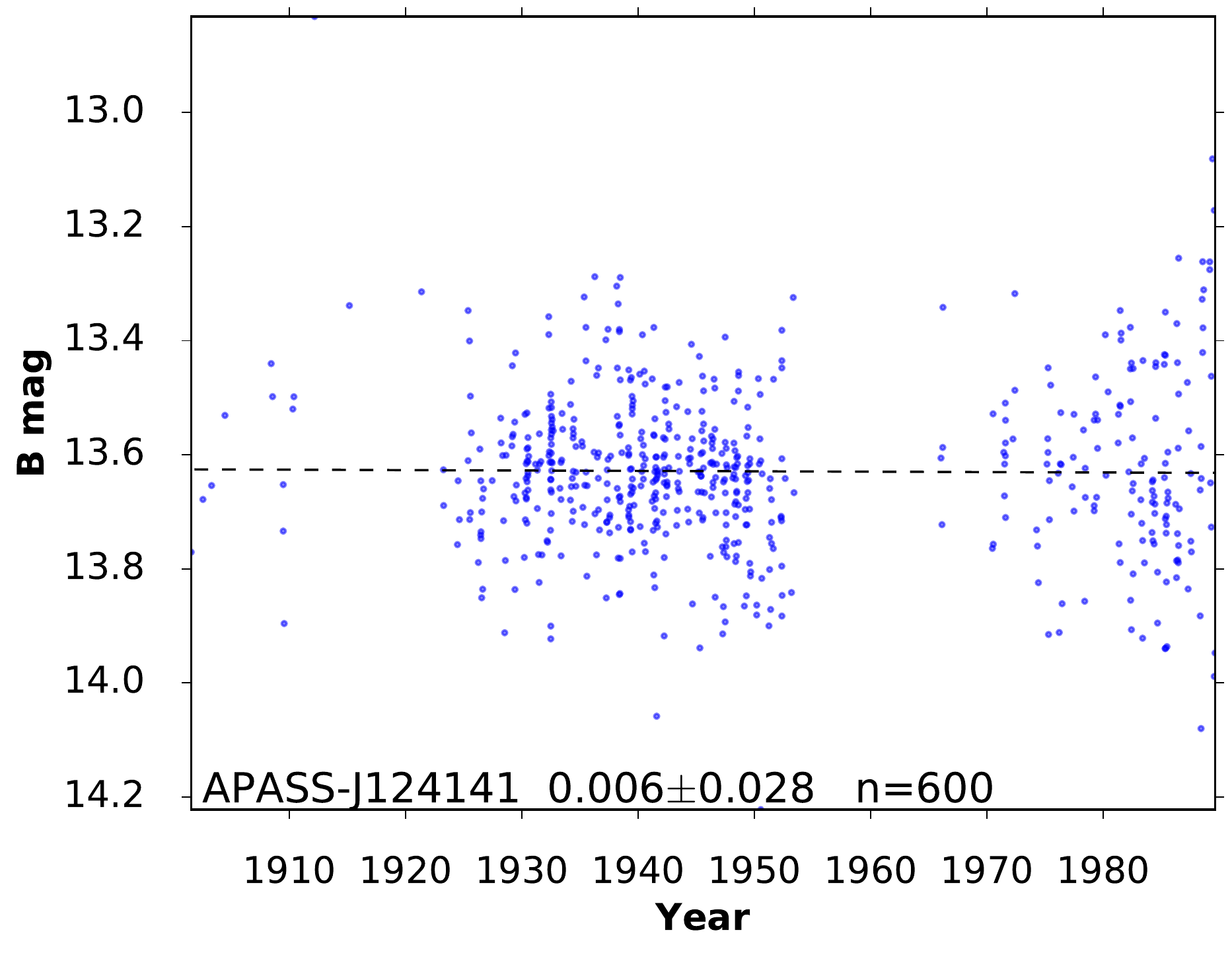}
\includegraphics[width=.5\linewidth]{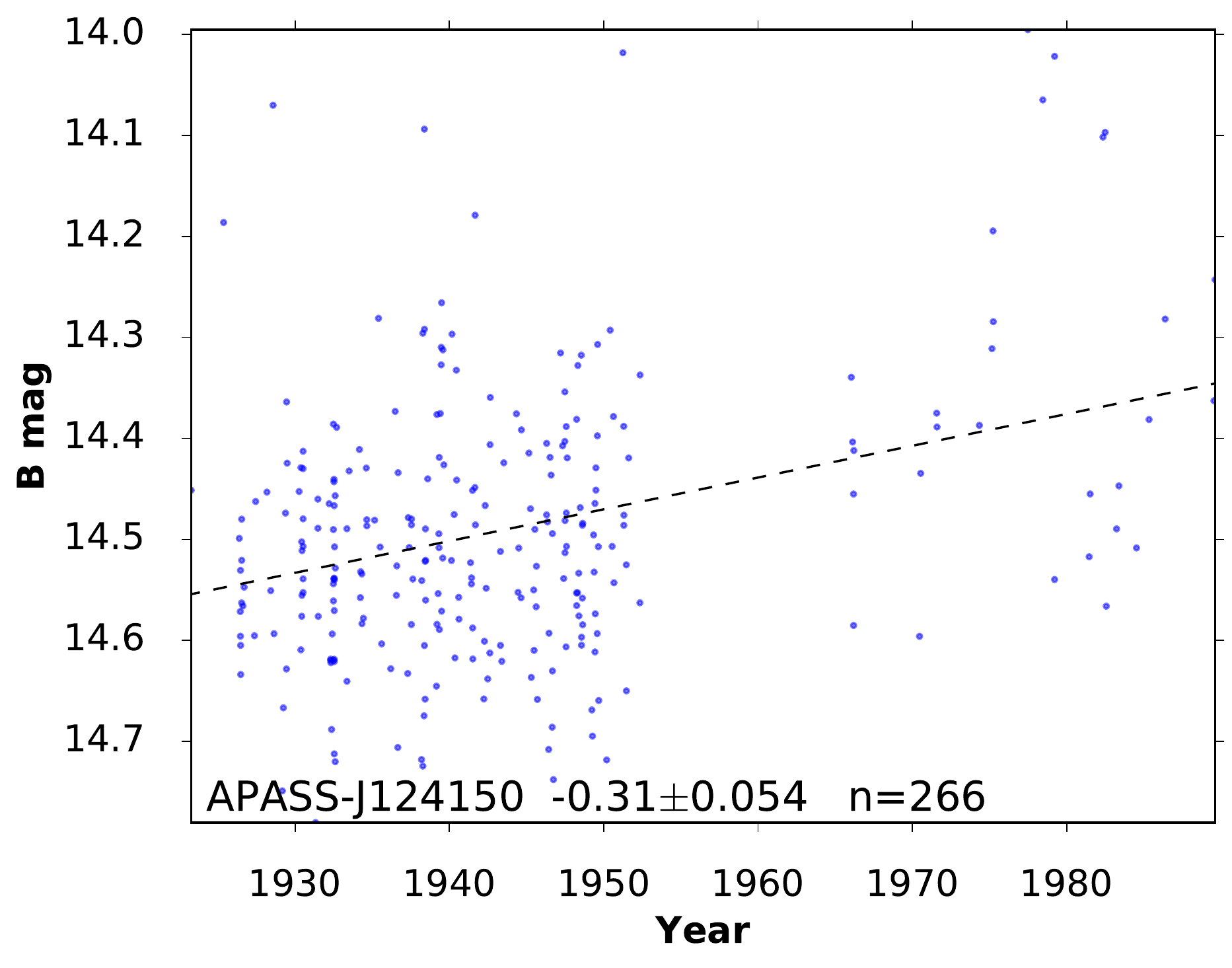}
\caption{\label{fig:land7}Photometry of Landolt standard stars with $n>100$ and all AFLAGS removed (continued).}
\end{figure*}

\begin{figure*}
\includegraphics[width=.5\linewidth]{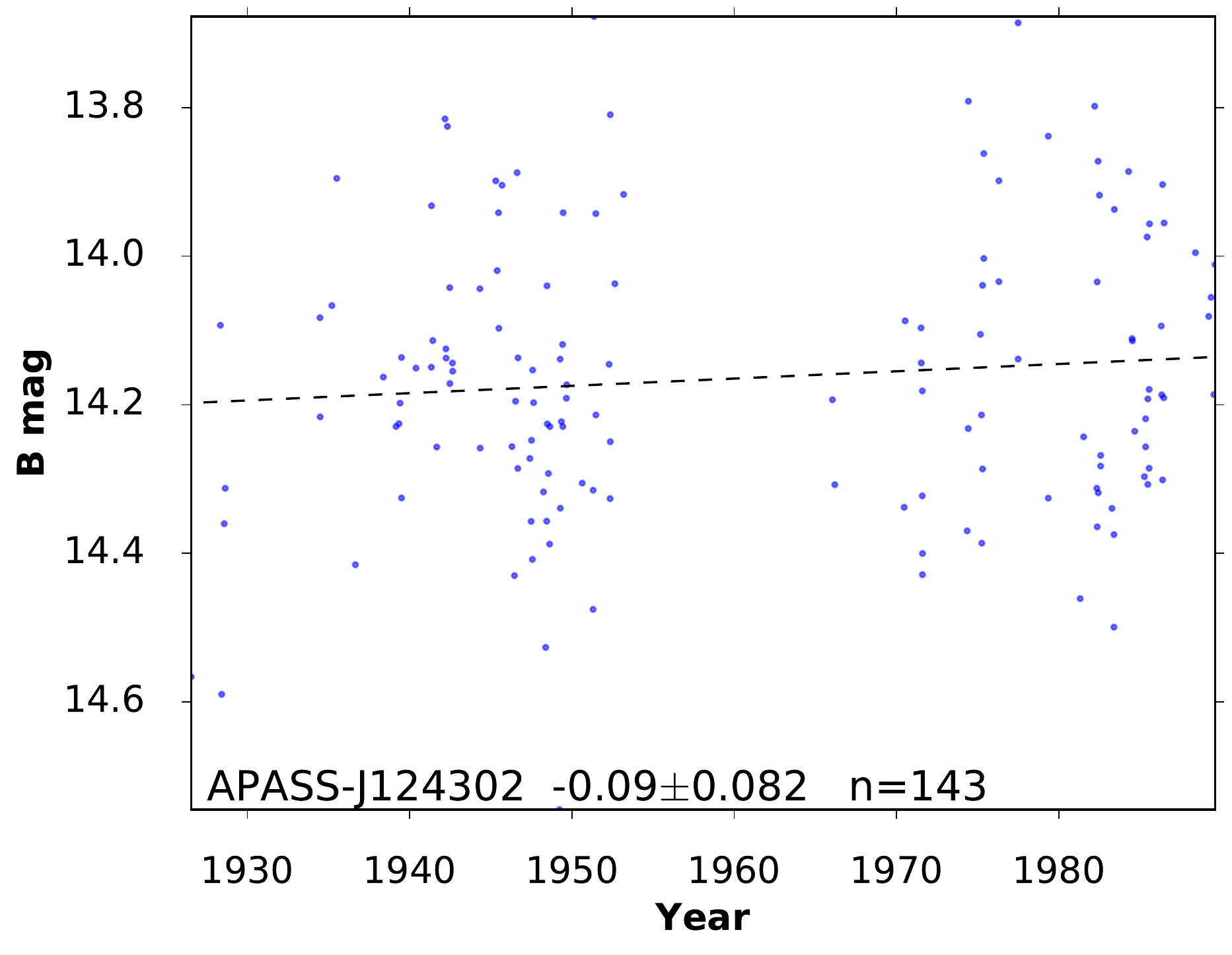}
\includegraphics[width=.5\linewidth]{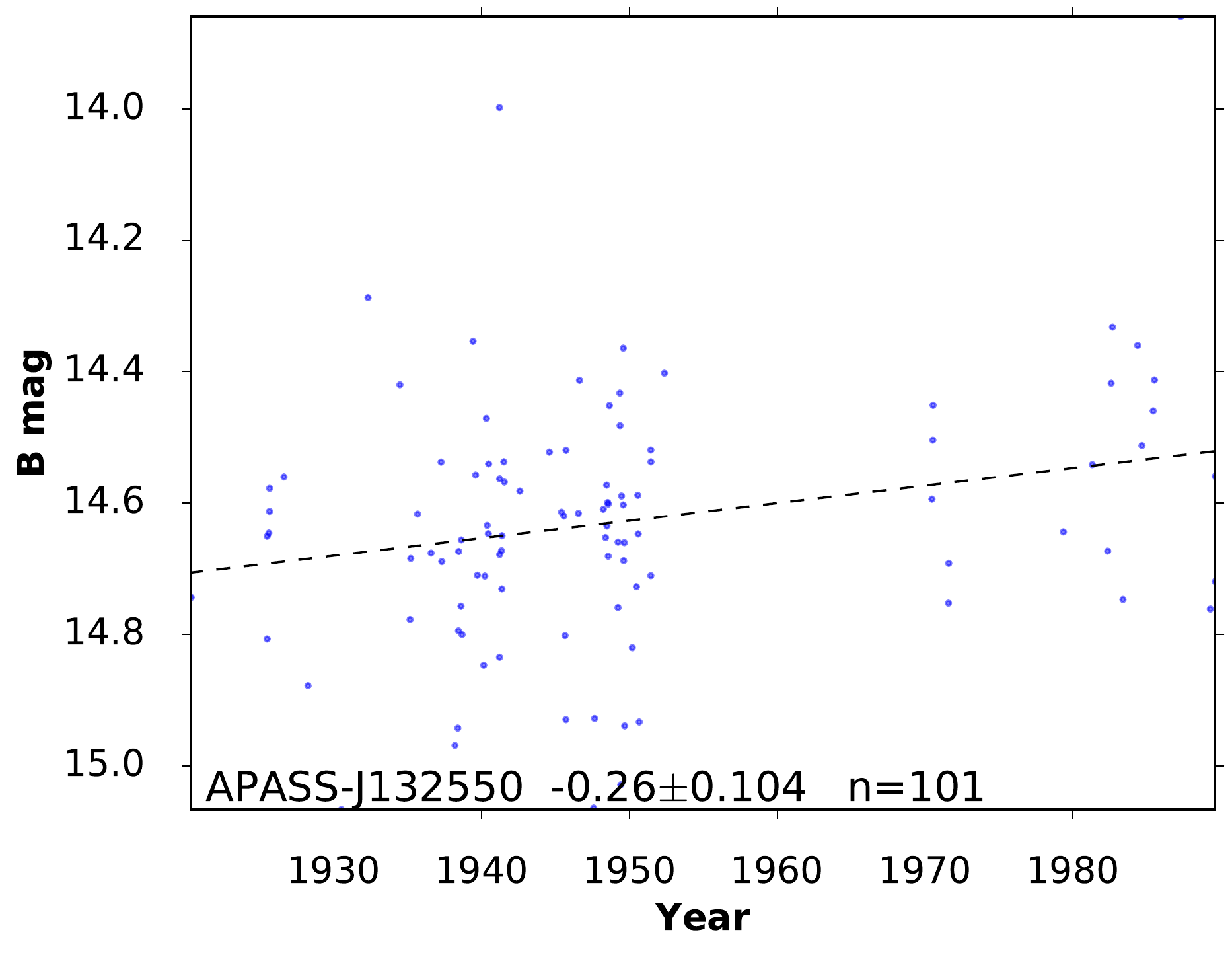}

\includegraphics[width=.5\linewidth]{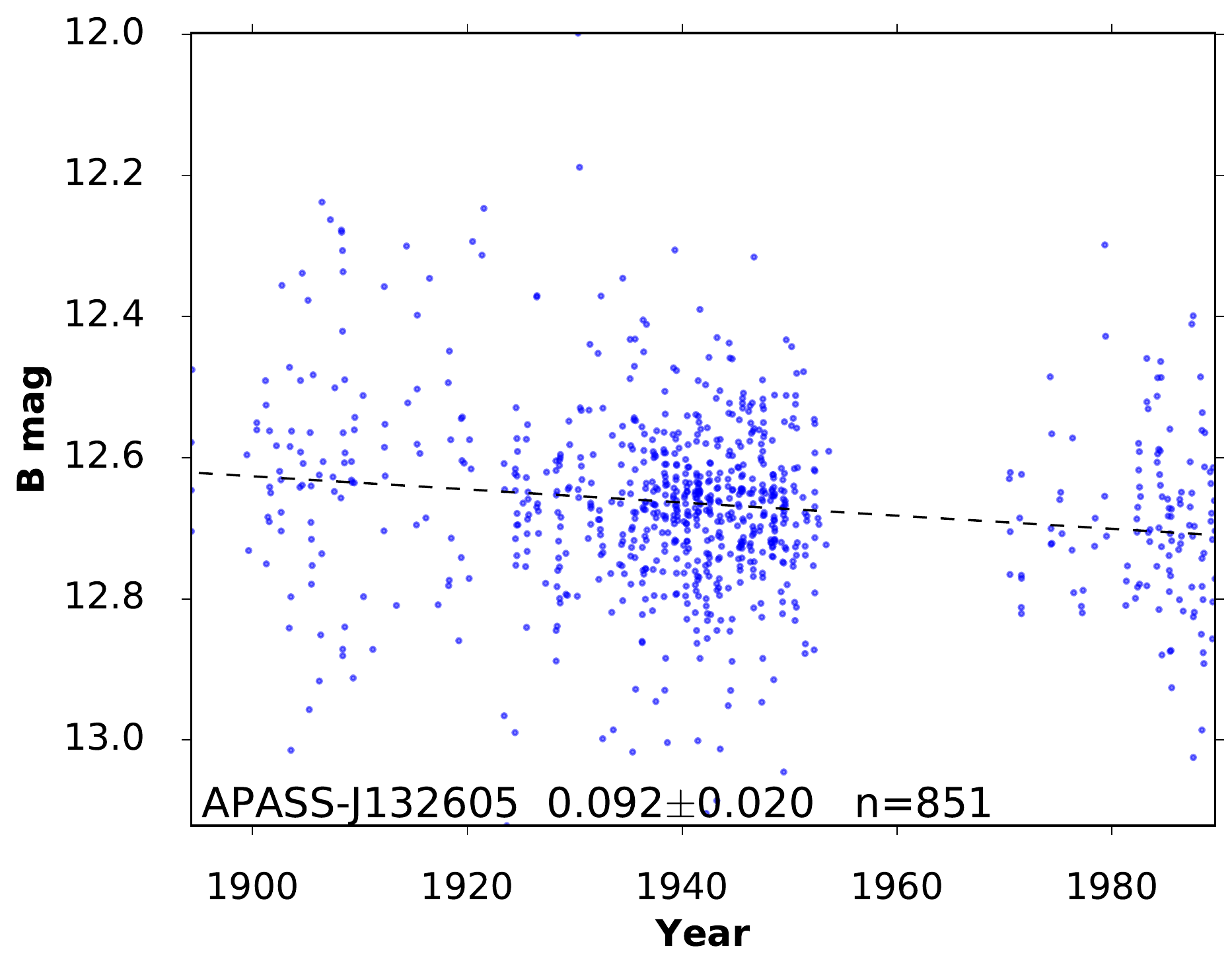}
\includegraphics[width=.5\linewidth]{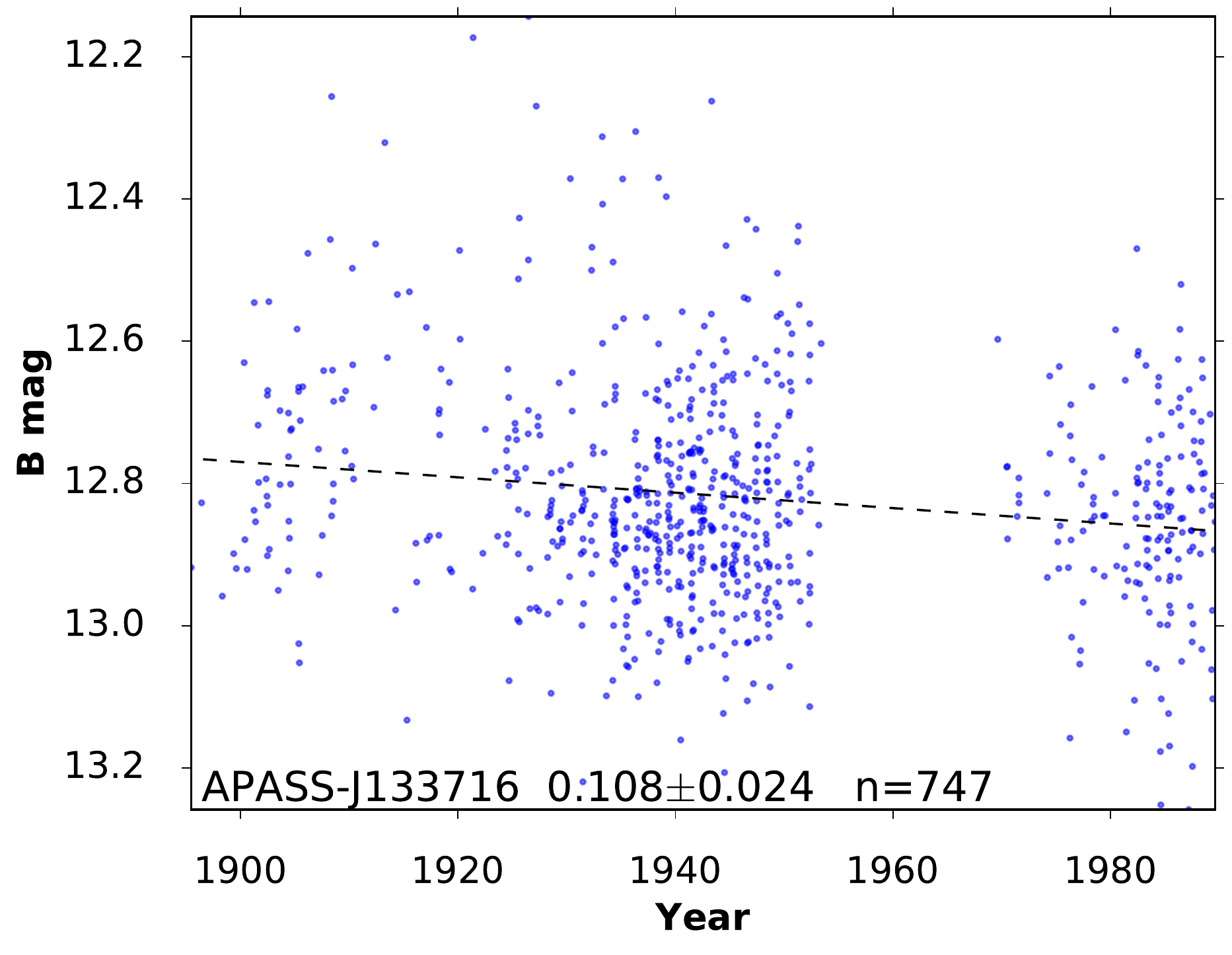}

\includegraphics[width=.5\linewidth]{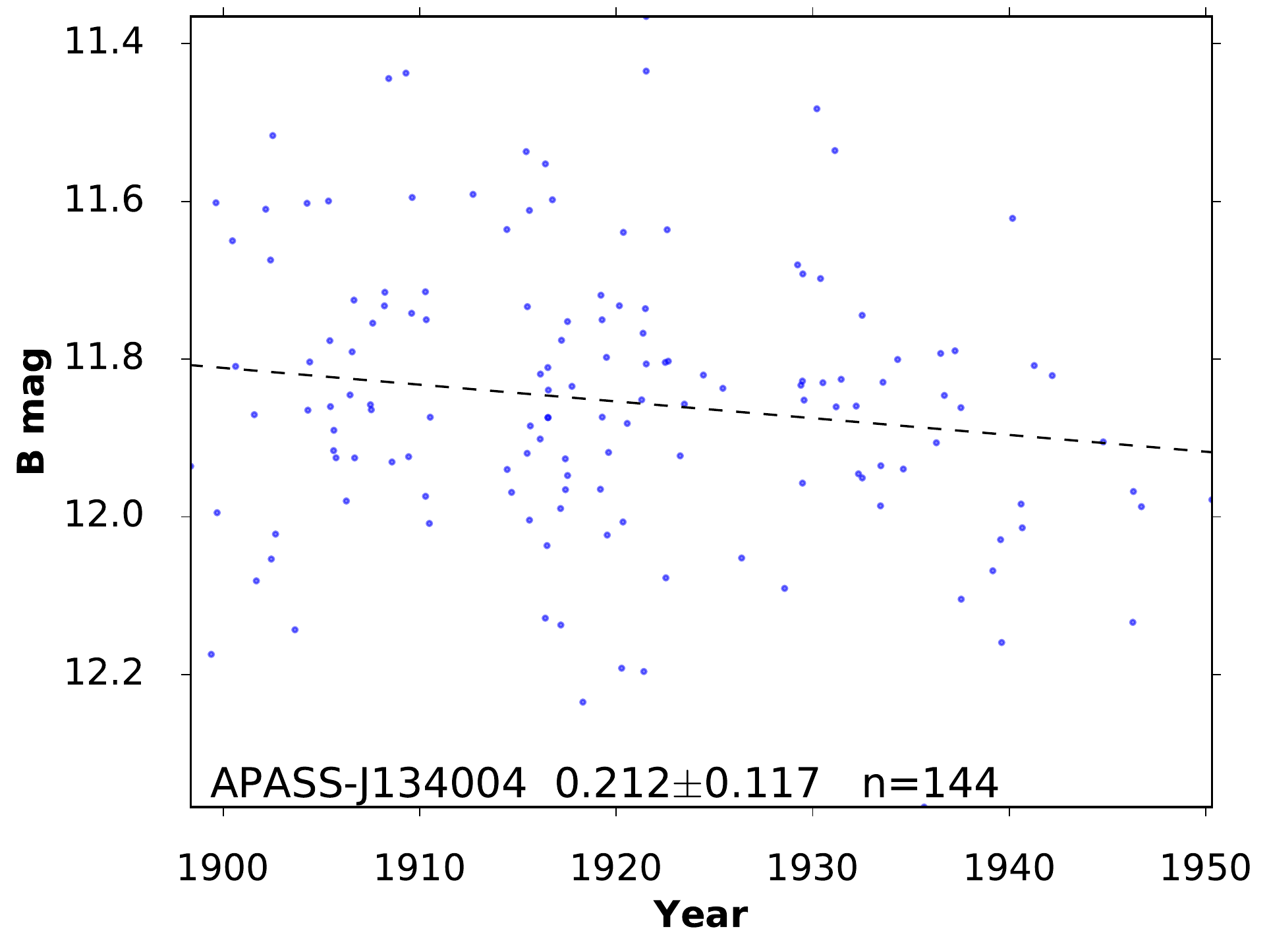}
\includegraphics[width=.5\linewidth]{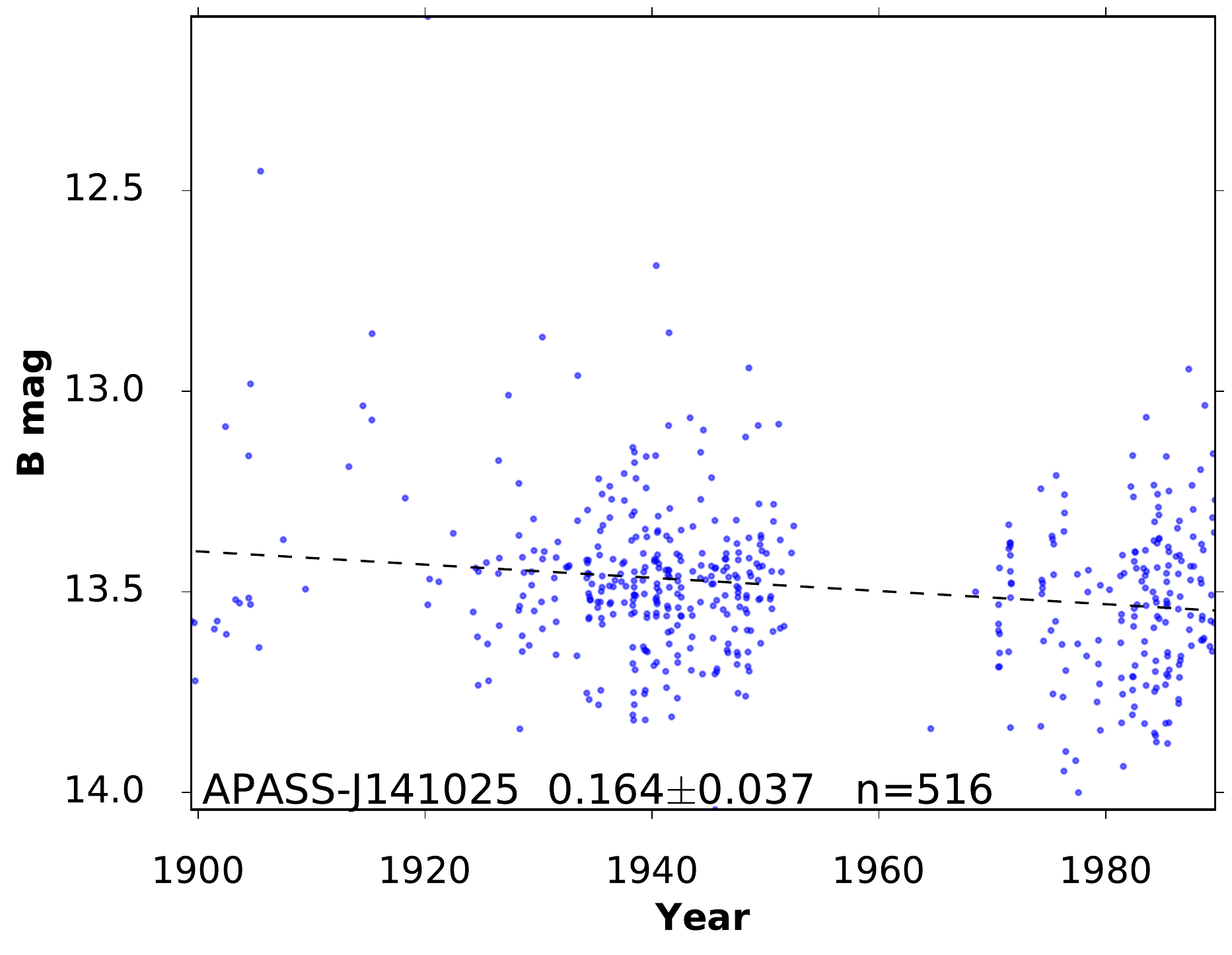}
\caption{\label{fig:land8}Photometry of Landolt standard stars with $n>100$ and all AFLAGS removed (continued).}
\end{figure*}

\begin{figure*}
\includegraphics[width=.5\linewidth]{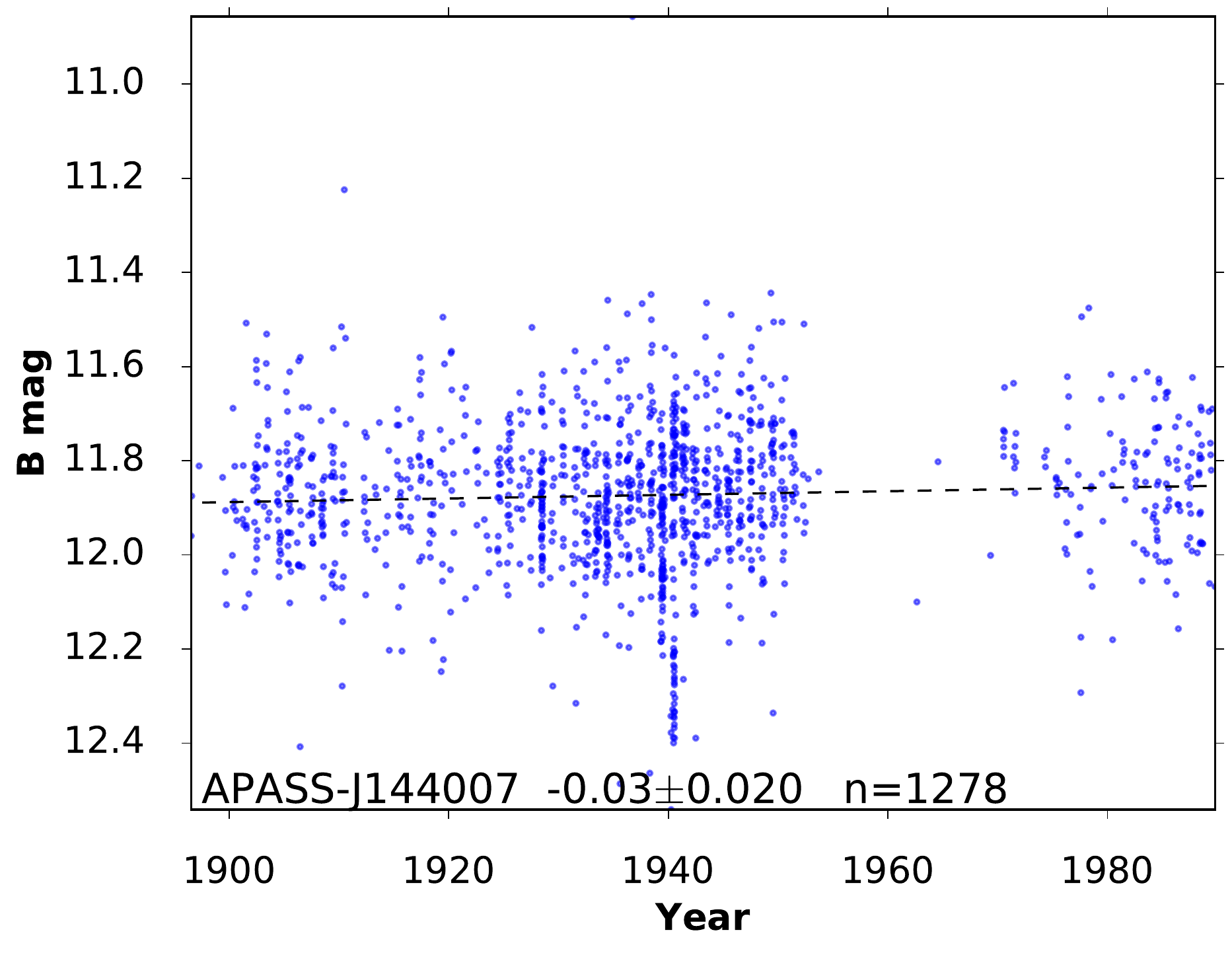}
\includegraphics[width=.5\linewidth]{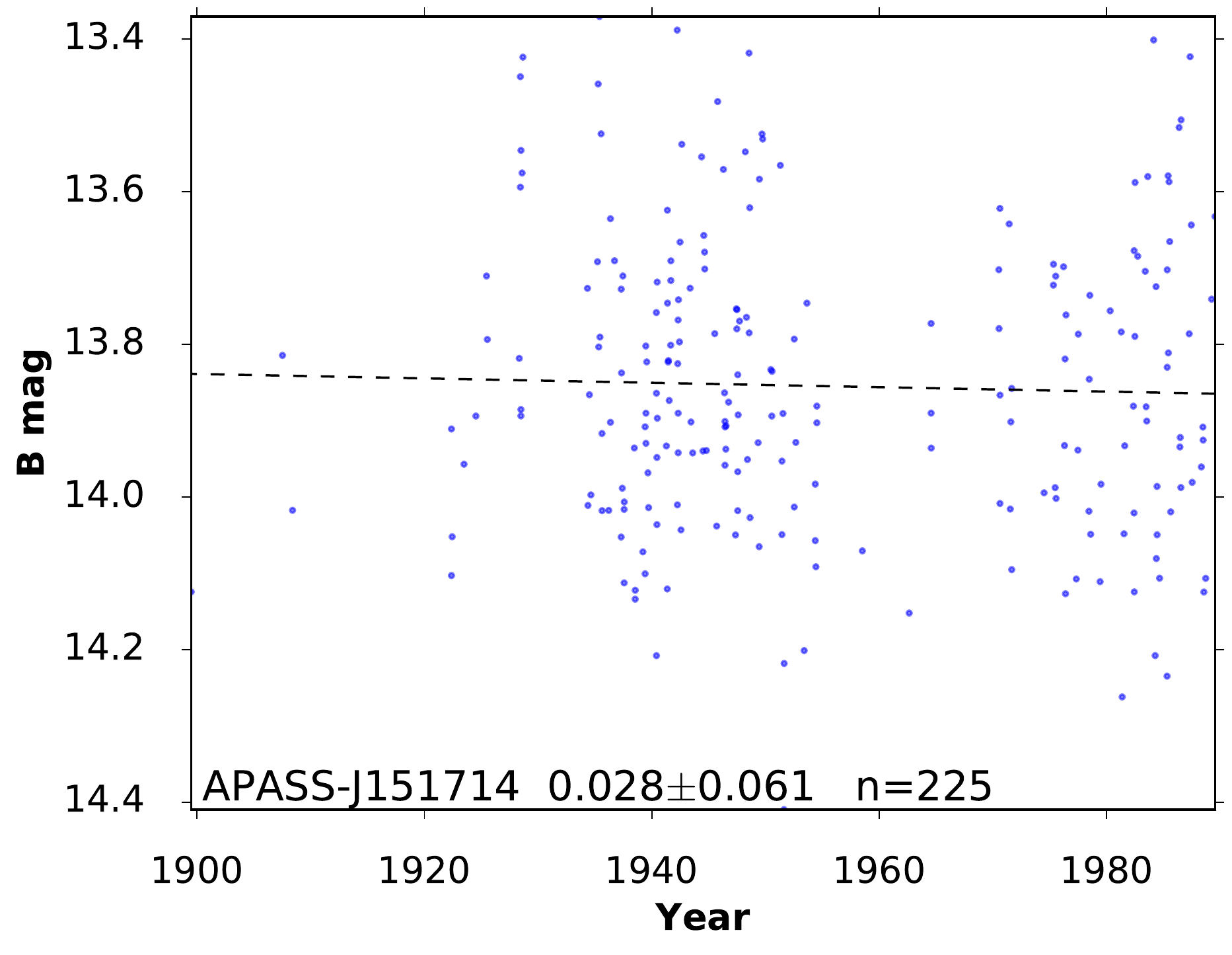}

\includegraphics[width=.5\linewidth]{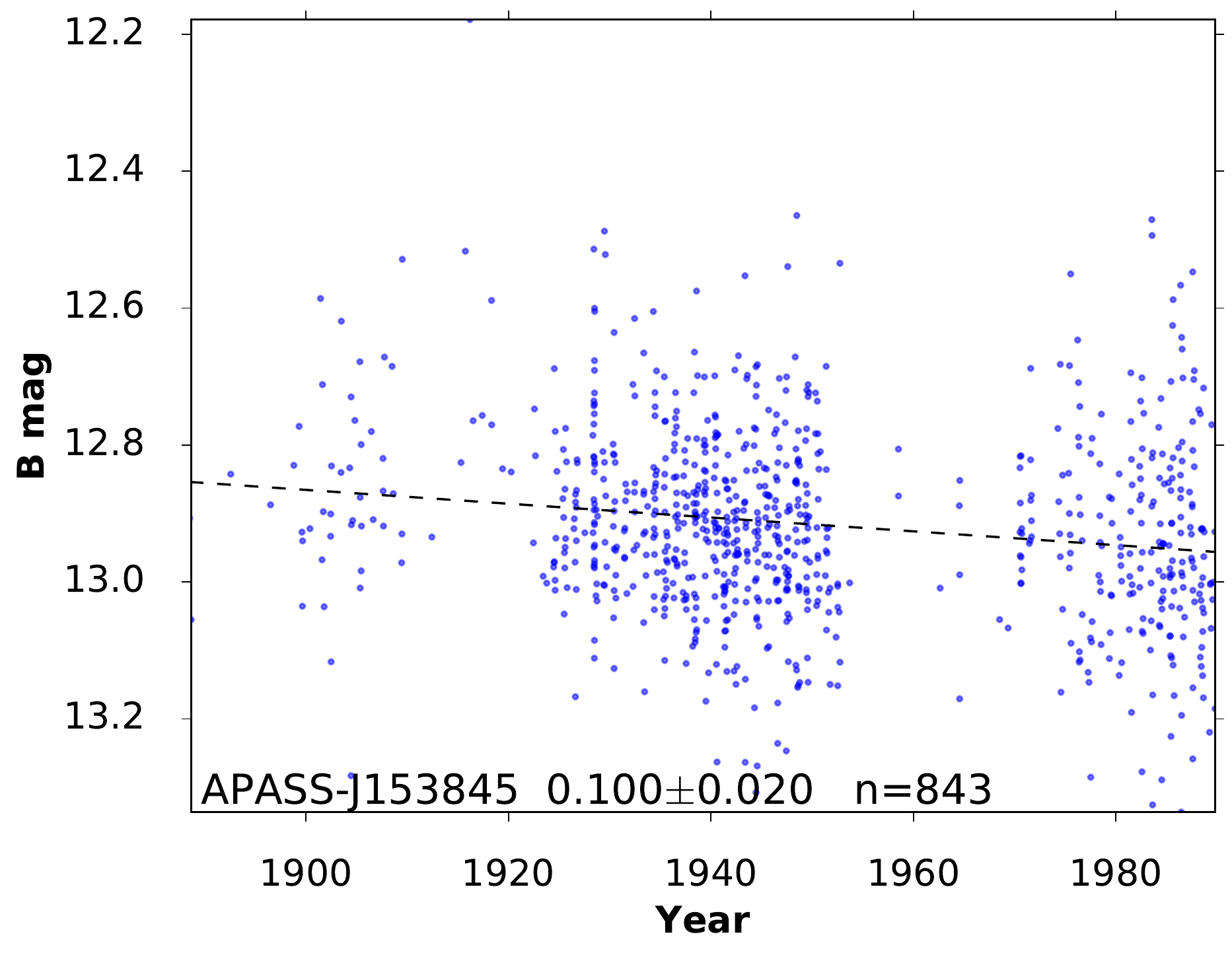}
\includegraphics[width=.5\linewidth]{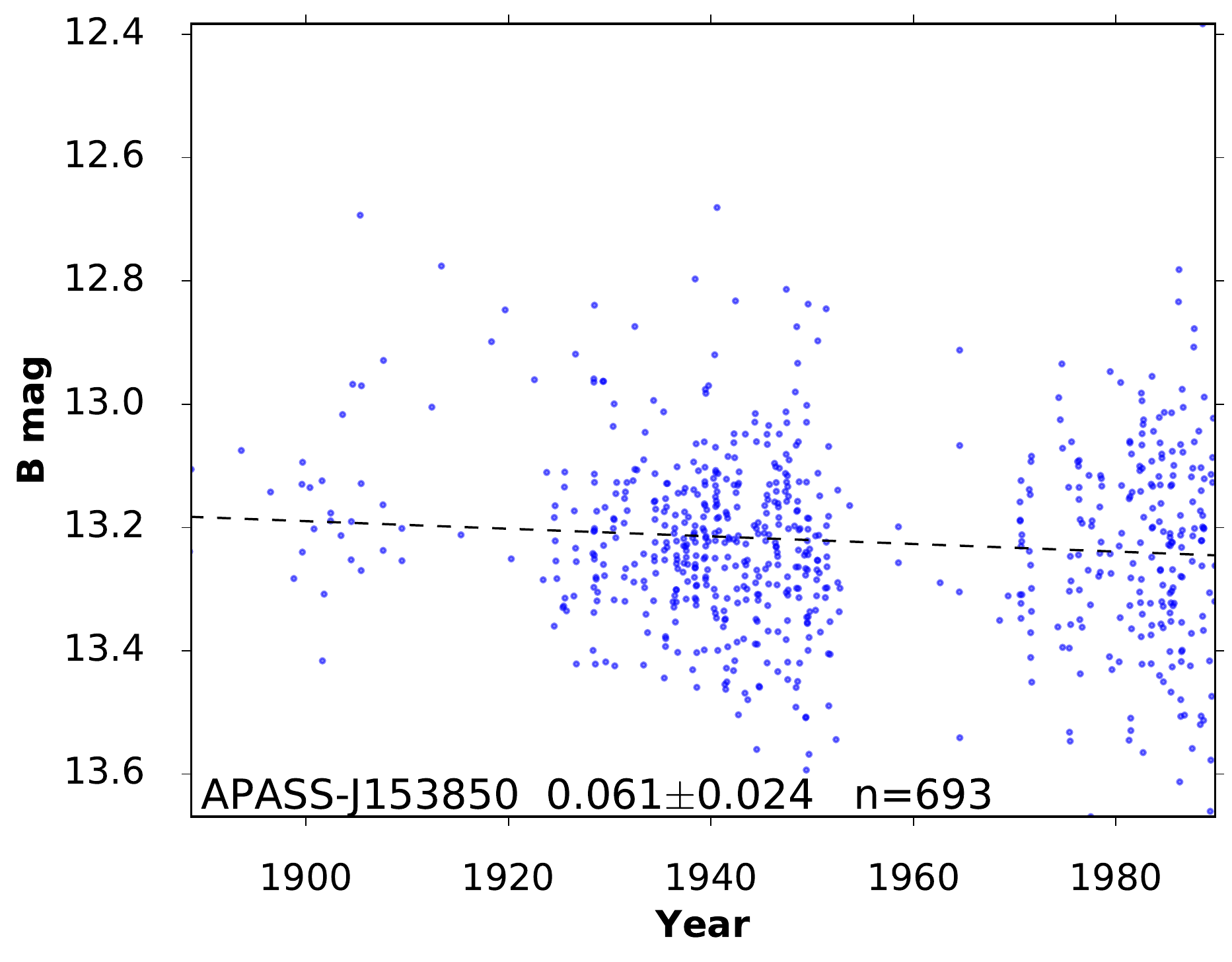}

\includegraphics[width=.5\linewidth]{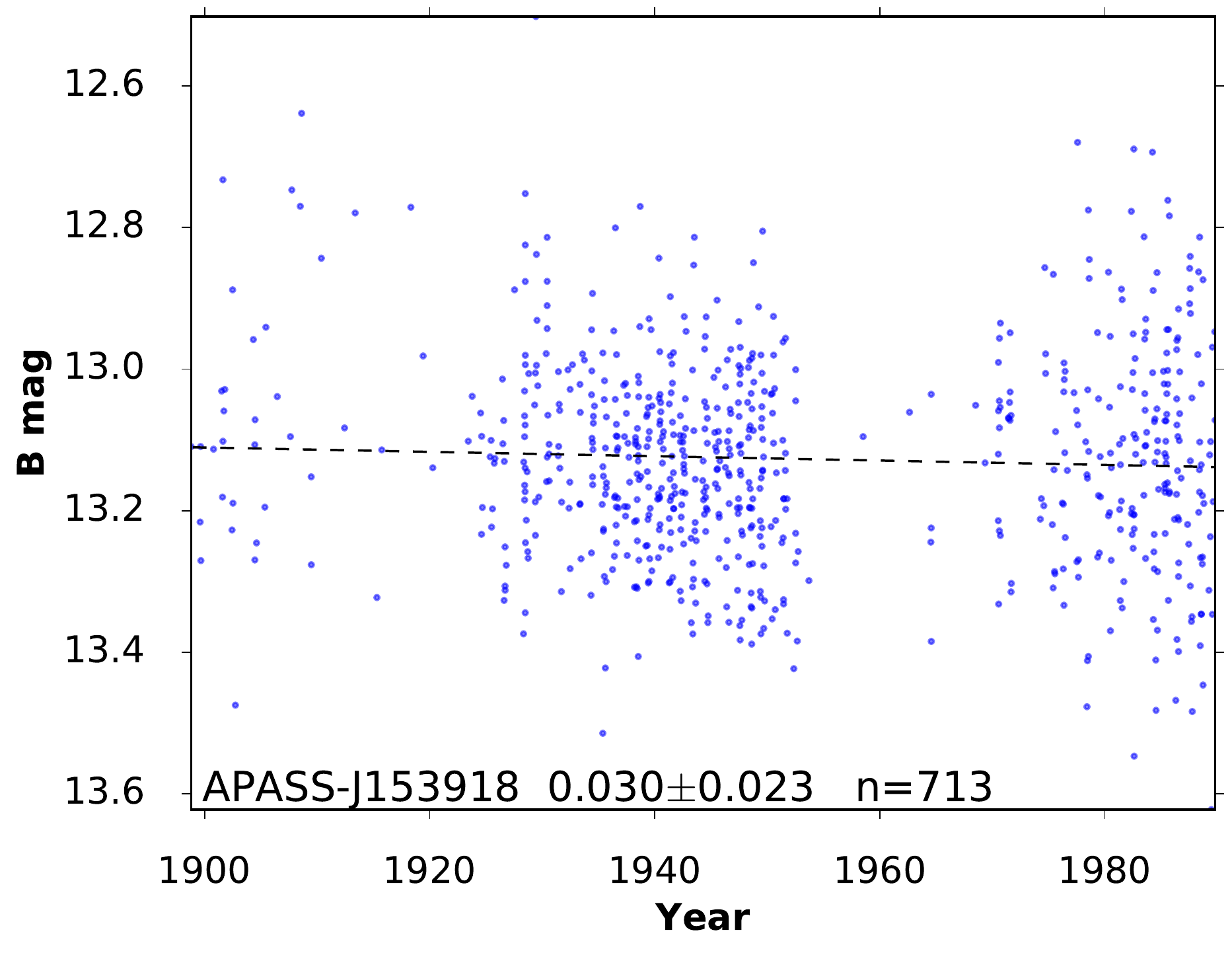}
\includegraphics[width=.5\linewidth]{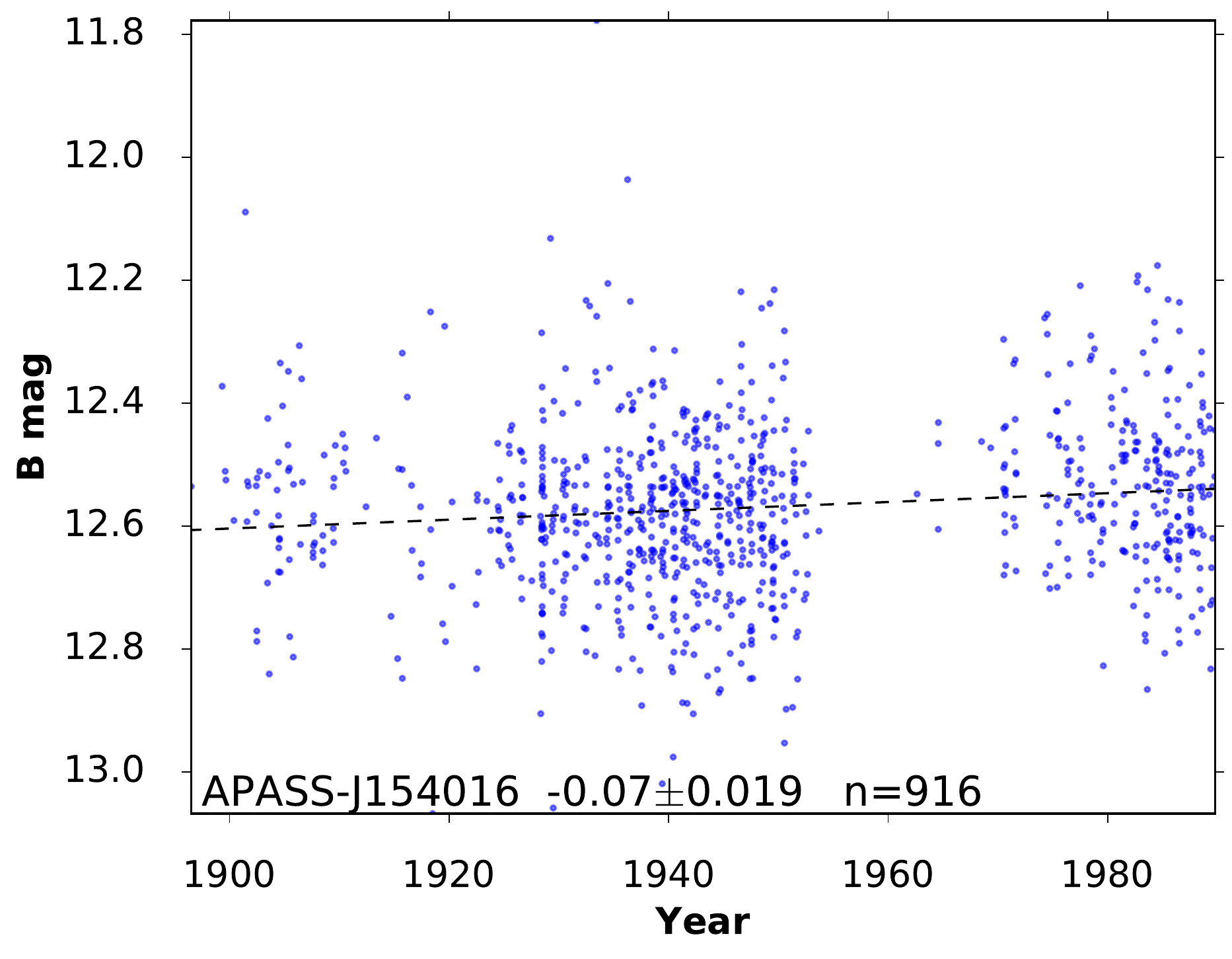}
\caption{\label{fig:land9}Photometry of Landolt standard stars with $n>100$ and all AFLAGS removed (continued).}
\end{figure*}

\begin{figure*}
\includegraphics[width=.5\linewidth]{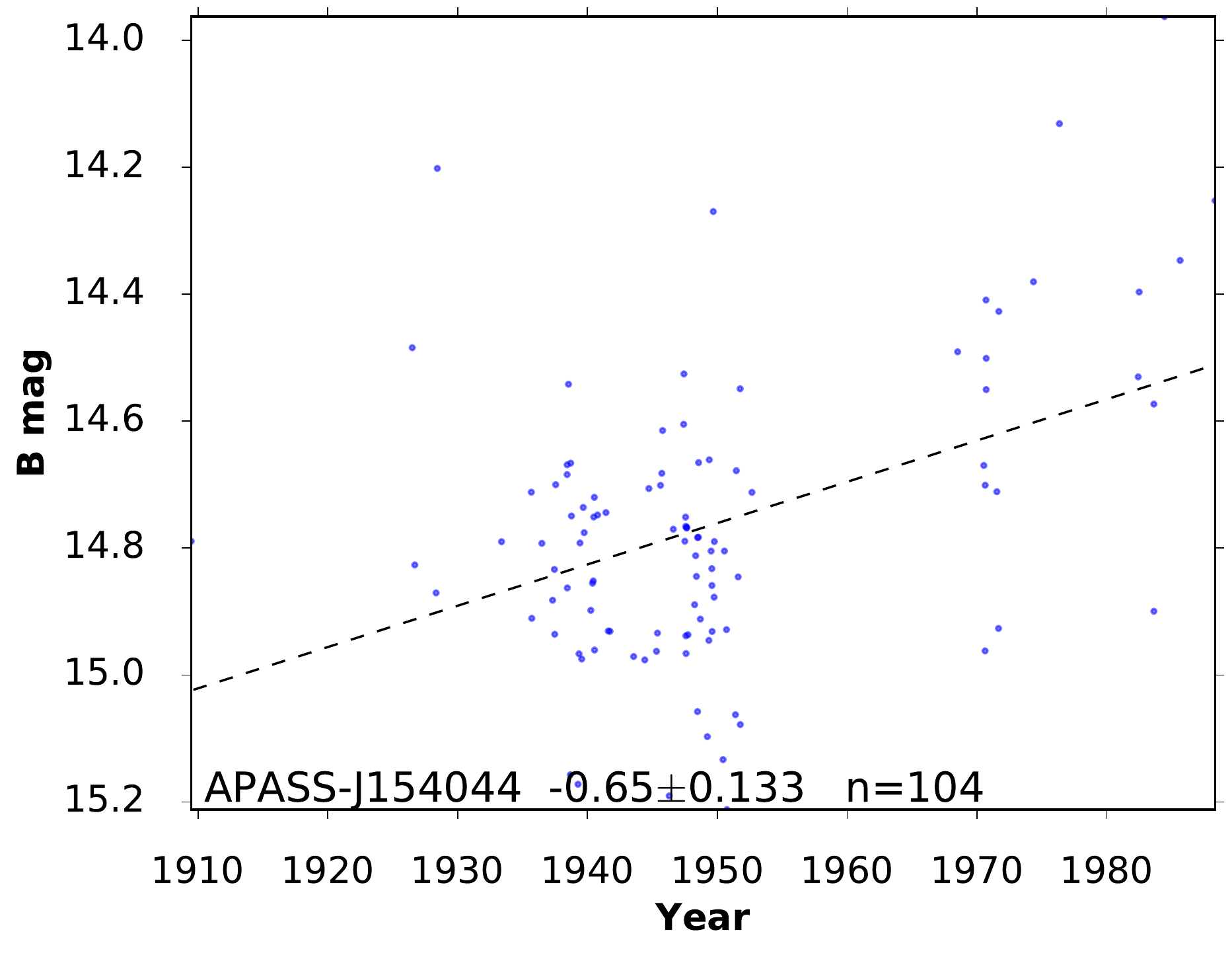}
\includegraphics[width=.5\linewidth]{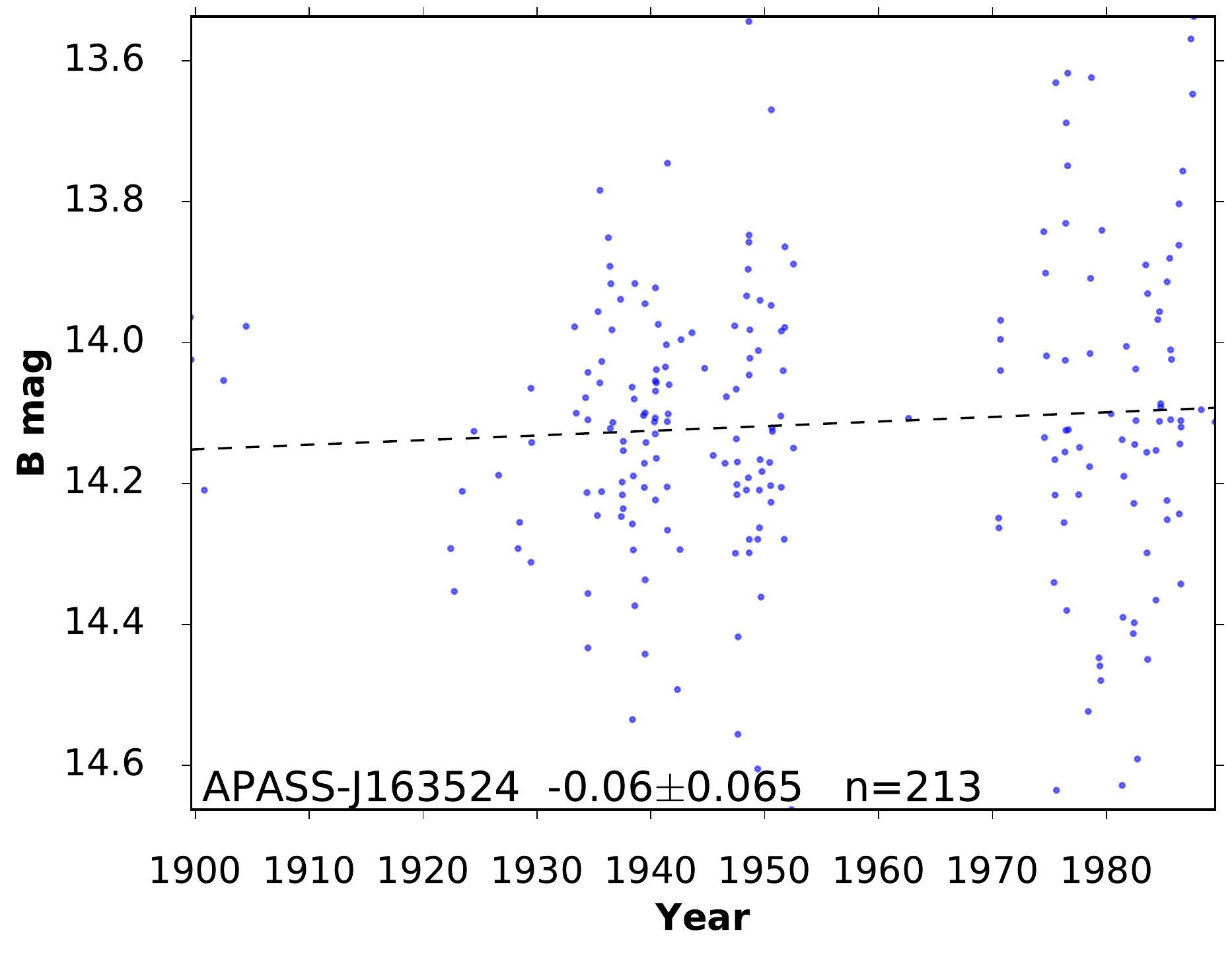}

\includegraphics[width=.5\linewidth]{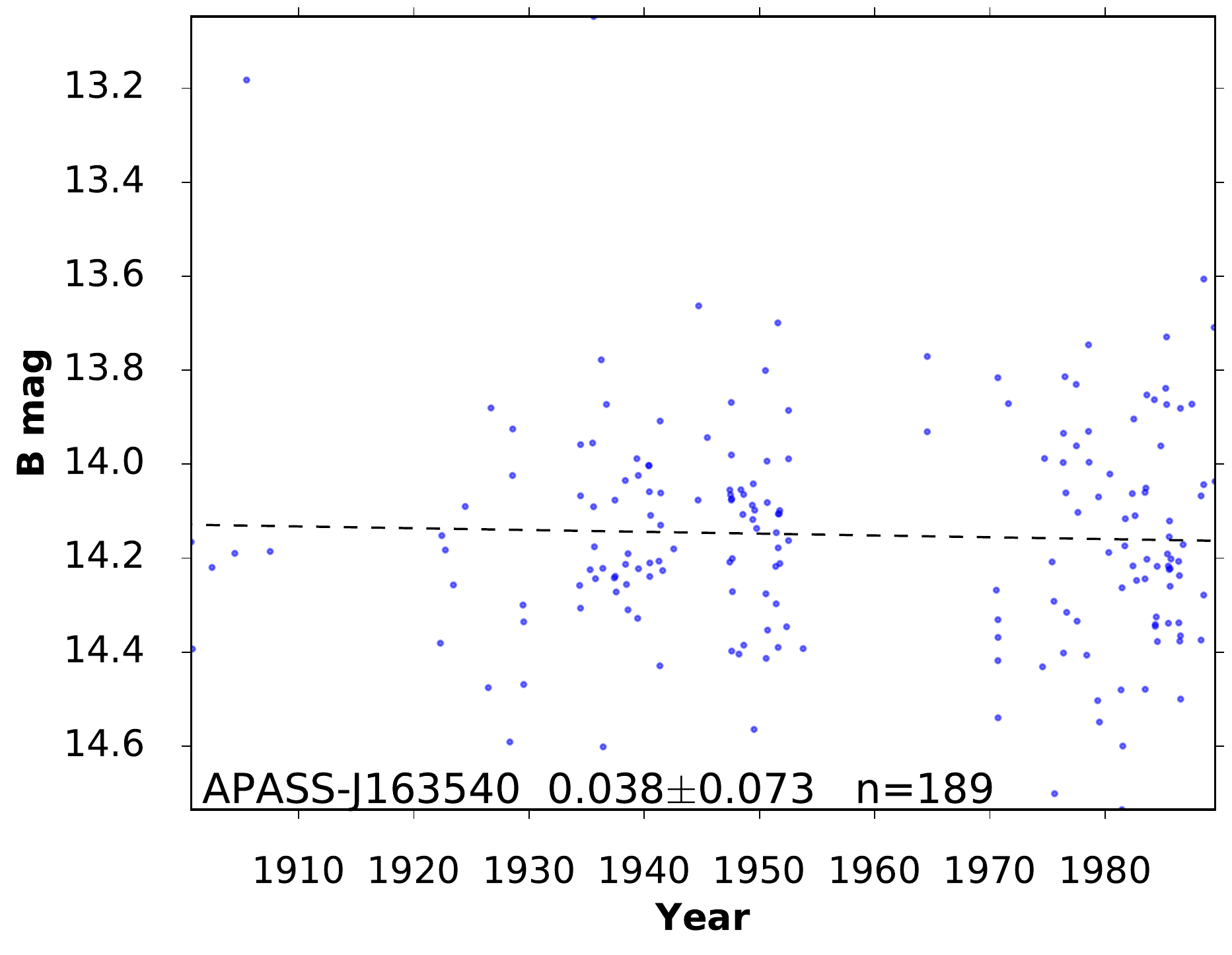}
\caption{\label{fig:land10}Photometry of Landolt standard stars with $n>100$ and all AFLAGS removed (continued).}
\end{figure*}

\end{document}